\documentclass[aps, prd, secnumarabic, showpacs, showkeys, amsmath, amssymb, 
twocolumn, 
nobibnotes, superscriptaddress]{revtex4-1}
\usepackage{graphicx}
\usepackage{dcolumn}   
\usepackage{cancel}
\usepackage{hyperref}
\hypersetup{colorlinks=true}
\usepackage{epstopdf}
\usepackage{multirow}

\newcommand{\nuebar}{$\overline{\nu}_{e}$}

\newcommand{\nGdResult}{$0.084~\pm~0.005$} 
\newcommand{\nHresult}{$0.071~\pm~0.011$} 
\newcommand{\combin}{0.082~\pm~0.004} 

\begin{document}
\title{New measurement of $\mathbf{\theta_{13}}$ via neutron capture on hydrogen at Daya Bay}

\newcommand{\ECUST}{\affiliation{Institute of Modern Physics, East China University of Science and Technology, Shanghai}}
\newcommand{\IHEP}{\affiliation{Institute~of~High~Energy~Physics, Beijing}}
\newcommand{\Wisconsin}{\affiliation{University~of~Wisconsin, Madison, Wisconsin 53706, USA}}
\newcommand{\Yale}{\affiliation{Department~of~Physics, Yale~University, New~Haven, Connecticut 06520, USA}}
\newcommand{\BNL}{\affiliation{Brookhaven~National~Laboratory, Upton, New York 11973, USA}}
\newcommand{\NTU}{\affiliation{Department of Physics, National~Taiwan~University, Taipei}}
\newcommand{\NUU}{\affiliation{National~United~University, Miao-Li}}
\newcommand{\Dubna}{\affiliation{Joint~Institute~for~Nuclear~Research, Dubna, Moscow~Region}}
\newcommand{\CalTech}{\affiliation{California~Institute~of~Technology, Pasadena, California 91125, USA}}
\newcommand{\CUHK}{\affiliation{Chinese~University~of~Hong~Kong, Hong~Kong}}
\newcommand{\NCTU}{\affiliation{Institute~of~Physics, National~Chiao-Tung~University, Hsinchu}}
\newcommand{\NJU}{\affiliation{Nanjing~University, Nanjing}}
\newcommand{\TsingHua}{\affiliation{Department~of~Engineering~Physics, Tsinghua~University, Beijing}}
\newcommand{\SZU}{\affiliation{Shenzhen~University, Shenzhen}}
\newcommand{\NCEPU}{\affiliation{North~China~Electric~Power~University, Beijing}}
\newcommand{\Siena}{\affiliation{Siena~College, Loudonville, New York  12211, USA}}
\newcommand{\IIT}{\affiliation{Department of Physics, Illinois~Institute~of~Technology, Chicago, Illinois  60616, USA}}
\newcommand{\LBNL}{\affiliation{Lawrence~Berkeley~National~Laboratory, Berkeley, California 94720, USA}}
\newcommand{\UIUC}{\affiliation{Department of Physics, University~of~Illinois~at~Urbana-Champaign, Urbana, Illinois 61801, USA}}
\newcommand{\RPI}{\affiliation{Department~of~Physics, Applied~Physics, and~Astronomy, Rensselaer~Polytechnic~Institute, Troy, New~York  12180, USA}}
\newcommand{\SJTU}{\affiliation{Department of Physics and Astronomy, Shanghai Jiao Tong University, Shanghai Laboratory for Particle Physics and Cosmology, Shanghai}}
\newcommand{\BNU}{\affiliation{Beijing~Normal~University, Beijing}}
\newcommand{\WM}{\affiliation{College~of~William~and~Mary, Williamsburg, Virginia  23187, USA}}
\newcommand{\Princeton}{\affiliation{Joseph Henry Laboratories, Princeton~University, Princeton, New~Jersey 08544, USA}}
\newcommand{\VirginiaTech}{\affiliation{Center for Neutrino Physics, Virginia~Tech, Blacksburg, Virginia  24061, USA}}
\newcommand{\CIAE}{\affiliation{China~Institute~of~Atomic~Energy, Beijing}}
\newcommand{\SDU}{\affiliation{Shandong~University, Jinan}}
\newcommand{\NanKai}{\affiliation{School of Physics, Nankai~University, Tianjin}}
\newcommand{\UC}{\affiliation{Department of Physics, University~of~Cincinnati, Cincinnati, Ohio 45221, USA}}
\newcommand{\DGUT}{\affiliation{Dongguan~University~of~Technology, Dongguan}}
\newcommand{\XJTU}{\affiliation{Xi'an Jiaotong University, Xi'an}}
\newcommand{\UCB}{\affiliation{Department of Physics, University~of~California, Berkeley, California  94720, USA}}
\newcommand{\HKU}{\affiliation{Department of Physics, The~University~of~Hong~Kong, Pokfulam, Hong~Kong}}
\newcommand{\UH}{\affiliation{Department of Physics, University~of~Houston, Houston, Texas  77204, USA}}
\newcommand{\Charles}{\affiliation{Charles~University, Faculty~of~Mathematics~and~Physics, Prague, Czech~Republic}} 
\newcommand{\USTC}{\affiliation{University~of~Science~and~Technology~of~China, Hefei}}
\newcommand{\TempleUniversity}{\affiliation{Department~of~Physics, College~of~Science~and~Technology, Temple~University, Philadelphia, Pennsylvania  19122, USA}}
\newcommand{\CUC}{\affiliation{Instituto de F\'isica, Pontificia Universidad Cat\'olica de Chile, Santiago, Chile}} 
\newcommand{\CGNPG}{\affiliation{China General Nuclear Power Group}}
\newcommand{\NUDT}{\affiliation{College of Electronic Science and Engineering, National University of Defense Technology, Changsha}} 
\newcommand{\IowaState}{\affiliation{Iowa~State~University, Ames, Iowa  50011, USA}}
\newcommand{\ZSU}{\affiliation{Sun Yat-Sen (Zhongshan) University, Guangzhou}}
\newcommand{\CQU}{\affiliation{Chongqing University, Chongqing}} 
\author{F.~P.~An}\ECUST
\author{A.~B.~Balantekin}\Wisconsin
\author{H.~R.~Band}\Yale
\author{M.~Bishai}\BNL
\author{S.~Blyth}\NTU\NUU
\author{D.~Cao}\NJU
\author{G.~F.~Cao}\IHEP
\author{J.~Cao}\IHEP
\author{W.~R.~Cen}\IHEP
\author{Y.~L.~Chan}\CUHK
% ...

\collaboration{The Daya Bay Collaboration}\noaffiliation

\date{\today}

\begin{abstract}
\noindent This article reports an improved independent measurement of neutrino mixing angle $\theta_{13}$ at the Daya Bay Reactor Neutrino Experiment. Electron antineutrinos were identified by inverse $\beta$-decays with the emitted neutron captured by hydrogen, yielding a data set with principally distinct uncertainties from that with neutrons captured by gadolinium. 
With the final two of eight antineutrino detectors installed, this study used 621 days of data including the previously reported 217-day data set with six detectors.  
The dominant statistical uncertainty was reduced by 49\%.  Intensive studies of the cosmogenic muon-induced $^9$Li and fast neutron backgrounds and the neutron-capture energy selection efficiency, resulted in a reduction of the systematic uncertainty by 26\%. 
The deficit in the detected number of antineutrinos at the far detectors relative to the expected number based on the near detectors yielded $\sin^2$2$\theta_{13}$ = \nHresult~in the three-neutrino-oscillation framework.  
The combination of this result with the gadolinium-capture result is also reported.  
\end{abstract}


\maketitle

\section{Introduction}
Precise measurements of neutrino mixing parameters are 
crucial to searches for {\it CP}-symmetry violation among neutral leptons and tests of neutrino oscillation theory.  In particular, the precision of neutrino mixing angle $\theta_{13}$ is of key significance in constraining the leptonic {\it CP} phase $\delta$~\cite{T2K_CPvsT13, NOvA, MINOS, LBNE}.  
Prior to 2012, many experimental efforts had been made to determine $\theta_{13}$~\cite{DC2012, T2K2011, MINOS2011, Chooz, PaloVerde, K2K}.  
The first measurement of $\theta_{13}$ with a significance greater than five standard deviations was reported by the Daya Bay Reactor Neutrino Experiment in 2012~\cite{DYB1}.
The most recent determinations of $\theta_{13}$ from reactor and accelerator experiments~\cite{nGd8AD, RENO, DC_nH16, DYB_nH, DC, T2K, MINOS_t13} are consistent.  

The three reactor antineutrino experiments, Double Chooz~\cite{DC_TDR}, RENO~\cite{RENO_TDR}, and Daya Bay~\cite{DYB_TDR}, currently provide the most precise measurements of the mixing angle.  They use gadolinium-doped liquid scintillator to identify electron antineutrinos through inverse $\beta$-decay (IBD) reactions (\nuebar$\ +\ p \rightarrow n+e^{+}$) with the neutron capturing on gadolinium ($n$Gd). A surrounding volume of undoped liquid scintillator improves the efficiency of detecting $\gamma$'s that escape from the doped volume, and has been used (in conjunction with the doped volume) by each of the three reactor experiments 
to determine $\sin^{2}$2$\theta_{13}$ independently through IBD reactions with the neutron captured by hydrogen ($n$H)~\cite{DC_nH, DYB_nH, RENO_nH, DC_nH16}.  The KamLAND experiment has used $n$H IBDs to measure the disappearance of reactor \nuebar~\cite{KamLAND} and the flux of geo-\nuebar~\cite{KamLANDgeo}.  The Super-Kamiokande experiment has used $n$H IBDs to search for relic supernova \nuebar~\cite{SuperK}.  
Future projects, including the medium-baseline reactor experiments JUNO~\cite{JUNO} and RENO-50~\cite{RENO50}, and LENA~\cite{LENA}, will also make use of $n$H IBDs.  
Techniques developed for this analysis may be useful for these future experiments.

The previous analysis of $n$H IBDs from Daya Bay~\cite{DYB_nH} is improved in this article with 3.6 times the number of detected IBDs and with reduced uncertainties of backgrounds and the neutron-capture energy selection efficiency.  
This statistically-independent measurement is also largely systematically independent from the $n$Gd-IBD analysis, and improves the overall uncertainty of $\sin^{2}$2$\theta_{13}$ from Daya Bay.  

This article is organized as follows.  Section~\ref{sec:Exp} describes the Daya Bay experiment.  The calculation of reactor antineutrino flux is described in Section~\ref{sec:reactor}.  Analysis of the data, including event reconstruction and IBD selection, is described in Section~\ref{sec:Selection}.  Section~\ref{sec:ACC} describes the accidental background, and Section~\ref{sec:CorrBkg} describes correlated backgrounds.  The IBD selection efficiency is discussed in Section~\ref{sec:DetEff}.  The fit for $\sin^{2}$2$\theta_{13}$ and its combination with the $n$Gd-IBD result are presented in Section~\ref{sec:Results}.  Section~\ref{sec:Future} briefly discusses the impact of the results and improvements expected in the future.

\section{Experiment}
\label{sec:Exp}
Located in Guangdong province, China, the Daya Bay experiment measures electron antineutrinos emitted from three pairs of nuclear reactors, each reactor nominally producing 2.9 GW of thermal power.  Inside the adjacent mountains, two $near$ experimental halls (EH1 and EH2) are located roughly 360-470~m from their nearest reactor, and one $far$ experimental hall (EH3) is located 1.52-1.93~km from all six reactors.  

Each far (near) experimental hall contains 4 (2) antineutrino detectors (ADs) submerged in a two-zone water Cherenkov detector~\cite{muon}.  
An inner and outer zone together provide each AD with $>$ 2.5~m of shielding against ambient radiation and spallation products of nearby cosmogenic muons.  
These inner and outer water shields (IWS and OWS) are independent cosmogenic muon detectors with 160 (121) and 224 (167) 20-cm photomultiplier tubes (PMTs), respectively, in the far (near) hall(s).  Detecting muons enables estimates of muon-induced backgrounds; particularly, $^9$Li/$^8$He decay products and spallation neutrons.  

\begin{figure}[b] 
\includegraphics[trim=220 130 450 60,clip,width=\columnwidth]{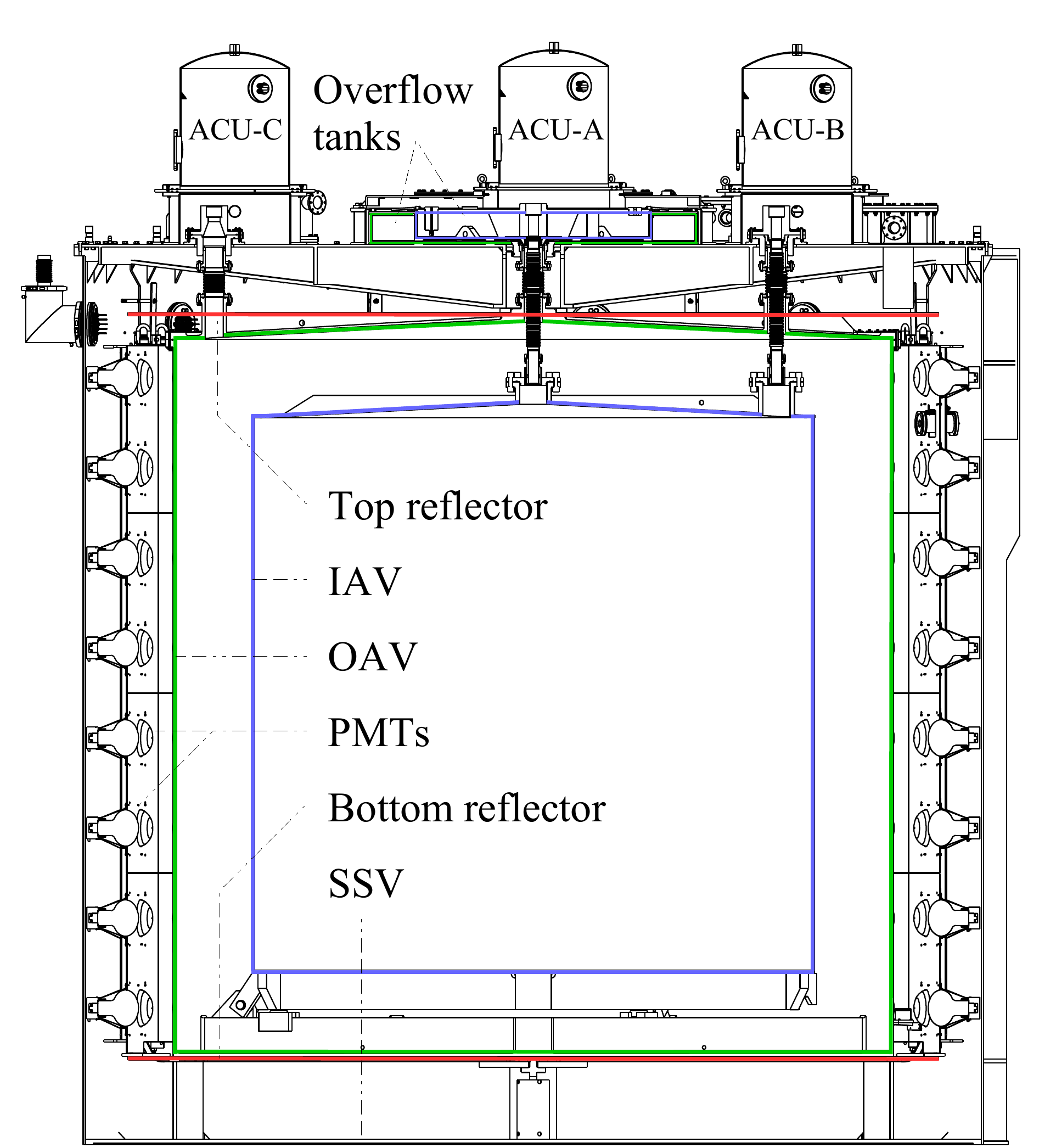}
\caption{Schematic of an antineutrino detector.  See the text for definitions.}
\label{fig:AD}
\end{figure}

The ADs were identically designed and consist of three nested, coaxial cylindrical vessels: an inner and outer acrylic vessel (IAV and OAV)~\cite{AVs} and an outermost stainless steel vessel (SSV), as shown in Fig.~\ref{fig:AD}.  
For future reference, the $z$ coordinate is defined by the central axis of the cylinders and the $r$ coordinate is measured radially from the central axis.  
The IAV is about 3~m in both height and diameter, and holds 20~tons of gadolinium-doped (0.1\% by mass) liquid scintillator (GdLS)~\cite{GdLS}.  The surrounding OAV is about 4~m in both height and diameter, and holds 22~tons of undoped liquid scintillator (LS) to improve the efficiency of detecting $\gamma$'s that escape from the GdLS.  The surrounding SSV is about 5~m in both height and diameter, and holds 36~tons of mineral oil (MO) to shield against radiation from the PMTs and the SSV.  

Each AD contains 192 20-cm PMTs arranged in 24 columns and 8 rings at a fixed radius ($r \approx$ 2.19~m) in the MO.  Reflectors were installed above and below the OAV to improve light collection.  
Three automated calibration units (ACUs) are affixed atop each AD and house LEDs and various radioactive sources for calibrating the energy scale and position reconstruction of events in the ADs~\cite{ACU}.  The ACUs deploy vertically at three radial positions: ACU-A at the center ($r$ = 0), ACU-B near the wall of the IAV ($r$ = 1.35~m), and ACU-C near the wall of the OAV ($r$ = 1.77~m).  

ADs were triggered, and recorded the time and charge information of each PMT channel, when the number of PMTs with pulses above threshold ($N_\mathrm{PMT}$) was $\geq45$ or when the integrated sum of PMT pulses from all 192 PMTs ($Q_\mathrm{sum}$) was $\gtrsim$ 65~photoelectrons.  
Both trigger thresholds corresponded to approximately 0.4~MeV and accepted 100\% of IBD positrons with $>$~0.7 MeV of deposited energy~\cite{DYB_NIM}.  
Water shields triggered independently under analogous conditions~\cite{muon}.  
The trigger criteria were tested within each cycle of an 80-MHz clock, and if satisfied, the subsequent 1~$\mu$s (and preceding 200~ns) of data from all channels were recorded.  
The physical interactions that caused a single trigger in a given detector are referred to as an ``event''.  
The time of an event is defined as the time of the trigger.  

More detailed descriptions of the detector hardware are given in Ref.~\cite{DYB_Det}.  

The analysis presented in this article determines $\sin^{2}$2$\theta_{13}$ by counting interactions of reactor antineutrinos in each AD in the one far and two near experimental halls.  
Antineutrinos were identified in both the GdLS and LS volumes via IBD reactions (\nuebar$\ +\ p \rightarrow n+e^{+}$) in which the positron carried away 99.4\% of the kinetic energy of the final state on average.  The positron deposited energy within $O$(1)~ns and then annihilated with an electron, usually producing two back-to-back 0.511-MeV $\gamma$'s (several percent of the positrons annihilated in flight such that the sum of $\gamma$ energies was greater than $2~\times$ 0.511 MeV).  The neutron thermalized and was captured primarily by Gd or H, releasing an approximately 8-MeV $\gamma$ cascade or a single 2.22-MeV $\gamma$, respectively.  The time from production to capture was typically tens to hundreds of microseconds.  The temporal coincidence of the prompt positron and delayed neutron-capture clearly distinguishes antineutrinos from single-event backgrounds.

\section{Reactor antineutrino flux}
\label{sec:reactor}
The expected number of IBDs in an AD was calculated as the product of the number of IBDs per target proton $\Phi$ and the efficiency-weighted number of target protons $N_\varepsilon$: 
\begin{equation}
\label{eq:predIBD}
\overline{N}_\mathrm{IBD}=\Phi N_\varepsilon.  
\end{equation}
The latter is discussed in Section~\ref{sec:DetEff} and the former is defined for the $d$-th AD as 
\begin{equation}
\label{eq:Phi}
\Phi_d \equiv \sum_{r=1}^6 \frac{1}{4\pi L_{dr}^2} \iint_{\{t_d\}}
\!\! \! \sigma_\nu(E) \ P_{\mathrm{\nu}}\left(\tfrac{L_{dr}}{E}\right) \  
\frac{d^{2}N_r(E, t)}{dE dt} dE dt,
\end{equation}
where $L_{dr}$ is the baseline distance between the $d$-th AD and the $r$-th reactor core, 
$\sigma_\nu(E)$ is the IBD reaction cross section of an antineutrino with energy $E$, 
$P_\nu(L_{dr}/E)$ is the neutrino survival probability, 
and 
$d^{2}N_r(E, t)/dE dt$ is the number of antineutrinos emitted from the $r$-th reactor at time $t$ with energy $E$, which is integrated over the periods of data acquisition for the $d$-th AD $\{t_d\}$.  

The baselines $L_{dr}$~\cite{supp} were measured with negligible uncertainty~\cite{DYB_Det}.  The cross section $\sigma_\nu$ was evaluated according to Ref.~\cite{IBDcs} using physical parameters from Ref.~\cite{PDG2014}.  
In the three-neutrino-oscillation framework, the survival probability of electron (anti)neutrinos is expressed as 
\begin{equation}
\label{eq:Psur}
\begin{aligned}
P_\nu = 1
& \left. -\cos^{4}\theta_{13}\sin^{2}2\theta_{12}\sin^{2}\Delta_{21}\right.\\
& \left. -\sin^{2}2\theta_{13}\cos^{2}\theta_{12}\sin^{2}\Delta_{31} \right.\\
& \left. -\sin^{2}2\theta_{13}\sin^{2}\theta_{12}\sin^{2}\Delta_{32} \right., 
\end{aligned}
\end{equation}
where $\Delta_{ij} \equiv 1.267\Delta m_{ij}^{2}L/E$, $E$ [MeV] is the energy of the neutrino at production, $L$ [m] is the distance between the points of production and interaction of the neutrino, and $\Delta m_{ij}^{2}$ [eV$^2$] is the difference between the squared masses of mass eigenstates $\nu_i$ and $\nu_j$.  
The values of $\sin^{2}$2$\theta_{12} =$ 0.846 $\pm$ 0.021, $\Delta m_{21}^{2} =$ (7.53 $\pm$ 0.18)$\times$10$^{-5}$ eV$^2$, and $\Delta m_{32}^{2} =$ (2.44 $\pm$ 0.06)$\times$10$^{-3}$ eV$^2$ (for the normal hierarchy) [$\Delta m_{32}^{2} =$ (2.52 $\pm$ 0.07)$\times$10$^{-3}$ eV$^2$ (for the inverted hierarchy)] were taken from Ref.~\cite{PDG2014}.  These uncertainties were found to have negligible impact on the fit of $\sin^2$2$\theta_\mathrm{13}$ and its uncertainty.  
The reactor antineutrino emission rate was calculated as 
\begin{equation}
\label{eq:flux}
\resizebox{\hsize}{!}{$\displaystyle\frac{d^{2}N(E, t)}{dE \ dt} = \frac{W_\mathrm{th}(t)}{\sum_{i}f_{i}(t)e_{i}}\sum_{i}f_{i}(t)S_{i}(E)c^\mathrm{ne}_{i}(E, t) + S_\mathrm{snf}(E, t),$}
\end{equation}
where the sum is over the four primary fissile isotopes: $^{235}$U, $^{239}$Pu, $^{238}$U, $^{241}$Pu.  The thermal power of the reactor $W_\mathrm{th}(t)$ and fraction of fissions due to the $i$-th isotope $f_i(t)$ were supplied by the nuclear power plant, the average thermal energies released per fission $e_i$ were from Ref.~\cite{fissionEnergy}, 
the antineutrino yields per fission $S_{i}(E)$ from $^{238}$U, and from $^{235}$U, $^{239}$Pu, and $^{241}$Pu, were from Ref.~\cite{Mueller} and Ref.~\cite{Huber}, respectively.  The correction to the energy spectrum due to nonequilibrium effects of long-lived fission fragments $c^\mathrm{ne}_i(E,t)$ followed Ref.~\cite{Mueller}.  The contribution from spent nuclear fuel $S_{\mathrm{snf}}(E,t)$ was estimated following Refs.~\cite{FengpengSNF,ZhouSNF}.  Combining the uncertainties of these components gave a 0.9\% reactor-uncorrelated uncertainty of predicted IBD rate associated with a single reactor~\cite{DYB_reactor}.  Additional information is given in Refs.~\cite{DYB_CPC,DYB_reactor}. 
These quantities were estimated on a daily basis, weighted by the fractional data acquisition time of each day for each experimental hall, and then summed for each week.  
The accumulated predicted spectra $dN_r(E)/dE$ are provided~\cite{supp}.

\section{Data analysis}
\label{sec:Selection}

The data used in this analysis were recorded beginning on December 24, 2011, with two ADs in EH1, one in EH2, and three in EH3.  Recording was paused on July 28, 2012, to install the final two ADs in EH2 and EH3.  On October 19, 2012, recording resumed with the full-design configuration of eight ADs.  The first measurement with $n$H IBDs at Daya Bay~\cite{DYB_nH} used the 217 days of data recorded in the six-AD configuration while this study uses an additional 404 days of data recorded in the full eight-AD configuration until November 27, 2013.  
Data acquisition maintained an operational efficiency of $>$ 97\% with occasional pauses for maintenance.  Excluding weekly calibrations, special calibrations, and problematic data, the data acquisition (DAQ) time $T_\mathrm{DAQ}$ of each AD is listed in Table~\ref{tab:IBDsummary}.  With the $n$H selection criteria described in the following sections, about 780000 IBDs were observed.  

\subsection{Calibration and reconstruction}
\label{sec:Reconstruction}
The gain [analog-to-digital converter channel/photoelectron] of each PMT channel was calibrated $in~situ$ by fitting the single photoelectron peak in the PMT dark noise spectrum.  The peak was fit with a Poisson-Gaussian convolution~\cite{DYB_Det}.  
This gain calibration was validated by an independent method using low-intensity LED pulses.  
The energy scale [MeV/photoelectron] of each AD was calibrated $in~situ$ with muon-induced spallation neutrons that captured on Gd throughout the GdLS volume.  
The two isotopes $^{157}$Gd and $^{155}$Gd, which release $\gamma$-cascades of 7.94 and 8.54~MeV, respectively, were fit with two Crystal Ball functions~\cite{CrystalBall} as described in Ref.~\cite{DYB_NIM}.  
This energy scale calibration was validated by an independent method using weekly deployments of the $^{60}$Co $\gamma$ source of ACU A at the center of each AD.  

The energy scale of an AD increased by 10-15\% from the center of the detector to the wall of the OAV, and changed by 2-6\% between the bottom and the top of the OAV, depending on the radial position.  
Corrections of energy scale as a function of position were applied with two-dimensional maps ($z \ vs.\  r$) derived from spallation neutron-captures on Gd in each AD.  The maps were extrapolated to the LS volume using spallation neutron-captures on H throughout the GdLS and LS volumes.  
The energy after correction is referred to as the ``reconstructed'' energy $E_\mathrm{rec}$.  
Using $n$H $\gamma$'s, the standard deviation of $E_\mathrm{rec}$ across an AD was observed to be less than 1.0\% for all ADs.  
The energy resolution was measured to be roughly $9\%/\sqrt{E_\mathrm{rec}\mathrm{[MeV]}}$ at the center of an AD.  It improved by around 20\% (relative) from the center to the wall of the OAV.  

A single position associated with each event in an AD was ``reconstructed'' using charge-pattern templates derived from Monte Carlo simulation~\cite{DYB_NIM}.  
From a simulation of positrons, the average distribution of charge from the 192 PMT channels, or the charge-pattern, was determined for each of 9600 voxels within the OAV, corresponding to 20, 20, and 24 divisions in $r^2$, $z$, and $\phi$ (where symmetry of $\phi$ was assumed to decrease statistical uncertainty).  
For each event, a $\chi^2$ was calculated for each voxel using the expected (from the templates) and observed charges from each PMT channel.  
The voxel with the smallest $\chi^2$ was selected and, with its nearest-neighbor voxels, interpolated to obtain the reconstructed position.  The reconstructed positions of prompt events (see Section~\ref{sec:EventSelect}) are shown in Figs.~\ref{Fig:NF}(e) and \ref{Fig:NF}(f), where a residual voxel grid is apparent.  
The resolution for a 2.2-MeV $\gamma$ was about 12~cm in the $r$-$\phi$ plane and 13~cm along the $z$ axis, in the LS volume.  The position resolution improved by more than 40\% from the center of a detector to the wall of an OAV, and varied within a few percent vertically.  
Using the $^{60}$Co $\gamma$ sources of the ACUs, the bias of the reconstruction was found to be about four times smaller than the resolution, near the wall of an OAV.

\subsection{IBD Candidate Selection}
\label{sec:EventSelect}
IBD candidates were selected from pairs of successive events in an AD, excluding those within predefined time ranges of detected muons to suppress muon-induced backgrounds.  
The IBD selection criteria for the $n$Gd-~\cite{nGd8AD} and $n$H-IBD analyses are listed in Table~\ref{tab:criteria}.  
First, AD events caused by spontaneous light emission from PMTs (PMT flashes) were removed as described in Section~\ref{sec:PMTflash}.  
Then, for the $n$H-IBD analysis, AD events were required to have $E_\mathrm{rec} >$ 1.5 MeV to exclude low-energy backgrounds (see Section~\ref{sec:lowE}).  
The AD events remaining after muon-event vetoes (see Section~\ref{sec:muonVetoes}) were grouped within a time window to identify double coincidences (see Section~\ref{sec:DCselection}).  
The resulting prompt and delayed events were required to have $E_\mathrm{rec} <$ 12~MeV and $E_\mathrm{rec}$ within three standard deviations of the fitted $n$H $\gamma$ energy in each AD, respectively.  
Finally, the distance between the reconstructed positions of the prompt and delayed events was required to be within 50~cm to suppress uncorrelated double coincidences (accidentals), which dominated the set of double coincidences (see Section~\ref{sec:distanceCut}).  The resulting number of $n$H-IBD candidates ($N_\mathrm{DC}$) is listed in Table~\ref{tab:IBDsummary} for each AD.  Details of the selection criteria are described below.  
\begin{table}[t]
\begin{center}
\begin{tabular}[c]{l | c  c} \hline\hline
 & $n$H & $n$Gd \\ \hline
AD trigger & \multicolumn{2}{c}{$N_\mathrm{PMT} \geq$ 45 \textsc{or} $Q_\mathrm{sum} \gtrsim$ 65 p.e.} \\
20-cm PMT flash & \multicolumn{2}{c}{$Ellipse <$ 1} \\
5-cm PMT flash & \multicolumn{2}{c}{$Q < 100$\ p.e.} \\
Low energy & $>$ 1.5 MeV & $>$ 0.7 MeV \\
Detector latency & \multicolumn{2}{c}{\ $< 2\ \mu s$} \\
WS muon ($\mu_\mathrm{WS}$) \textsc{[iws/ows]} & $N_\mathrm{PMT} > 12/15$ & $N_\mathrm{PMT} > 12/12$ \\
AD muon ($\mu_\mathrm{AD}$) & \multicolumn{2}{c}{$> 20$\ MeV} \\
Showering AD muon ($\mu_\mathrm{sh}$) & \multicolumn{2}{c}{$> 2.5$\ GeV} \\
WS muon veto & (0, 400)\ $\mu s$ & (-2, 600)\ $\mu s$ \\
AD muon veto & (0, 800)\ $\mu s$ & (-2, 1000)\ $\mu s$ \\
Showering AD muon veto & (0 $\mu s$, 1 $s$) & (-2 $\mu s$, 1 $s$) \\
Coincidence time ($t_c$) & [1, 400]\ $\mu s$ & [1, 200]\ $\mu s$ \\
Prompt energy ($E_p$) & \multicolumn{2}{c}{$<$ 12\ MeV} \\
Delayed energy ($E_d$) & peak $\pm$ 3$\sigma$ & [6, 12] MeV \\
Coincidence distance ($d_c$) & $<$ 50\ cm & NA \\ \hline \hline
\end{tabular}
\caption{IBD selection criteria for the $n$H and $n$Gd~\cite{nGd8AD} analyses.  See text for details.  }
\label{tab:criteria}
\end{center}
\end{table}
\begin{table*}[t]
\begin{tabular}[c]{l | c c | c c | c c c c}\hline\hline
 & EH1-AD1& EH1-AD2& EH2-AD1& EH2-AD2 & EH3-AD1 & EH3-AD2 & EH3-AD3 & EH3-AD4\\ \hline
$T_\mathrm{DAQ}$ [d] & 565.436 & 565.436 & 568.019 & 378.407 & 562.414 & 562.414 & 562.414 & 372.685 \\ 
$\varepsilon_{\mu}$ & 0.7949 & 0.7920 & 0.8334 & 0.8333 & 0.9814 & 0.9814 & 0.9812 & 0.9814 \\ 
$\varepsilon_m$ & 0.9844 & 0.9845 & 0.9846 & 0.9846 & 0.9844 & 0.9841 & 0.9839 & 0.9845 \\ 
$R_{\mu}$ [Hz] & 200.32 & 200.32 & 150.08 & 149.80 & 15.748 & 15.748 & 15.748 & 15.757 \\ 
$R_{s}$ [Hz]  & 20.111 & 19.979 & 19.699 & 19.702 & 19.651 & 20.020 & 20.182 & 19.649 \\ 
$N_\mathrm{DC}$ & 217613 & 219721 & 208606 & 136718 & 56880 & 56106 & 59230 & 38037 \\ 
$N_\mathrm{Acc}$ & 26240$\pm$49 & 25721$\pm$49 & 25422$\pm$43 & 16365$\pm$29 & 29920$\pm$19 & 30065$\pm$20 & 32179$\pm$21 & 20427$\pm$15 \\ 
$N_\mathrm{Cor}$ & 191373$\pm$473 & 194000$\pm$475 & 183184$\pm$465 & 120353$\pm$449 & 26960$\pm$246 & 26041$\pm$244 & 27051$\pm$251 & 17610$\pm$196 \\ 
$R_\mathrm{Acc}$ [d$^{-1}$]& $59.31\pm0.11$ & $58.34\pm0.11$ & $54.54\pm0.09$ & $52.71\pm0.09$ & $55.07\pm0.04$ & $55.35\pm0.04$ & $59.27\pm0.04$ & $56.73\pm0.04$ \\ 
$R_\mathrm{Li9}$ [d$^{-1}$] & \multicolumn{2}{c |}{$2.36\pm1.02$} & \multicolumn{2}{c |}{$1.73\pm0.75$} & \multicolumn{4}{c}{$0.19\pm0.09$} \\ 
$R_\mathrm{FastN}$ [d$^{-1}$] & \multicolumn{2}{c |}{$2.11\pm0.18$} & \multicolumn{2}{c |}{$1.81\pm0.17$} & \multicolumn{4}{c}{$0.16\pm0.03$} \\ 
$R_\mathrm{AmC}$ [d$^{-1}$] & $0.07\pm0.04$ & $0.07\pm0.04$ & $0.07\pm0.03$ & $0.07\pm0.03$ & $0.03\pm0.02$ & $0.03\pm0.02$ & $0.03\pm0.02$ & $0.02\pm0.01$ \\ 
$R_\mathrm{IBD}$ [d$^{-1}$] & $428.01\pm1.48$ & $435.49\pm1.49$ & $389.41\pm1.25$ & $384.03\pm1.42$ & $49.24\pm0.45$ & $47.56\pm0.45$ & $49.44\pm0.46$ & $48.54\pm0.55$ \\ 
$n$H/$n$Gd & $0.993\pm0.007$ & $0.993\pm0.007$ & $0.995\pm0.007$ & $0.995\pm0.008$ & $1.015\pm0.012$ & $0.981\pm0.012$ & $1.019\pm0.012$ & $0.987\pm0.014$ \\ \hline \hline
\end{tabular}
\caption{Data summary for each AD. All per-day rates are corrected with $\varepsilon_{\mu}\varepsilon_m$.  $T_\mathrm{DAQ}$ is the DAQ time, $\varepsilon_{\mu}$ is the muon-veto efficiency, $\varepsilon_m$ is the multiplicity selection efficiency, $R_{\mu}$ is the muon rate, $R_{s}$ is the rate of uncorrelated single events, $N_\mathrm{DC}$ is the number of double-coincidence (DC) events satisfying all IBD selection criteria, $N_\mathrm{Acc}$ is the number of accidental DCs, $N_\mathrm{Cor}$ is the number of correlated DCs, $R_\mathrm{Acc}$, $R_\mathrm{Li9}$, $R_\mathrm{FastN}$, $R_\mathrm{AmC}$, and $R_\mathrm{IBD}$ are the rates of accidental, fast neutron, $^9$Li/$^8$He, Am-C, and IBD (with all the backgrounds subtracted) DCs, and $n$H/$n$Gd is the ratio of the efficiency- and target proton-corrected $R_\mathrm{IBD}$ for the $n$H- and $n$Gd-IBD analyses.  The differences in $R_\mathrm{IBD}$ among ADs in the same near hall are due primarily to differences in baselines to the reactors, and secondarily to differences in target mass.  
}
\label{tab:IBDsummary}
\end{table*}

\subsubsection{PMT Flashes}
\label{sec:PMTflash}
PMT flashes are spontaneous emissions of light from the voltage divider of a PMT.  
AD events caused by a flash from any one of the 192 20-cm PMTs were removed by requiring $Ellipse \equiv \sqrt{Quadrant^2+(q_\mathrm{max}/0.45)^2} < 1$, where $q_\mathrm{max}$ is the largest fraction of an AD event's total charge in a single PMT and $Quadrant$ is defined as $Q_3/(Q_2+Q_4)$ in which $Q_i$ is the total charge in AD azimuthal quadrant $i$ and quadrant 1 is approximately centered on the PMT with $q_\mathrm{max}$.  
The efficiency of this criterion to select IBDs in the combined GdLS plus LS volume was estimated with Monte Carlo simulation~\cite{DYB_CPC} to be $>$ 99.99\%.  
Flashes from six 5-cm calibration PMTs~\cite{DYB_Det} near the top and bottom reflectors were simply removed by requiring the charge output from each 5-cm PMT to be $<$ 100 photoelectrons. 

\subsubsection{Low-energy Criterion}
\label{sec:lowE}
AD events were required to have $E_\mathrm{rec} >$ 1.5~MeV to exclude events caused by correlated $\beta$-$\alpha$ decays from the ${}^{214}$Bi-${}^{214}$Po-${}^{210}$Pb and ${}^{212}$Bi-${}^{212}$Po-${}^{208}$Pb decay chains, which originate from naturally-occurring ${}^{238}$U and ${}^{232}$Th, respectively.  
Due to the greater quenching associated with $\alpha$'s, the 8.78-MeV $\alpha$ from the latter chain resulted in an apparent energy of $E_\mathrm{rec} =$ 1.26~MeV and the 7.68-MeV $\alpha$ from the former chain resulted in $E_\mathrm{rec} =$ 1.00~MeV.  
Excluding these decays reduced the uncertainty of the total rate of accidentals by an order of magnitude.   
This criterion rejected about 10\% of IBD prompt events.  

\subsubsection{Muon-event Vetoes}
\label{sec:muonVetoes}
To suppress backgrounds from muon-induced spallation neutrons (Section~\ref{sec:FastN}) and long-lived spallation products such as $^9$Li and $^8$He (Section~\ref{sec:Li9}), an AD event was excluded from the analysis if it occurred within predefined veto time windows after cosmogenic muon events identified by the water shields or ADs.  
Muon events from the ADs, IWS, and OWS that occurred within the 2-$\mu$s detector latency were grouped together for the accounting of all events associated with cosmogenic muons.  The muon event with the earliest time in the group defined the start of the muon-veto time window. 

A muon event in a water shield, referred to as a $\mu_{\mathrm{WS}}$, was defined by requiring $N_\mathrm{PMT} >$ 12 (15) in the IWS (OWS).  The muon-detection efficiency of these selections was essentially 100\%, as determined relative to the ADs~\cite{muon}.  
The higher threshold of the OWS in the $n$H-IBD analysis (see Table~\ref{tab:criteria}) removed correlated triggers that sometimes occurred $O$(100)~$\mu s$ after an OWS event, due to electronics noise.  
These triggers were handled in the $n$Gd-IBD analysis by slightly modifying the multiple-coincidence criteria (see Section~\ref{sec:DCselection}) to have no overlap with a muon-veto time window.  

An AD event that was grouped with a $\mu_{\mathrm{WS}}$ and with 20~MeV $< E_\mathrm{rec} <$ 2.5~GeV was defined as an AD muon event $\mu_{\mathrm{AD}}$.  If instead, $E_\mathrm{rec} >$ 2.5~GeV, the event was defined as a showering AD muon event $\mu_{\mathrm{sh}}$.  The total rate of muon events measured by each AD ($R_{\mu}$) is listed in Table~\ref{tab:IBDsummary}. 

An AD event was excluded if it occurred within a veto time window of 400~$\mu$s, 800~$\mu$s, or 1~s after a $\mu_{\mathrm{WS}}$, $\mu_{\mathrm{AD}}$, or $\mu_{\mathrm{sh}}$, respectively.  
The fraction of DAQ time remaining for IBD analysis after implementing these offline muon-vetoes is reported as $\varepsilon_\mu$ in Table~\ref{tab:IBDsummary}, with typical values of 79\%, 83\% and 98\% in EH1, EH2, and EH3, respectively.  
\begin{figure*}[t]
\includegraphics[angle=0,width=\textwidth]{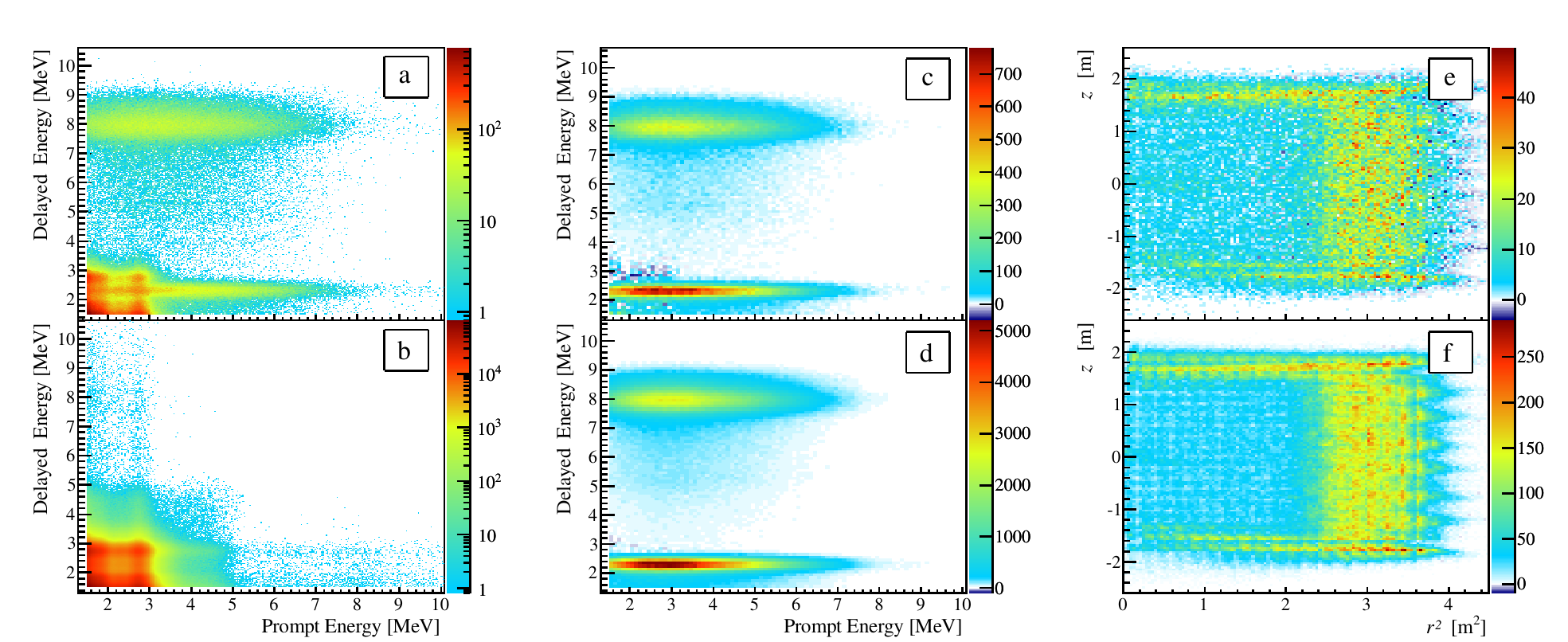}
\caption{(a) Distribution of prompt {\it vs.}\ delayed reconstructed energy for all double coincidences with a maximum 50-cm separation in all near-hall ADs,
(b) total (621-day) accidental background sample (ABS) for all ADs in the near halls, (c) and (d) are the distributions of prompt {\it vs.} delayed reconstructed energy after subtracting the total ABS for the far and near halls, respectively, (e) and (f) are the reconstructed positions of all prompt events after subtracting the total ABS for the far and near halls, respectively.  The sparser distribution of events at the bottoms of the ADs is due to the presence of acrylic supports below the IAV.}
\label{Fig:NF}
\end{figure*}

\subsubsection{Coincidence Time}
\label{sec:DCselection}
Correlated AD events were selected using a coincidence time window of [1, 400]~$\mu$s, which is about two times longer than the mean capture time of an IBD neutron on hydrogen in LS and about 14 times longer than that in GdLS.  
Given the data recording window of 1~$\mu$s, coincidence windows were initiated 1~$\mu$s after an event to ensure distinction of prompt and delayed events.  Lone events are denoted as ``singles'' and were used to construct accidental background samples (see Section~\ref{sec:ACC}).  Only pairs of events, denoted as double coincidences (DCs), were used to select IBD candidates.  
If more than two events occurred within [1, 400]~$\mu$s, they were excluded from further analysis.  
In addition, if the first, or prompt, event of a DC occurred within [1, 400]~$\mu$s of a preceding event or muon-veto time window, the DC was excluded (this requirement was also applied to singles).  
The fraction of DAQ time remaining for IBD analysis after implementing these multiple-coincidence criteria was about 98.4\% for each AD, and is reported as $\varepsilon_m$ in Table~\ref{tab:IBDsummary}.
This multiplicity selection efficiency was derived as described in Ref.~\cite{acc}, and calculated using the duration of the coincidence time window $T_c =$~399~$\mu$s and the rate of uncorrelated single events $R_s$ (which are uncorrelated events that satisfy the criteria of Sections~\ref{sec:PMTflash}-\ref{sec:muonVetoes}; not singles, which exclude events involved in coincidences): 
\begin{equation}
\label{eq:mult}
\begin{aligned}
\varepsilon_m = ~ e^{-R_sT_c} 
& \left\{ e^{-(R_{s}+R_{\mu})T_{c}} \right.\\
& \left.+ \frac{R_{\mu}}{R_{s}+R_{\mu}}[1-e^{-(R_{s}+R_{\mu})T_{c}}] \right.\\
& \left.+ \frac{R_s}{R_s+R_{\mu}}e^{-R_{\mu}T_c}[1-e^{-(R_s+R_{\mu})T_c}] \right.\\
& \left.- \frac{R_s}{2R_s+R_{\mu}}e^{-R_{\mu}T_c}[1-e^{-(2R_s+R_{\mu})T_c}] \right\} . 
\end{aligned}
\end{equation}

\subsubsection{Coincidence Distance}
\label{sec:distanceCut}
The set of DCs was largely comprised of accidental coincidences (whose positions are uncorrelated throughout the detector); therefore, the spatial separation of the reconstructed positions of the prompt and delayed events $d_c$ was required to be within 50~cm.  
This rejected 98\% of the accidental coincidences at a loss of 25\% of the IBDs.

Figure~\ref{Fig:NF}(a) shows the distribution of prompt energy {\it vs.}\ delayed energy for all DCs in all near-hall ADs after applying the coincidence-distance criterion.  
Bands for both the 2.22-MeV $n$H and 8-MeV $n$Gd delayed events are apparent, with a large background of low-energy DCs around the $n$H band.  
The clusters around 1.5 and 2.7 MeV are due to $\gamma$'s from ${}^{40}$K and ${}^{208}$Tl decays, respectively.  
The bands between these clusters are dominated by the decay products of ${}^{238}$U.  
The measured $n$H $\gamma$ energy was around 2.33~MeV, which is offset from the true value of 2.22~MeV because of nonlinear detector response and the calibration of the energy scale with $n$Gd events.  
The $n$H delayed events were fit as described in Section~\ref{sec:Ecuts}, providing a mean and a standard deviation $\sigma$ for each AD.  Delayed events were required to have $E_\mathrm{rec}$ within 3$\sigma$ ($\approx$0.42~MeV) of the mean for each AD, which excludes $\gamma$'s from ${}^{40}$K.  
The accidental background from the remaining decays was effectively removed by the subtraction described in Section~\ref{sec:ACC}.  
Backgrounds from correlated events are described in Section~\ref{sec:CorrBkg}.  
Efficiencies and uncertainties of the IBD selection criteria are described in Section~\ref{sec:DetEff}.

\section{Accidental Background}
\label{sec:ACC}
Accidental backgrounds were caused by two uncorrelated AD events that satisfied the IBD selection criteria, and were almost entirely due to natural radioactivity in the materials around and within the detectors.  The energy spectra of this background are visible below 3~MeV in Fig.~\ref{Fig:NF}(a).  Because the delayed event of an $n$H IBD is from a 2.22-MeV $\gamma$, which overlaps with this background spectrum, the accidental background rate relative to the IBD rate was typically $>$ 50 times that of the $n$Gd-IBD analysis for the ADs in EH3 after applying all IBD selection criteria.  

The background was estimated for each AD within each run (about 2-3 days) by constructing accidental background samples (ABSs) from the singles in a run.  
An ABS was constructed by sequentially pairing singles from the first half of the run with singles from the second half of the run.  
The resulting ABS consisted of $N_\mathrm{ABS-tot}$ accidentals, and after applying the remaining IBD selection criteria (distance and energy), the ABS consisted of $N_\mathrm{ABS-cut}$ accidentals.  
To obtain the true value for $\varepsilon_\mathrm{ABS} \equiv N_\mathrm{ABS-cut}$/$N_\mathrm{ABS-tot}$, the calculation of $\varepsilon_\mathrm{ABS}$ was repeated for several hundred different pairing sequences of the singles, and the Gaussian mean of the resulting distribution was taken as $\varepsilon_\mathrm{ABS}$.  
Figure~\ref{Fig:NF}(a) shows the energy distribution of all DCs (621 days) of all near-hall ADs without applying the delayed-energy criterion, and Fig.~\ref{Fig:NF}(b) shows the energy distribution of the total ABS (621 days) of all near-hall ADs after applying the coincidence-distance criterion.  
Each ABS was scaled to a calculated number of accidentals ($N_\mathrm{Acc}$) and subtracted from its corresponding number of DCs ($N_\mathrm{DC}$) to obtain the energy distribution of correlated DCs ($N_\mathrm{Cor}$), which are dominantly due to IBDs: 
\begin{equation}
\label{eq:sub}
\begin{aligned}
& \left. N_\mathrm{Cor} = N_\mathrm{DC} - N_\mathrm{Acc}, \right. \\
& \left. N_\mathrm{Acc} \equiv R_\mathrm{Acc} \cdot T_\mathrm{DAQ}\cdot\varepsilon_{\mu}\cdot\varepsilon_\mathrm{ABS}
, \right.
\end{aligned}
\end{equation}
where $T_\mathrm{DAQ}$ is the DAQ time, $\varepsilon_{\mu}$ is the muon-veto efficiency, and $R_\mathrm{Acc}$ is the rate of coincidence of uncorrelated single events, which is expressed as~\cite{acc} 
\begin{equation}
\label{eq:Acc}
\begin{aligned}
R_\mathrm{Acc} 
& \left. = R_s^2 \cdot T_c \cdot \varepsilon_m \right. \\
& \left. \approx R_s \cdot e^{-R_s T_c} \cdot R_sT_c e^{-R_sT_c}, \right.
\end{aligned}
\end{equation}
where $R_s$ is the rate of uncorrelated single events and $\varepsilon_m$ is the multiplicity selection efficiency, both defined in Eq.~(\ref{eq:mult}).  
The approximation of Eq.~(\ref{eq:mult}) used in the second line ($\varepsilon_m \approx e^{-R_s T_c} \cdot e^{-R_sT_c}$)  results from the condition $(R_s + R_{\mu})T_c \ll 1$ and is valid to within 0.1\% for $T_c =$ 399~$\mu$s, $R_s =$ 20~Hz, and the $R_\mu$ in Table~\ref{tab:IBDsummary}.  
This approximation is not used in this analysis, but is shown here to illustrate the basic components of the calculation: $e^{-R_sT_c}$ is the probability of no prior event within $T_c$ and 
$R_sT_c e^{-R_sT_c}$ is the probability of a subsequent event within $T_c$.  
$N_\mathrm{DC}$, $N_\mathrm{Acc}$, and $N_\mathrm{Cor}$ are listed for each AD in Table~\ref{tab:IBDsummary}.  

\begin{figure}[!b]
\begin{center}
\includegraphics[angle=0,width=\columnwidth]{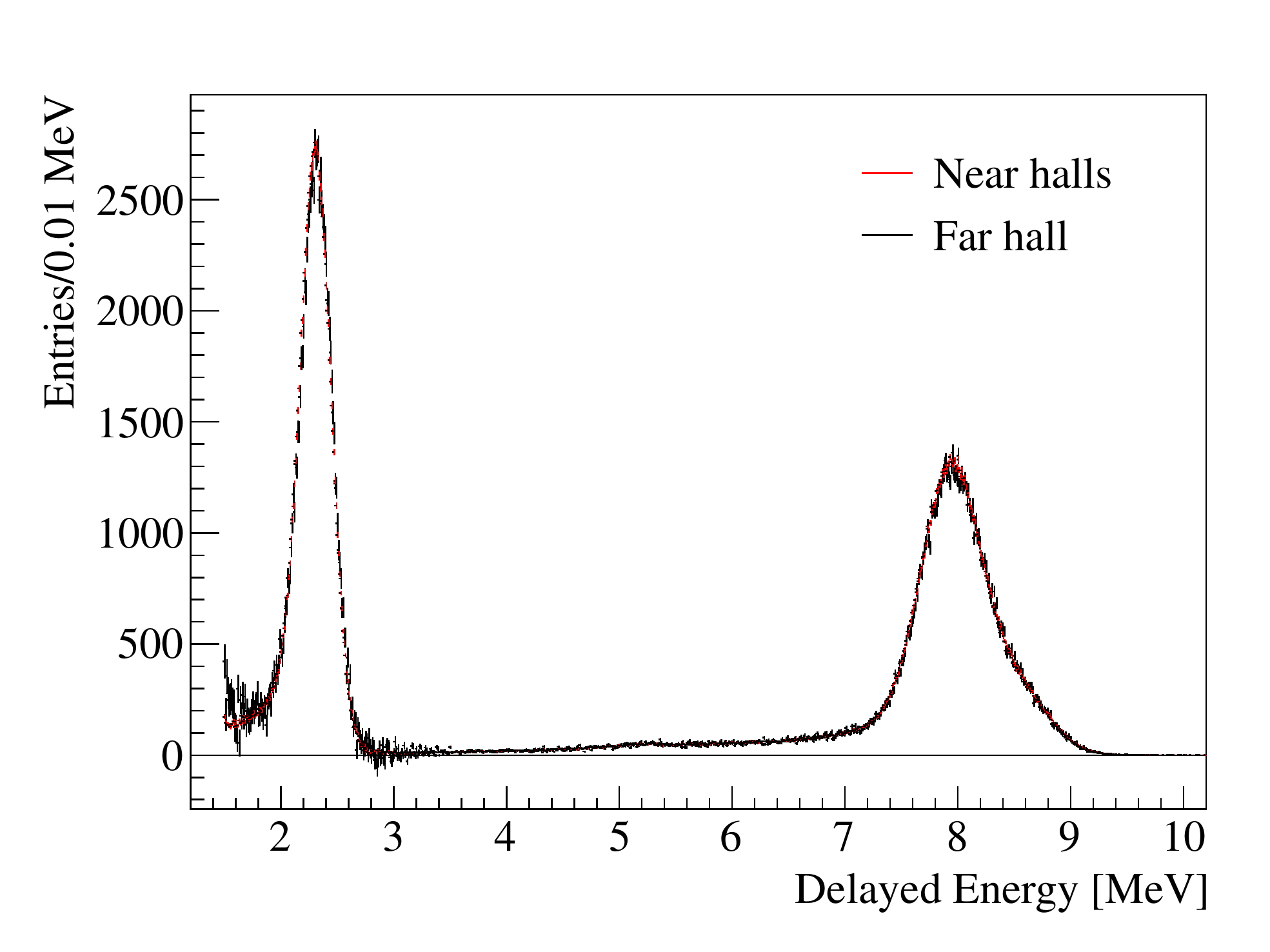}
\caption{Reconstructed delayed-energy distribution after subtracting the accidental background for all four ADs in EH3 (black) and all four ADs in EH1 and EH2 (red), where the far-hall spectrum has been normalized to the area of the near-halls spectrum.  (621 days of data.)}
\label{Fig:Ed}
\end{center}
\end{figure}

Figure~\ref{Fig:NF}(d) shows the energy distribution of $N_\mathrm{Cor}$ for all near-hall ADs [Fig.~\ref{Fig:NF}(c) shows $N_\mathrm{Cor}$ for the far-hall ADs], where the $n$H $\gamma$ peak is cleanly isolated from the accidental-dominated DCs shown in Fig.~\ref{Fig:NF}(a).  
The effectiveness of the subtraction is also illustrated in Fig.~\ref{Fig:Ed}, which shows the energy spectrum of the delayed events after subtracting the accidental background for all near-hall ADs and all far-hall ADs.  
Both the $n$H and $n$Gd peaks are very similar between the two groups of ADs.  
Figures~\ref{Fig:NF}(e) and \ref{Fig:NF}(f) show the reconstructed positions of $N_\mathrm{Cor}$ prompt events after subtracting the accidental background for all ADs in the far and near halls, respectively.  The positions are generally uniform throughout the GdLS and LS volumes.  The smaller concentration of events in the GdLS volume ($r^2 < 2.40$~m$^2$ and $|z| < 1.50$~m) is due to the greater fraction of neutron-captures on Gd.  

The uncertainty of $N_\mathrm{Cor}$ is composed of the statistical uncertainties of $N_\mathrm{DC}$ and $N_\mathrm{ABS-cut}$, 
and the systematic uncertainty of $R_{Acc}$, which is determined by the uncertainty of $R_s$.  The uncertainty from $\varepsilon_m$ was negligible: using Eq.~(\ref{eq:mult}) and $R_s=40$~Hz, $R_{\mu}=200$~Hz, and $T_c=399~\mu$s (which are conditions similar to those in EH1), $d\varepsilon_m = 3\times10^{-6} d{R_{\mu}} - 6\times10^{-3} d{R_s}$.  
By taking the average over a run, the induced systematic uncertainty from variations in $R_s$ or $R_{\mu}$ was negligible.  

\begin{figure}[!b]
\begin{center}
\includegraphics[angle=0,width=\columnwidth]{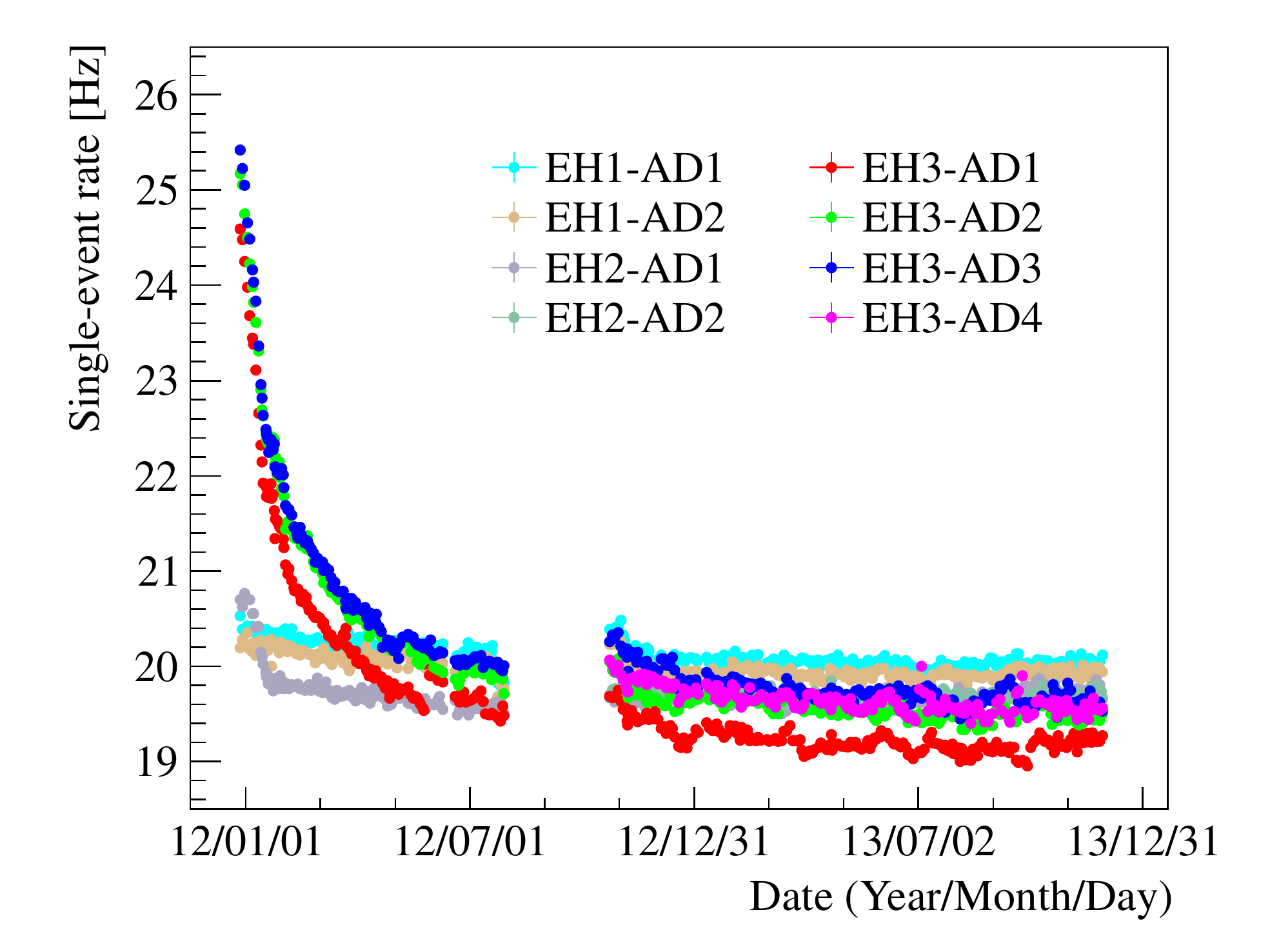}
\caption{Rate of uncorrelated single events {\it vs.} time for each AD.  Rates stabilized several months after water shields were filled (EH3 was filled less than a month before data-recording began).}
\label{Fig:SingleRate}
\end{center}
\end{figure}

$R_s$ was estimated as the average of an upper and lower limit.  The upper limit was derived from the total number of AD events after applying muon-event vetoes.  These events were dominantly singles but included DCs and multiple coincidences.  The lower limit was derived from the number of singles plus DCs that did not satisfy the coincidence-distance criterion.  These DCs were dominantly accidentals. 
Time-averaged values of $R_{s}$ are listed in Table~\ref{tab:IBDsummary} for each AD.   
The difference between the two limits was assigned as the systematic uncertainty of $R_{s}$ and propagated to $R_{Acc}$, resulting in 0.18\%, 0.16\% and 0.05\% uncertainties of the accidental rate in EH1, EH2, and EH3, respectively.  The larger uncertainties for the near halls are due to the higher rates of IBD reactions from reactor antineutrinos, which enlarged the upper limits.  
Figure~\ref{Fig:SingleRate} shows $R_{s}$ as a function of time for each AD, where a downward trend began after the water shields were filled.  
During the first few weeks, $R_s$ decreased by $<$~0.05 Hz per day for near-hall ADs and by $<$~0.08 Hz per day for far-hall ADs.  The near-hall water shields were filled earlier and so, the AD rates stabilized earlier.  Considering that $R_s$ was calculated every 2-3~days, the uncertainty introduced to $R_{Acc}$ by these trends was estimated to be $< 2\times10^{-5}$, which is more than an order of magnitude smaller than the uncertainty in EH3.  
There were also instantaneous increases of $R_s$, which were caused by muon-generated spallation products such as $^9$Li and $^8$He (Section~\ref{sec:Li9}), and spallation neutrons (Section~\ref{sec:FastN}).  From a study of $R_s$ {\it vs.} time after muon-event vetoes, the impact of these products was estimated to be negligible.  

\begin{figure}[!b]
\begin{center}
\includegraphics[angle=0,width=\columnwidth]{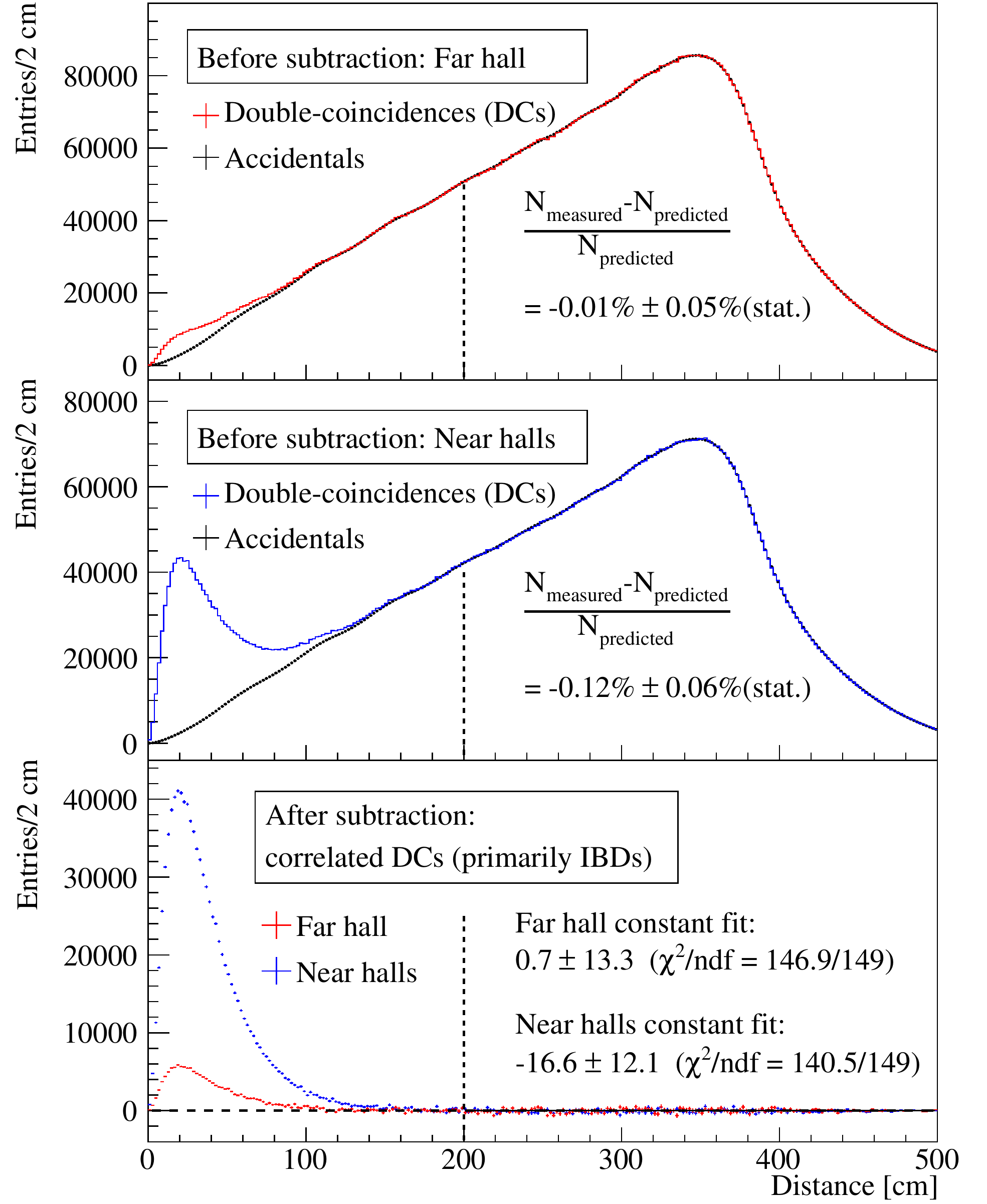}
\caption{Distributions of the distance between the prompt and delayed events of all measured double coincidences and of the predicted accidental backgrounds (black points) in the far hall (top panel) and near halls (middle panel).  The bottom panel shows the distance distributions after subtracting the accidental backgrounds for the near halls (blue) and the far hall (red).  See the text for details.  
}
\label{Fig:DistIBD}
\end{center}
\end{figure}

Two methods were used to validate the subtraction of the accidental background.  
The first method used the distribution of distance between the prompt and delayed events, which was dominated by accidental coincidences at large separations.  After subtracting the accidental background, the resulting number of correlated DCs with large separations is expected to be zero.  
Figure~\ref{Fig:DistIBD} shows the distribution of distance between the prompt and delayed events for DCs, accidentals, and correlated DCs.  
The two upper panels of Fig.~\ref{Fig:DistIBD} contain calculations of the relative difference between the measured number of double coincidences ($N_\mathrm{DC}$) and the predicted number of accidentals ($N_\mathrm{Acc}$), beyond 200~cm.  These differences are consistent with zero with respect to their statistical uncertainties.  
A constant fit in the bottom panel also shows that the distribution of selected $n$H IBD candidates ($N_\mathrm{Cor}$) beyond 200~cm is consistent with an expected fraction of about 0.05\%, which was determined from Monte Carlo simulation.  This fraction corresponds to an expected fitted constant of about 0 (3) entries/2~cm for the far (near) hall(s).  

\begin{figure}[!b]
\begin{center}
\includegraphics[angle=0,width=\columnwidth]{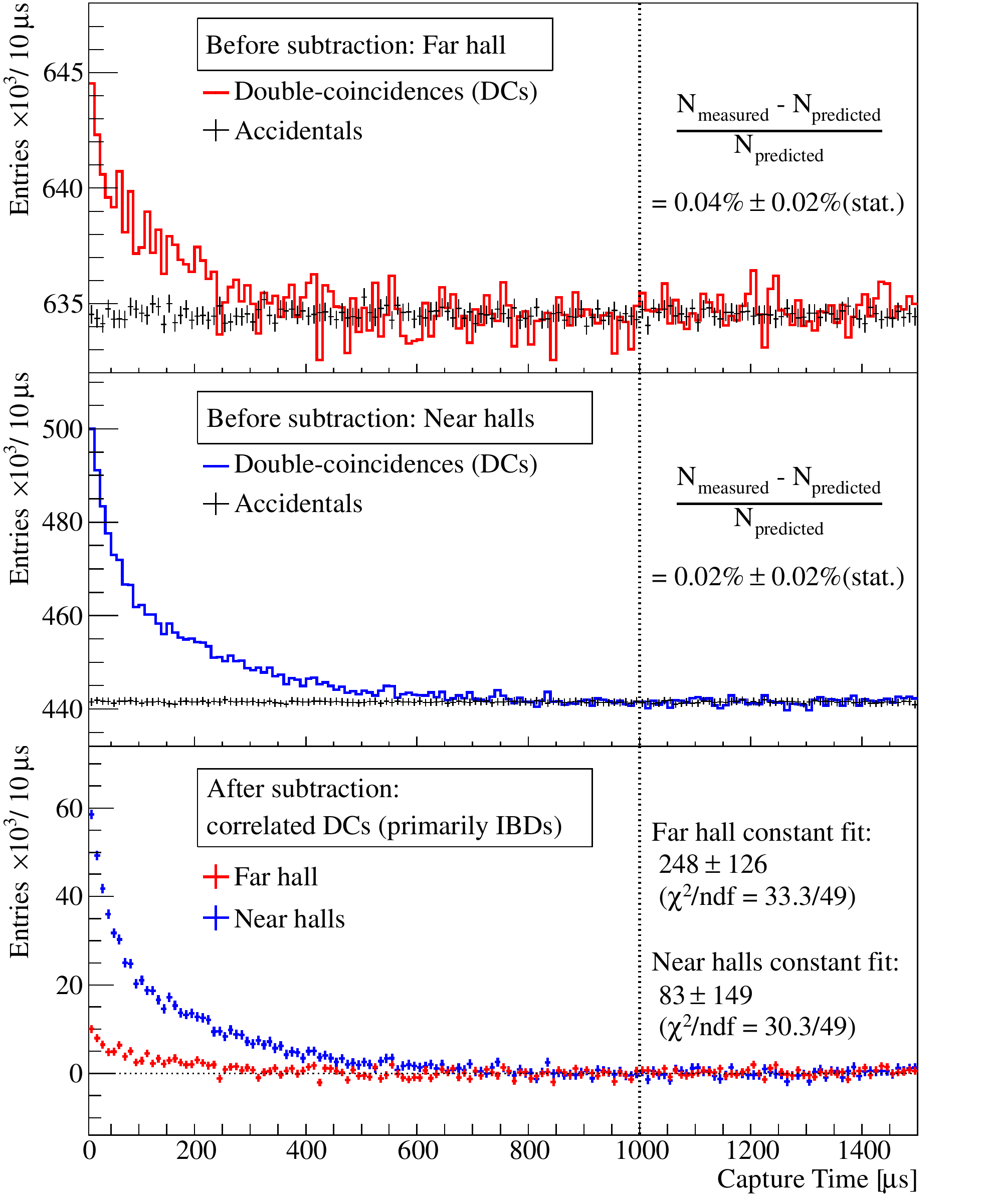}
\caption{Distributions of the time between the prompt and delayed events of all measured double coincidences and of the predicted accidental backgrounds (black points) in the far hall (top panel) and near halls (middle panel).  The bottom panel shows the time distributions after subtracting the accidental backgrounds for the near halls (blue) and the far hall (red).  See the text for details.  }
\label{Fig:Time}
\end{center}
\end{figure}

The subtraction of the accidental background was also validated by the distribution of time between prompt and delayed events. 
Figure~\ref{Fig:Time} shows the distribution of time between prompt and delayed events for DCs, accidentals, and correlated DCs.  
The two upper panels of Fig.~\ref{Fig:Time} contain calculations of the relative difference between the measured number of double coincidences ($N_\mathrm{DC}$) and the predicted number of accidentals ($N_\mathrm{Acc}$), beyond 1000~$\mu$s.  These differences are consistent with zero with respect to their statistical uncertainties.  
A constant fit in the bottom panel also shows that the distribution of selected $n$H IBD candidates ($N_\mathrm{Cor}$) beyond 1000~$\mu$s is consistent with an expected fraction of 0.7\%, which was determined from Monte Carlo simulation.  This fraction corresponds to an expected fitted constant of about 16 (110) entries/10~$\mu$s for the far (near) hall(s).

\section{Correlated Backgrounds}
\label{sec:CorrBkg}
After the accidental background was subtracted to obtain $N_\mathrm{Cor}$, correlated backgrounds were subtracted to obtain the number of measured $n$H IBDs ($N_\mathrm{IBD}$).  In EH3 (EH1), $N_\mathrm{IBD}/N_\mathrm{Cor} =$~99.2\% (99.0\%).  Correlated backgrounds consist of prompt and delayed events that originate from a single source and satisfy the IBD selection criteria.  These backgrounds are primarily from cosmogenic muon-induced $^9$Li/$^8$He isotopes and spallation neutrons, and neutrons from $^{241}$Am-$^{13}$C calibration sources interacting with the SSV and its appendages.  
The $^{13}$C($\alpha$,n)$^{16}$O background is less significant for the $n$H-IBD analysis than for the $n$Gd-IBD analysis and is briefly discussed.

\subsection{ $^9$Li/$^8$He Background}
\label{sec:Li9}
Cosmogenic muons and their spallation products interact with the $^{12}$C in organic liquid scintillators, producing neutrons and isotopes via hadronic or electromagnetic processes.  Among the muon-induced isotopes, $^9$Li and $^8$He $\beta^-$-decay to neutron-unstable excited states, immediately followed by the ejection of a neutron.  These $\beta^-$-neutron decays mimic the prompt and delayed events of IBD reactions.
The lifetimes of $^9$Li and $^8$He (257.2 and 171.7~ms, respectively) are longer than the muon-veto windows for a $\mu_{\mathrm{WS}}$ or $\mu_{\mathrm{AD}}$ (see Section~\ref{sec:EventSelect}), leading to a contamination of the IBD candidate sample. The temporal relation between $^9$Li/$^8$He decays and prior detected muons was used to estimate the collective yield of the $^9$Li and $^8$He background $N_\mathrm{Li/He}$ in each hall. The distribution of the time between the prompt event of a DC and its preceding muon was described by a formula following Ref.~\cite{Li}: 
\begin{equation}
\label{eq:Equ_Li9He8}
\begin{aligned}
  N(t)~=~ & N_\mathrm{Li/He} \left[ r\cdot\lambda_\mathrm{Li}\cdot e^{-\lambda_\mathrm{Li}t}+(1-r)\cdot\lambda_\mathrm{He}\cdot e^{-\lambda_\mathrm{He}t} \right] \\
  & + N_\mathrm{BB}\cdot \lambda_\mathrm{BB}\cdot e^{-\lambda_\mathrm{BB}t} \\
  & + N_{\mathrm{DC}\cancel{\mu}}\cdot R_{\mu}\cdot e^{-R_{\mu}t},
\end{aligned}
\end{equation}
where $\lambda_\mathrm{isotope} \equiv R_{\mu}+1/\tau_\mathrm{isotope}$ and $\tau_\mathrm{isotope}$ is the lifetime of the specific isotope ($^9$Li or $^8$He), $R_{\mu}$ is the muon rate (which depends on the muon selection criteria), $r$ is the fraction of $^9$Li decays among the $^9$Li and $^8$He decays, $\lambda_\mathrm{BB} \equiv R_{\mu}+2/\tau_\mathrm{B}$, and $N_\mathrm{BB}$ and $N_{\mathrm{DC}\cancel{\mu}}$ are the numbers of $^{12}$B-$^{12}$B coincidences and all other double coincidences (excluding those from cosmogenically-produced isotopes), respectively.  

The beta-decaying isotope $^{12}$B was produced with a yield about one order of magnitude greater than the combined yield of $^9$Li and $^8$He.  With its lifetime of $\tau_\mathrm{B}~\approx~$29~ms, double coincidences of $^{12}$B-$^{12}$B originating from a single muon contributed mainly within the first $\approx$50~ms of the time since the preceding muon distribution. 
The fitted value of $N_\mathrm{Li/He}$ changed by up to 10\% when including and excluding the $^{12}$B term.  

The fraction of $^9$Li $r$ could not be reliably determined because of the similar lifetimes of $^9$Li and $^8$He.  Measurements of $^9$Li and $^8$He yields from Ref.~\cite{KamLANDisotopes} indicate that $r$ should be between roughly 85\% and 100\% at Daya Bay.  Varying $r$ in this range resulted in a 4\% variation in the fitted value of $N_\mathrm{Li/He}$ in all halls.  

To obtain a better estimate of $N_\mathrm{Li/He}$, $N_{\mathrm{DC}\cancel{\mu}}$ was reduced by suppressing accidentals among the double coincidences.  This was done by augmenting the prompt-energy criterion from 1.5 $<E_{p}<$ 12.0 MeV to 3.5 $<E_{p}<$ 12.0 MeV.  The measured number of $^9$Li/$^8$He was corrected with the efficiency of the augmented criterion with respect to the nominal criterion.  
This ratio was determined to be 74\% by averaging measurements from all three halls with visible muon energy $E^\mathrm{vis}_{\mu}>$ 1~GeV ($E^\mathrm{vis}_{\mu}$ is the detected energy that was deposited by a muon traversing the detector).  
The weighted average of the three measurements had a statistical uncertainty of 5\%. 
The systematic uncertainty was estimated as the difference between the average and a Monte Carlo simulation, and therefore accounted for backgrounds in the measurements.  
The simulation used $\beta$ spectra of $^9$Li/$^8$He decays calculated as those in Ref.~\cite{DYB3}.  The resulting prompt-energy spectrum from the simulation is shown in Fig.~\ref{Fig:DCbkgSpectra}, where it has been normalized to $N_\mathrm{Li/He}$.  
The difference in efficiency between the measurement and simulation was 6\%, giving a total uncertainty of 8\% for the efficiency of the augmented $E_p$ criterion.  

The $^9$Li/$^8$He background was determined for three ranges of $E^\mathrm{vis}_{\mu}$: 0.02-1.0~GeV, 1.0-2.5~GeV, and $>$~2.5~GeV.  
The highest energy range was defined as such because it identically defines a $\mu_\mathrm{sh}$, which was vetoed for 1~s (see Table~\ref{tab:criteria}) and therefore contributed only $O$(1)\% of the total $^9$Li/$^8$He background. 
The lowest energy range was defined as such because it could not provide a reliable fit of $^9$Li/$^8$He due to its 
higher $R_{\mu}$ and lower signal-to-background ratio: relative to the middle energy range, $R_{\mu}$ was 14 (11) times greater and $N_\mathrm{Li/He} / N_{\mathrm{DC}\cancel{\mu}}$ was about 5 (10) times lower, in EH1 (EH3).  

To obtain a more reliable estimate of the $^9$Li/$^8$He background of the lowest energy range, $R_{\mu}$ was reduced and the signal-to-background ratio was increased, by isolating the muons that produced $^9$Li/$^8$He.  Under the assumption that the isotopes were produced along with neutrons, every $\mu_\mathrm{AD}$ without a subsequent neutron (defined as a 1.8-12~MeV event within 20-200 $\mu$s) was excluded.  
The measured number of $^9$Li/$^8$He was corrected with the efficiency of this altered $\mu_\mathrm{AD}$ definition with respect to the nominal definition.  
Since this ratio could not be determined for the lowest energy range, the ratio for the middle energy range was used as a proxy. 
This ratio was determined to be about 69\% (66\%) in the far (near) hall(s).  
A 100\% uncertainty was assigned to the background for the lowest energy range, corresponding to a 1$\sigma$ lower bound of 35\% (33\%) for the efficiency of the altered $\mu_\mathrm{AD}$ definition in the far (near) hall(s).

The number of $^9$Li/$^8$He for both the middle and lowest energy ranges in EH1 and EH2 were determined with the combined data samples of EH1 and EH2.  
The energy spectra of muons in EH1 and EH2 are similar~\cite{muon} such that their yields of $^9$Li/$^8$He per muon are expected to agree to $O$(1)\%~\cite{KhalchukovPowerLaw,HagnerPowerLaw}.  
The $E^\mathrm{vis}_{\mu}$ spectra of the two near halls were observed to differ in scale by about 7\%.  
This was due to a 7\% lower average gain of the high-charge range~\cite{FEE2} of the EH2 electronics.  
After scaling the $E^\mathrm{vis}_{\mu}$ spectrum of EH2 by 7\%, the difference between the near-hall spectra was $O$(1)\% across the two energy ranges.  
This scaling introduced a negligible uncertainty to the fitted number of $^9$Li/$^8$He.  
The muon rate $R_{\mu}$ of the combined fit was fixed to the DC-weighted average of the measured muon rates in the two near halls. 
Combining the uncertainties of the measured muon rates (0.3\%) and numbers of DCs (1\%), the weighted average had a 0.2\% uncertainty.  
This 0.2\% uncertainty of $R_{\mu}$ corresponded to a 27\% change in the number of $^9$Li/$^8$He via Eq.~(\ref{eq:Equ_Li9He8}) for the middle energy range. 
The 0.2\% uncertainty had a negligible impact on the lowest energy range because its muon rate was reduced as described above. 
The fitted number of $^9$Li/$^8$He was divided among the near halls according to their measured muon rates (after scaling EH2) multiplied by their DAQ times.  

\begin{figure}[!h]
\centering
\includegraphics[width=\columnwidth]{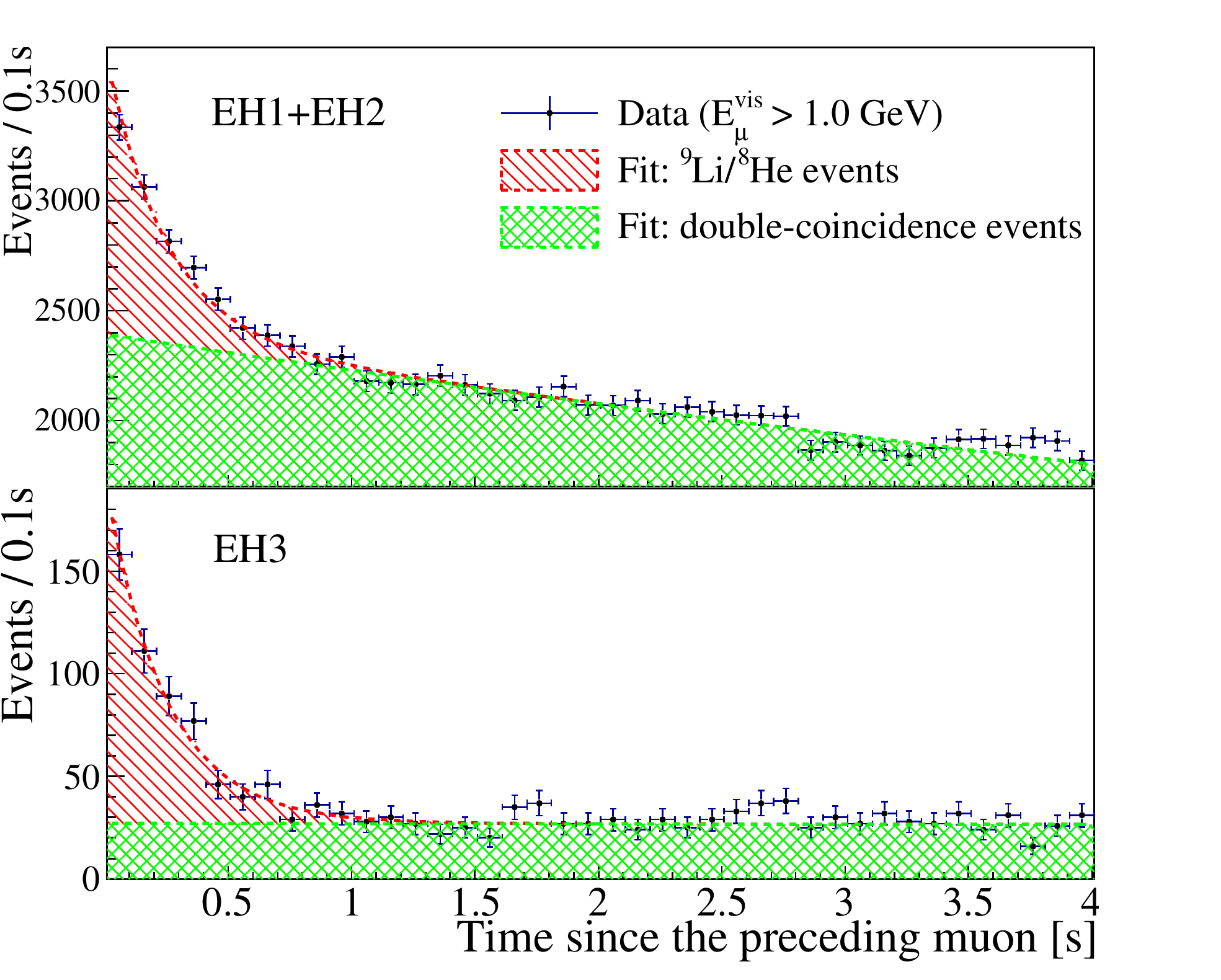}
\caption{Examples of fits of the time since the preceding muon in EH1+EH2 (top) and EH3 (bottom) for $E^\mathrm{vis}_{\mu}>$ 1.0~GeV. The green area is the noncosmogenic double-coincidence component and the red area is the $^9$Li/$^8$He component.}
\label{fig:Li9fit}
\end{figure}

Examples of fits to the time since the preceding muon without the $^{12}$B term for $E^\mathrm{vis}_{\mu}>$ 1.0~GeV are shown in Fig.~\ref{fig:Li9fit}.  
The green areas represent the noncosmogenic DCs and the red areas represent the $^9$Li/$^8$He DCs.  For presentation purposes, the plots use wider bins than the actual fits. 

Uncertainties were from statistics, the $^9$Li fraction $r$, the contribution of $^{12}$B, the augmented $E_p$ selection criterion, the altered $\mu_\mathrm{AD}$ definition for the lowest energy range, and binning effects. 
The total uncertainty of the $^9$Li/$^8$He background was determined from the combination of all components of uncertainty, and was dominated by statistical uncertainty.  

Table~\ref{tab:IBDsummary} lists the determined rate of background DCs due to $^9$Li/$^8$He in each hall.  The rate was calculated by dividing the estimated $N_\mathrm{Li/He}$ by $T_\mathrm{DAQ}\varepsilon_{\mu}\varepsilon_m$ and correcting for the efficiencies of the altered definitions of the $E_{p}$ and $\mu_\mathrm{AD}$ criteria. 

Since the $n$H- and $n$Gd-IBD analyses used different data samples, and the efficiencies were determined with distinct methods, there was no correlation of the $^9$Li/$^8$He background determinations between the $n$H- and $n$Gd-IBD analyses.

\subsection{Fast-neutron Background}
\label{sec:FastN}
In addition to producing radioactive isotopes such as $^9$Li and $^8$He, cosmogenic muon interactions can generate energetic neutrons via spallation.  
Upon reaching an AD, a neutron may scatter off a proton and then capture on hydrogen, creating a prompt-delayed coincidence.  
Given the high efficiency with which $\mu_{\mathrm{WS}}$'s are detected, the neutrons that contribute to this background predominantly originate from the rock surrounding an OWS.  
Because the LS volume is more accessible than the GdLS volume to the externally-produced neutrons, this background is significantly higher than for the $n$Gd-IBD analysis.  

\begin{figure}[!b]
\begin{center}
\includegraphics[width=\columnwidth]{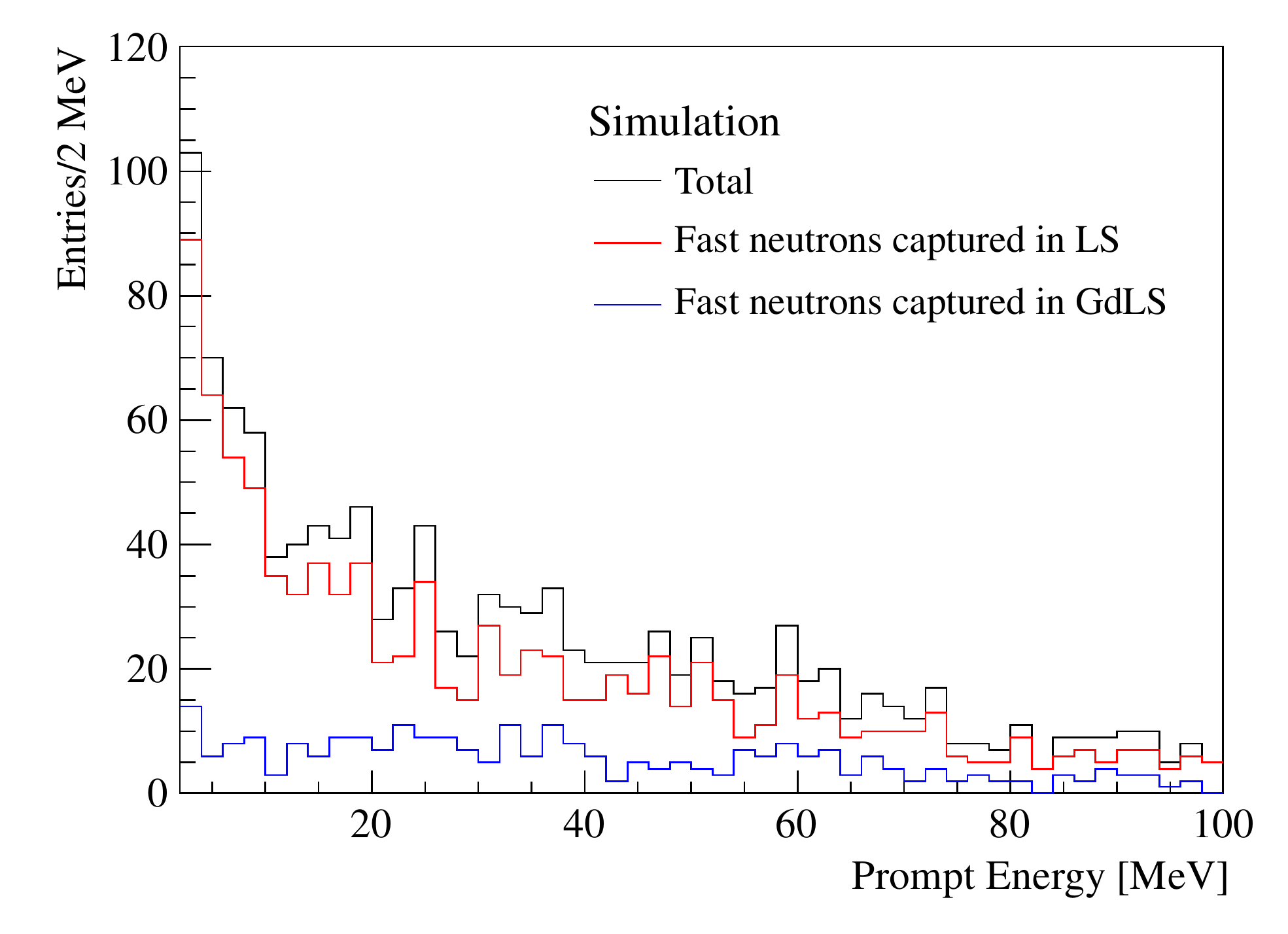}
\caption{Simulated prompt-recoil-energy spectra of spallation neutrons produced in the IWS or OWS by cosmogenic muons.  See text for details.}
\label{Fig:fastN_MC}
\end{center}
\end{figure}

A Monte Carlo simulation of neutrons induced from muons in the water shields was performed. 
An empirical parameterization for neutron production from cosmogenic muons~\cite{Spallation_Neutron_Production} and the estimated average muon energy in an experimental hall~\cite{muon} were used to generate the initial kinetic energy and zenith angle distributions of the neutrons.  
The resulting prompt-energy spectra of the simulated neutrons are shown in Fig.~\ref{Fig:fastN_MC}. 
The increase of events with decrease of energy in the LS volume is due to the lesser containment of the recoil protons within the LS volume: the protons that recoil from fast neutrons that capture in the LS volume are closer to the boundary of the scintillating region compared to those associated with fast neutrons that capture in the GdLS volume, and thus, are more likely to deposit less energy in scintillator.  

\begin{figure}[!b]
\begin{center}
\includegraphics[width=\columnwidth]{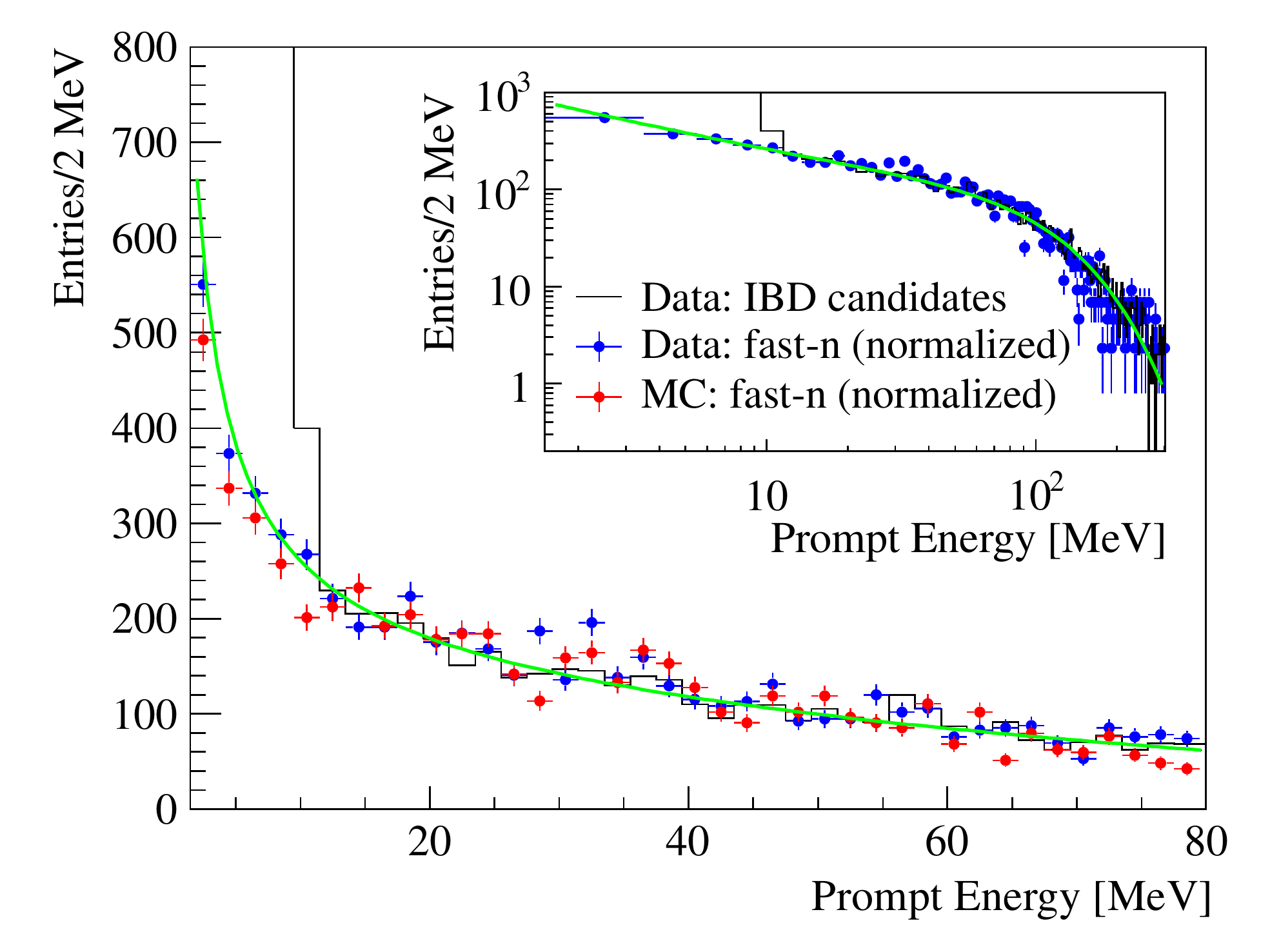}
\caption{Reconstructed prompt recoil-energy spectra of fast spallation neutrons from IBD candidates in EH1 with the upper $E_p$ limit removed (black line), OWS-identified muons (blue points), and simulation (red points). 
The latter two spectra were normalized to the area of the extended IBD spectrum.
The green curve is the fit of the extended IBD spectrum using a first-order power law (see the text).  
The inset is a log-log scaling of the plot.  }
\label{Fig:fastNdata}
\end{center}
\end{figure}

To determine the fast neutron background spectrum, a sample of spallation neutrons was obtained by slightly modifying the nominal IBD selection criteria: the upper prompt-energy criterion was removed and the OWS muon-event veto was excluded.  
Muons identified with the IWS were still vetoed to avoid confusing a spallation neutron with a muon event in an AD.  
In addition, the prompt event was required to occur within 300~ns after an OWS-identified muon and the delayed event at least 15~$\mu s$ after the muon to exclude muon decays.  The OWS-identified muon events were required to occur at least 1200~$\mu$s after any muon events in an AD or the IWS.  
The prompt recoil-energy spectrum of the OWS-identified spallation neutrons from EH1 is shown in Fig.~\ref{Fig:fastNdata}.
Figure~\ref{Fig:fastNdata} also shows the prompt-energy distribution of IBD candidates without the upper $E_{p}$ criterion and the spectrum obtained from the simulation.  The OWS-identified and simulated spectra were normalized to the IBD candidates above 12~MeV, revealing consistent shapes.  

Plotting the prompt recoil-energy spectrum in a log-log scale (see the inset of Fig.~\ref{Fig:fastNdata}) shows that the low-energy portion of the spectrum up to several tens of MeV is consistent with a power law [$N(E)=N_0E^{-a}$], while there is a distinct energy-dependence at higher energies.  The entire spectrum could be fit after adding one degree of freedom to the power law; namely, extending the exponent to have a first-order dependence on energy: 
\begin{equation}
\label{eq:powerLaw}
N(E) = N_0 \left(\frac{E}{E_0}\right)^{-a-\frac{E}{E_0}}.  
\end{equation}
The fit of Eq.~(\ref{eq:powerLaw}) resulted in a $\chi^2$ per degree of freedom close to 1 for each hall.  Bin widths of 2~MeV were selected for the near halls based on the stability of the fit parameters and the $\chi^2$ per degree of freedom.  Due to the lower statistics of EH3, the corresponding bin width was 3~MeV.  The value of $a$ was consistent among the three halls, yielding an average of 0.690~$\pm$~0.023.  The value of $E_0$ averaged to (101.7~$\pm$~2.1)~MeV for the near halls and was (110~$\pm$~10)~MeV for the far hall.  

The fast neutron background and its uncertainty were both estimated as in Ref.~\cite{nGd8AD}.  The background was estimated as the number of events within the nominal prompt-energy selection window (1.5 $< E_\mathrm{rec} <$ 12~MeV) in the normalized OWS-identified spectrum of each hall.  The spectrum was normalized to the extended IBD spectrum from all the ADs in a hall, between 12 and 300~MeV.  
The systematic uncertainty was estimated using both the OWS-identified and extended IBD spectra.  First, the extended IBD spectrum of each hall was fit between 12 and 300~MeV with the power law given in Eq.~(\ref{eq:powerLaw}).  Then, the difference was taken between the integral of the function and the number of events in the normalized OWS-identified spectrum, with $E_\mathrm{rec}$ between 1.5 and 12~MeV.  The largest relative difference among the three halls (6\% in EH3) was assigned to be the systematic uncertainty for each hall.  In addition, each hall had a distinct fit uncertainty, which included the statistical uncertainty and was about 6\%, 7\%, and 18\% for EH1, EH2, and EH3, respectively.  
The results are listed for each experimental hall in Table~\ref{tab:IBDsummary}.  

There was no significant correlation between the $n$H- and $n$Gd-IBD fast neutron analyses because of their different selection criteria and independent event samples.

\subsection{Am-C Calibration Source Background}
One of the calibration sources deployed from each of the three ACUs atop an AD was an $^{241}$Am-$^{13}$C neutron source with a detected rate of 0.7~Hz~\cite{AmC}.  Neutrons from these sources could inelastically scatter with the nuclei in the surrounding steel (SSV, ACU enclosures, {\it etc.}) and then capture on Fe, Cr, Ni, or Mn within the steel, producing $\gamma$'s that could enter the scintillating regions and satisfy the IBD selection criteria. 
During the pause to install the final two ADs in the summer of 2012, two of the three Am-C sources were removed (from ACU-B and -C) from each AD in EH3, reducing this background in EH3 by about 40\% relative to the previous analysis~\cite{DYB_nH}. 

\begin{figure}[!t]
\begin{center}
\includegraphics[width=\columnwidth]{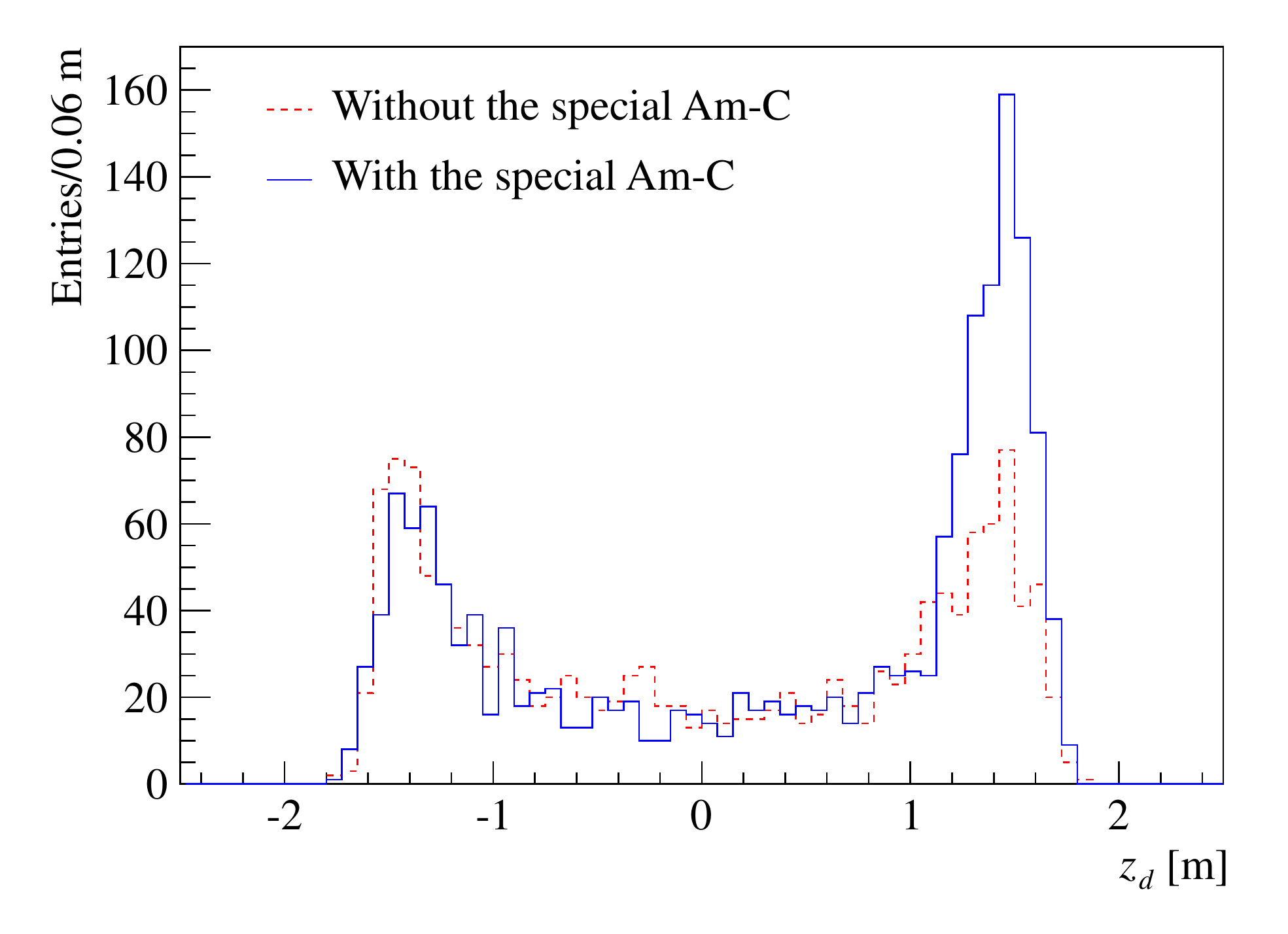}
\caption{Distribution of vertical position of delayed events for EH3-AD2 with both its Am-C calibration source and the special Am-C source (solid blue line), and EH3-AD1 with only its Am-C calibration source (dashed red line).   All sources were located at the tops of the detectors: $z \approx$~2.5~m. }
\label{Fig:LongPaper_AmC_1}
\end{center}
\end{figure}

This background was estimated using a special Am-C source~\cite{AmCbkgd} whose neutron emission rate was approximately 80 times higher than the Am-C calibration sources.  
The special source was positioned on the top of EH3-AD2 near ACU-B for about 10 days during the summer of 2012.
Figure~\ref{Fig:LongPaper_AmC_1} shows the resulting distribution of the reconstructed vertical position of delayed events, which exhibits an excess at positive $z$ (the top half of the AD).  
For comparison, the distribution from the adjacent EH3-AD1 (which had only an Am-C calibration source in ACU-A) is shown over the same period, exhibiting no apparent asymmetry.  
The distributions of the vertical position of prompt events are similar.  

The number of background DCs from the special Am-C source $N_{\mathrm{Special}}$ was estimated by subtracting $N_\mathrm{DC}$ of EH3-AD1 from $N_\mathrm{DC}$ of EH3-AD2 during the same period, resulting in $N_{\mathrm{Special}}$ = 137~$\pm$~41.6.  The vertical positions of both the prompt and delayed events were required to be in the top half of each AD ($z_p>0$ and $z_d>0$).

The intensity of the special Am-C source was scaled to the intensities of the Am-C calibration sources of each AD using ``delayed-type'' events, which are defined as singles that satisfy the delayed-energy criterion.  
The relatively low energy of the $n$H $\gamma$ selection admitted significant radioactive contamination into this sample of events.  To avoid this contamination, the higher-energy $n$Gd delayed-type events were used.  
In Ref.~\cite{AmCbkgd}, the number of $n$Gd delayed-type events due to an Am-C source $[N_{\mathrm{AmC-dtype}}]_{n\mathrm{Gd}}$ was estimated by the asymmetry of the vertical position distribution, which was similar to that in Fig.~\ref{Fig:LongPaper_AmC_1}. 
The number of background DCs from each Am-C calibration source $N_{\mathrm{AmC}}$ was estimated as 
\begin{equation}
\label{eq:AmC_corr_cal_edit}
N_{\mathrm{AmC}} = N_{\mathrm{Special}}\left[\frac{N_{\mathrm{AmC-dtype}}}{N_{\mathrm{Special-dtype}}}\right]_{n\mathrm{Gd}},
\end{equation}
where $N_{\mathrm{AmC-dtype}}$ is counted over the entire 621-day data period.  
The $n$Gd ratio in Eq.~(\ref{eq:AmC_corr_cal_edit}) was about 0.12 for the far hall and 0.23 for the near halls.  
The uncertainty of $N_{\mathrm{AmC}}$ was comprised of the 30\% statistical uncertainty of $N_{\mathrm{Special}}$ and an approximate 40\% systematic uncertainty shared with the $n$Gd-IBD analysis from a difference in delayed-type event rates among the near- and far-hall ADs.  
This gives a total uncertainty of 50\% for the Am-C background.  
Table~\ref{tab:IBDsummary} lists the rate of Am-C background DCs, which is $N_\mathrm{AmC}$ divided by $T_\mathrm{DAQ}\varepsilon_{\mu}\varepsilon_m$, for each AD.  The prompt-energy spectrum of the Am-C background was modeled with an exponential, which was determined from both the simulation and the data with the special Am-C source.  The spectrum is shown in Fig.~\ref{Fig:DCbkgSpectra}.  

For the $n$Gd-IBD analysis, this background had a 45\% total uncertainty.  Considering the common 40\% systematic uncertainty, the Am-C background determination was found to have a correlation coefficient of about 0.7 between the $n$H- and $n$Gd-IBD analyses:
\begin{equation}
\label{eq:AmCcorrelation}
\frac{40\% \cdot 40\%}{50\% \cdot 45\%} = 0.7.
\end{equation}

\subsection{$^{13}$C($\alpha$,n)$^{16}$O Background}
\label{sec:alphan}
The $^{13}$C($\alpha$,~n)$^{16}$O background is from four dominant sources of alpha decays in the liquid scintillator: the $^{227}$Ac (in the GdLS), $^{238}$U, and $^{232}$Th decay chains and $^{210}$Po, which is produced in the decay of $^{222}$Rn.  The ($\alpha$, n) background rate was roughly estimated using the rates from the $n$Gd-IBD analysis~\cite{nGd8AD} and the ratio of the $n$H/$n$Gd IBD selection efficiencies.  
The estimate in EH3 was approximately 0.02~$\pm$~0.01 DCs per AD per day. 
This estimate is expected to be conservative because of the lower activity of the LS relative to the GdLS: using the selection criteria outlined in Ref.~\cite{DYB_CPC}, 
the concentration of $^{232}$Th was determined to be a few hundred times greater in the GdLS while that of $^{238}$U was estimated to be similar.  
The uncertainty of the $^{13}$C($\alpha$,n)$^{16}$O background contributed negligibly to the total uncertainty of $\sin^2$2$\theta_{13}$ (see Table~\ref{tab:errorBudget}) and therefore, this background was neglected in this analysis.

\subsection{Summary of Correlated Backgrounds}
\begin{figure}[!t]
\begin{center}
\includegraphics[angle=0,width=\columnwidth]{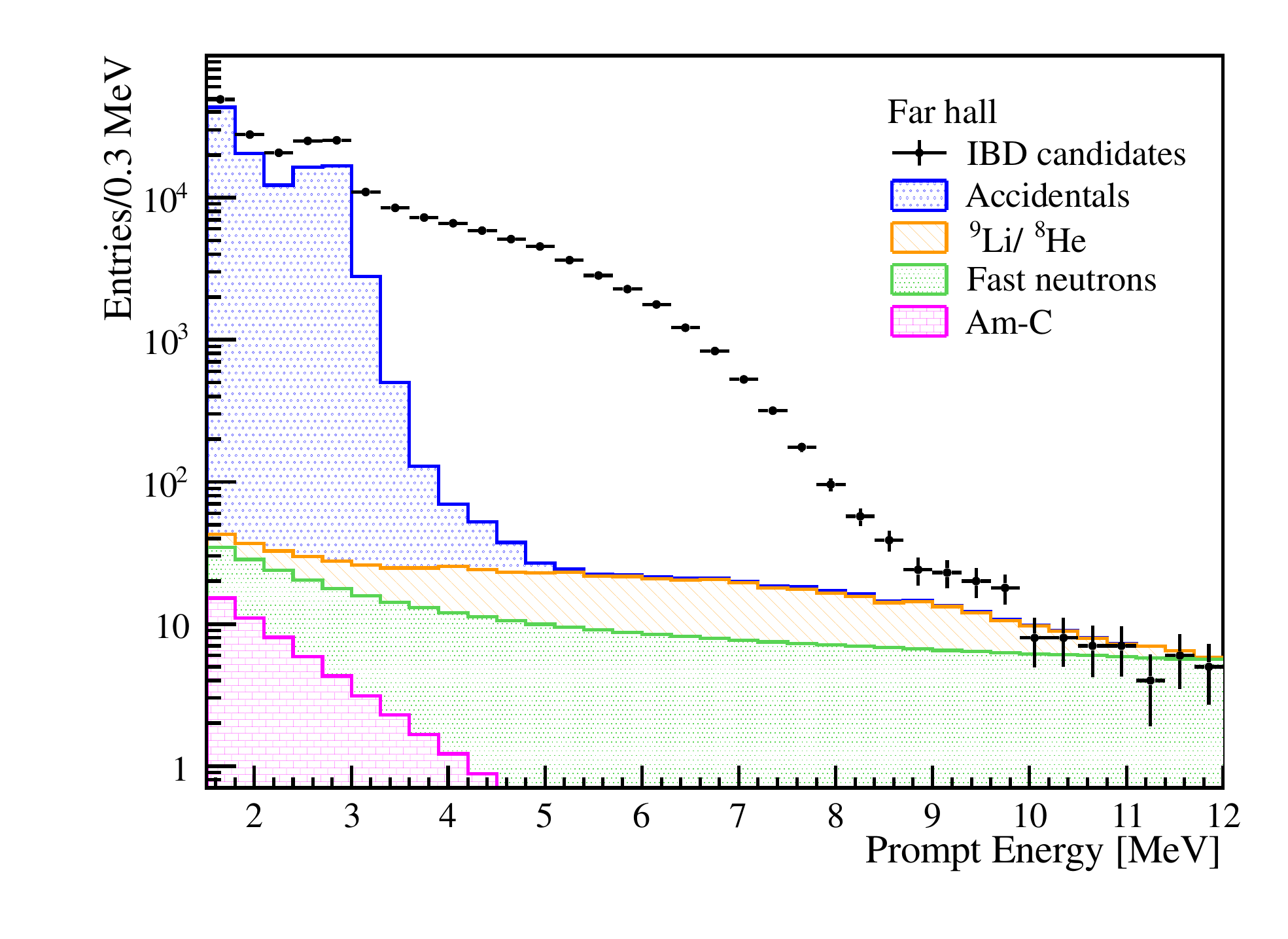}
\caption{Reconstructed prompt-energy distributions of the measured double coincidences after IBD selection (black points) and estimated backgrounds, for the sum of all ADs in EH3.  }
\label{Fig:DCbkgSpectra}
\end{center}
\end{figure}

The rates of the correlated backgrounds are summarized in Table~\ref{tab:IBDsummary} and their prompt-energy distributions are illustrated in Fig.~\ref{Fig:DCbkgSpectra} for EH3.  
The rates of $n$H IBDs after subtracting all the backgrounds are listed for each AD in Table~\ref{tab:IBDsummary}.  

With respect to the previous $n$H-IBD analysis~\cite{DYB_nH}, the absolute uncertainty of the dominant $^9$Li/$^8$He background was reduced by about 30\% because of increased statistics and various improvements in the method.  
Reductions in the uncertainties of the fast neutron and Am-C backgrounds resulted primarily from an improved method of estimation and a fit of the full spectrum, and the removal of some Am-C sources, respectively.  The overall uncertainty of backgrounds was reduced by 30\%.  

Comparing to the $n$Gd-IBD analysis, the fast neutron background was about four to five times larger relative to the IBD rate in EH3, while the $^9$Li/$^8$He and $^{241}$Am-$^{13}$C backgrounds were equal within uncertainties, and the $^{13}$C($\alpha$,n)$^{16}$O background was about half as large.  
The absolute uncertainty of the fast neutron background was about four to five times larger relative to the IBD rate in EH3, while the uncertainties of the $^9$Li/$^8$He and $^{241}$Am-$^{13}$C backgrounds were similar, 
and the uncertainty of the $^{13}$C($\alpha$,n)$^{16}$O background was about half that of the $n$Gd-IBD analysis.  
The impact of the uncertainties of the background estimations on the uncertainty of $\sin^2$2$\theta_{13}$ is described at the end of Section~\ref{sec:Fit}.  

Due to the sharing of uncertainty components between the $n$Gd- and $n$H-IBD analyses, the Am-C background determinations had a correlation coefficient of about 0.7, while the $^9$Li/$^8$He 
and fast neutron background determinations were uncorrelated, and the $^{13}$C($\alpha$,n)$^{16}$O background was neglected in this analysis.

\section{Detection Efficiency}
\label{sec:DetEff}

The expected number of selected IBDs from one AD was determined according to Eq.~(\ref{eq:predIBD}), in which the efficiency-weighted number of target protons was calculated considering antineutrino interactions in the GdLS, LS, and acrylic volumes $v$: 
\begin{equation}
\label{eq:eff}
N_{\varepsilon}= \varepsilon_{\mu}\varepsilon_m \left[\sum^\mathrm{GdLS, LS, acry.}_{v}N_{p,v}\varepsilon_{E_p,v}\varepsilon_{T,v}\varepsilon_{E_d,v}\right]\varepsilon_{D},
\end{equation}
where 
$\varepsilon_{\mu}$ and $\varepsilon_m$ are the muon-veto and multiplicity selection efficiencies of the AD, $N_p$ is the number of target protons of the AD, 
$\varepsilon_{E_p}$ and $\varepsilon_{E_d}$ are the prompt- and delayed-energy selection efficiencies, and $\varepsilon_{T}$ and $\varepsilon_{D}$ are the coincidence-time and -distance selection efficiencies, respectively.  The PMT flash selection efficiency (Section~\ref{sec:PMTflash}) is not included due to its negligible inefficiency.  

The number of target protons was determined for each AD from measurements made prior to AD deployment.  
The muon-veto, multiplicity, and distance selection efficiencies were determined with data.  The prompt- and delayed-energy, and time selection efficiencies were determined with a simulation using a predicted antineutrino spectrum such as described in Section~\ref{sec:reactor}.  
The simulation framework of Daya Bay is based on \textsc{Geant4}~\cite{GEANT4} and has been validated with comparisons to data~\cite{DYB_CPC}.  

In comparing the IBD rates among the far hall and near halls, efficiencies and uncertainties common to all the ADs are irrelevant.  The AD-uncorrelated uncertainties of the efficiencies, which reflect the identicalness of the ADs, were determined by comparing data among all eight ADs.  
The uncertainties of $\varepsilon_{\mu}$ and $\varepsilon_m$ were negligible (see Section~\ref{sec:EventSelect}).  The remaining quantities in Eq.~(\ref{eq:eff}) and their uncertainties, are discussed in this section.
The contribution from IBDs in the MO is described in Section~\ref{sec:spill}.

\subsection{Prompt-Energy Selection}
\label{sec:Eprompt}
The first selection criterion applied to AD events (after rejecting PMT flashes) was $E_\mathrm{rec} >$ 1.5~MeV.  Ultimately, this selection affected only prompt events because of the more stringent requirement applied to delayed events.  The prompt-energy selection efficiency and its uncertainty were determined with simulation in which the energy scale was aligned to that of the data (see Section~\ref{sec:Reconstruction}).  The efficiency was defined as the number of IBD reactions $N$ that satisfied the prompt-energy criterion divided by the total number of IBD reactions: 
\begin{equation}
\varepsilon_{E_p} = \frac{N(E_p >\ 1.5\ \mathrm{MeV})}{N_\mathrm{IBD}}.  
\end{equation}
The higher-energy requirement of $E_p <$ 12~MeV was estimated to contribute negligibly to the inefficiency and uncertainty, as suggested by Fig.~\ref{Fig:DCbkgSpectra}.
The efficiency in the LS volume was lower than that in the GdLS volume because a larger fraction of the annihilation $\gamma$'s deposited energy outside the scintillating volumes.  This fraction was largest for IBDs occurring in the acrylic elements.  The net efficiency of all volumes was about 90\%.  

The AD-uncorrelated uncertainty of the efficiency was estimated as the change in efficiency after shifting the energy scale by 0.5\%.  The relative change in efficiency was about 0.1\%.  
The 0.5\% shift is an estimate of the AD-uncorrelated uncertainty of the energy scale that was determined by comparing the fitted means of the $n$H-IBD $\gamma$ and $^{212}$Bi $\alpha$ peaks of all eight ADs.  
For reference, the estimated uncertainty of the energy scale in the GdLS volume was 0.2\%~\cite{nGd8AD}.  

\subsubsection{Variation with Baseline}
\label{sec:promptVar}
\begin{figure}[!h]
\begin{center}
\includegraphics[angle=0,width=\columnwidth]{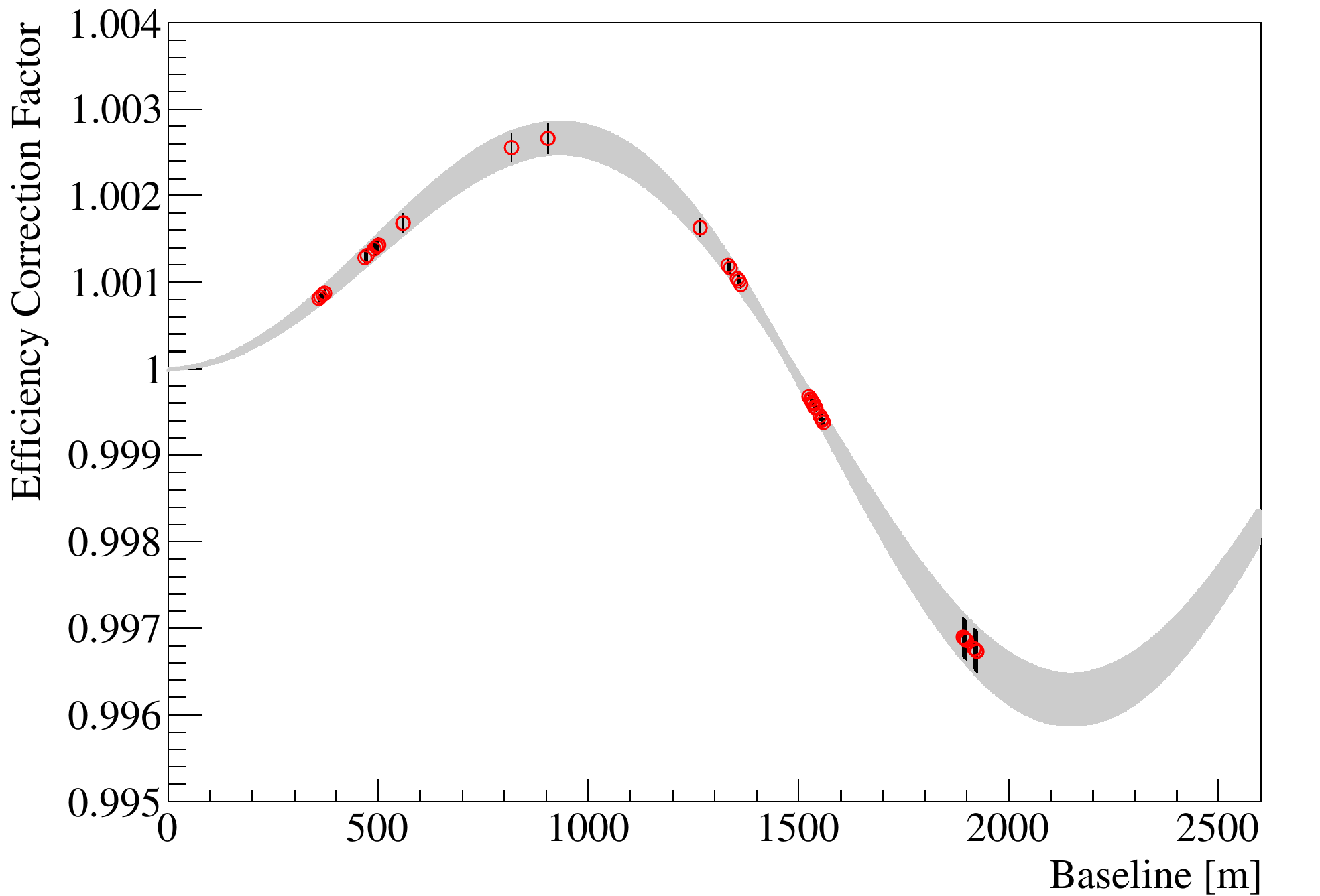}
\caption{An example of the relative variation of the IBD selection efficiency with baseline using the value of $\sin^{2}$2$\theta_{13}$ presented in this article.  This correction curve is for the LS region.  The red circles denote the 48 reactor-detector pairs. Their error bars and the error band are identically defined by the uncertainty of $\sin^{2}$2$\theta_{13}$.}
\label{Fig:EpVSbaseline}
\end{center}
\end{figure}

The $L/E$-dependence of neutrino oscillation [see Eq.~(\ref{eq:Psur})] implies that the shape of the neutrino energy spectrum changes with baseline $L$.  Therefore, the efficiency of the prompt-energy criterion varies with baseline.  The impact of this dependence on the multiple reactor-detector pairs at Daya Bay was estimated by applying oscillation to a predicted reactor antineutrino spectrum as a function of baseline.  At each baseline~\cite{supp}, the IBD selection efficiency was determined with simulation samples for the GdLS, LS, and acrylic volumes.  
The simulation accounted for energy deposited outside the scintillator volumes, and the nonlinearity~\cite{nGd8AD}, nonuniformity, and resolution of the detector energy-response.  The oscillation parameter values were the same as those in Section~\ref{sec:reactor}.  The resulting variation in the IBD selection efficiency as a function of baseline is illustrated for the LS region in Fig.~\ref{Fig:EpVSbaseline}.  
The shape of the curve is due to the span of the data in the $L/E$ domain.  For the near halls (smaller $L$), more oscillation occurred for lower energy antineutrinos, which decreased the number of IBD reactions with prompt energy below threshold and thus, increased the efficiency.  
For illustration, the mean energy of a prompt event without oscillation was 3.626~MeV while the corresponding energy in EH1 (EH2) due to antineutrinos from the two (four) nearby reactors with oscillation was 3.630 (3.632)~MeV.  
These numbers are representative of the first 4 (8) points in Fig.~\ref{Fig:EpVSbaseline}.  
For the far hall (larger $L$), more oscillation occurred at median antineutrino energies and about equally at higher and lower energies, resulting in a net decrease in efficiency.  

In the fit for $\sin^{2}$2$\theta_{13}$ (Section~\ref{sec:Fit}), the IBD selection efficiencies in the GdLS, LS, and acrylic volumes of each AD were multiplied by a correction factor for each reactor baseline (6 reactors $\times$ 8 ADs = 48 baselines)~\cite{supp}.  
The fit was first performed without correction factors.  The resulting value of $\sin^{2}$2$\theta_{13}$ was then used to generate a set of correction factors and then fit again.  
This iterative approach was tested using Asimov data samples generated according to Eq.~(\ref{eq:predIBD}) with known values of $\sin^{2}$2$\theta_{13}$.  Several values of $\sin^{2}$2$\theta_{13}$ were tested and all fits converged consistently with negligible bias.  No additional uncertainty was assigned.  
Although several iterations were performed, the value of $\sin^{2}$2$\theta_{13}$ converged within the precision reported in this article after only one iteration.  
The results of the fits without corrections were about 4\% larger than the true values for the Asimov data samples and the converged value for the measured data.  
This variation of the IBD selection efficiency was estimated to be an order of magnitude smaller for the $n$Gd-IBD analysis, which required $E_p >$ 0.7~MeV.

\subsection{Coincidence-Time Selection}
\label{sec:capTime}
The efficiency of the coincidence-time selection was different for each detector volume $v$ due to the different densities and neutron-capture cross sections of the materials. The efficiency was defined as 
\begin{equation}
\label{eq:EffLifetime}
\varepsilon_{T} = \frac{N(1 < t_c < 400\ \mu s; E_p >\ 1.5\ \mathrm{MeV})}{N(E_p >\ 1.5\ \mathrm{MeV})}, 
\end{equation}
and was determined with simulation.  The efficiency in the LS volume was about 85\% and that in the GdLS volume was about 99\% due to the shorter neutron-capture time of $n$Gd.  
These values were validated with data.  

The neutron-capture time was studied in the GdLS and LS volumes by fitting for the mean neutron-capture time with the following formulas: 
\begin{equation}
\label{eq:EffTime_TimeFit}
\begin{split}
N_\mathrm{Gd}(t)& = N_{0,\mathrm{Gd}} \cdot [(1+\alpha){\textstyle\frac{1}{\tau_\mathrm{Gd}}} e^{-t/\tau_\mathrm{Gd}} - \alpha {\textstyle\frac{1}{\tau_{0}}} e^{-t/\tau_{0}} ]+C_1 \\
N_\mathrm{LS}(t)  & = N_{0,\mathrm{LS}} \cdot {\textstyle\frac{1}{\tau_\mathrm{LS}}} e^{-t/\tau_\mathrm{LS}} +C_2 ,
\end{split}
\end{equation}
where $\alpha$ balances two terms, the first corresponding to the capture of a neutron at thermal energies [$O$(0.025)~eV] with time constant $\tau_\mathrm{Gd}$, and the second representing the difference in capture cross section between thermal and IBD neutron energies [$O$(0.015)~MeV], with effective time constant $\tau_{0}$.  
The capture-time spectrum in LS is due almost solely to $n$H which can be represented by a single exponential.  This is because the number of captures per volume per time, which is proportional to the product of capture cross section and neutron velocity, is essentially independent of energy below IBD neutron energies.  For $n$Gd, this product is much larger at thermal energies than at IBD energies (see, {\it e.g.} Ref.~\cite{ENDF}), effectively yielding two distinct time constants with $\tau_{0} < \tau_\mathrm{Gd}$.  
The capture-time constant in LS is denoted by $\tau_\mathrm{LS}$, and $C_1$ and $C_2$ account for accidentals.  

\begin{figure}[!b]
\includegraphics[width=\columnwidth]{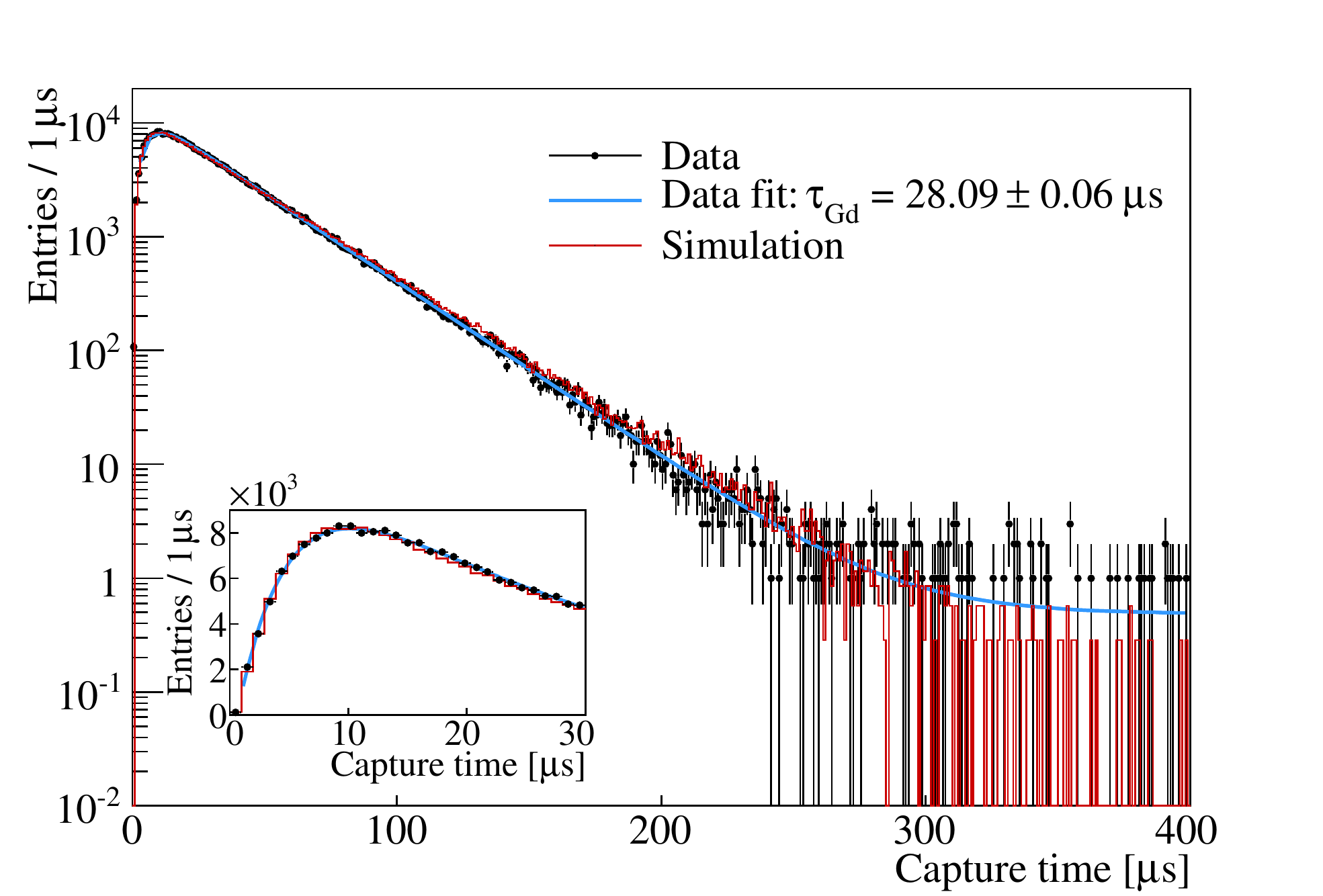}
\caption{Time separation for double coincidences selected with $n$Gd-IBD criteria in the GdLS volume from the data of all ADs (black points) and from simulation (red histogram).  The spectra are normalized by the number of coincidences between 6 and 150~$\mu$s.  The fit to data (blue curve) and fitted capture-time constant $\tau_\mathrm{Gd}$ are shown.}
\label{Fig:nGdTimeFit}
\end{figure}

\begin{figure}[!b]
\includegraphics[angle=0,width=\columnwidth]{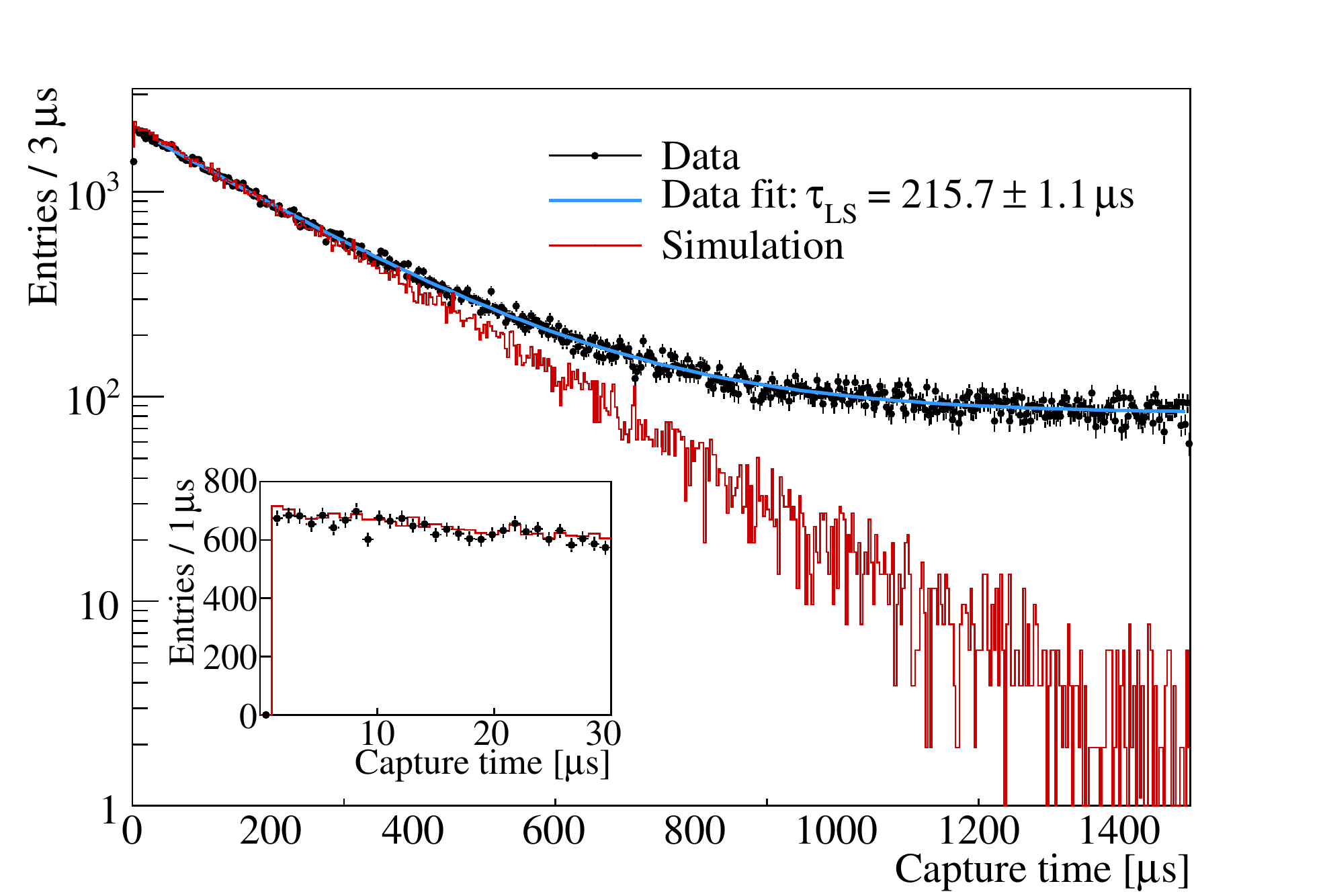}
\caption{Time separation for double coincidences selected with $n$H-IBD criteria in the LS volume from the data of all ADs (black points) and from simulation (red histogram).  The spectra are normalized by the number of coincidences between 30 and 300~$\mu$s.  The fit to data (blue curve) and fitted capture-time constant $\tau_\mathrm{LS}$ are shown.}
\label{Fig:nHTimeFit}
\end{figure}

The neutron-capture times for the GdLS and LS regions were studied using $n$Gd- and $n$H-IBDs, respectively.  The selection criteria were slightly modified from the nominal IBD criteria: the $n$Gd delayed events were selected between 6 and 10~MeV, while the $n$H prompt-energy lower limit was increased to 3.5~MeV to minimize the accidental background, and the $n$H delayed-energy criterion was fixed to 1.8-2.8~MeV.  When fitting the $n$Gd-IBD spectrum, the reconstructed positions of the prompt events were required to satisfy $|z| <$ 1~m and $r <$ 1~m to minimize the fraction of neutrons that originated from, or had any interactions, outside GdLS.  Similarly, when fitting the $n$H-IBD spectrum, a constraint of $r >$ 1.7~m was applied to minimize the fraction of neutrons that originated from GdLS. 
The fit results using the data from all ADs are shown in Figs.~\ref{Fig:nGdTimeFit} and \ref{Fig:nHTimeFit}.  Good agreement in the slopes is observed between the data and simulation.  The fitted capture-time constants were about 28.1 and 216~$\mu s$ for the GdLS and LS volumes, respectively. 
For reference, Fig.~\ref{Fig:Time} shows the total capture-time spectra of the far- and near-hall ADs for the nominal $n$H-IBD selection criteria before and after subtracting the accidental background.  

\begin{figure}[!b]
\begin{center}
\includegraphics[angle=0,width=\columnwidth]{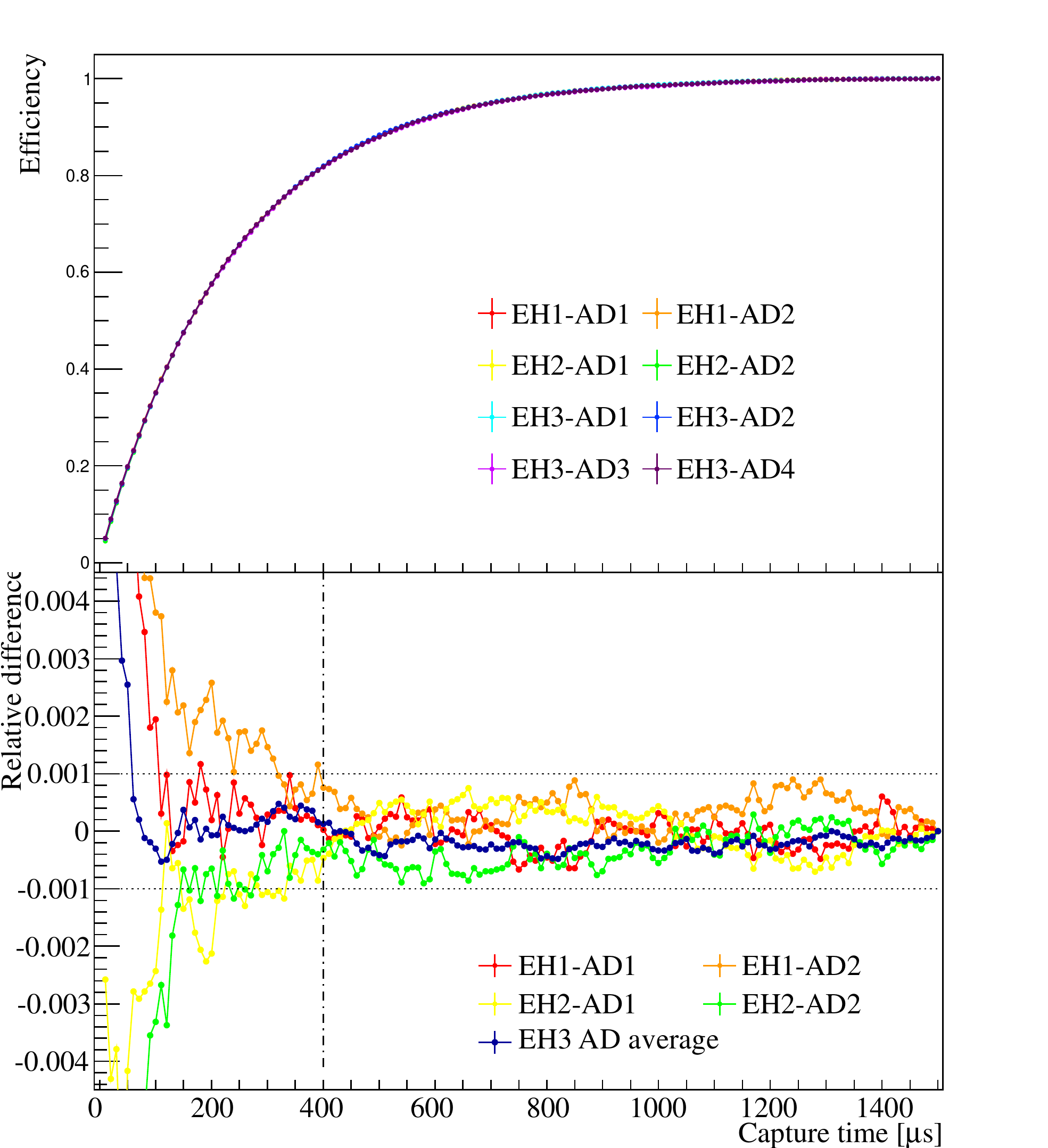}
\caption{Efficiency (top panel) and relative difference to the average (bottom panel) {\it vs.} event time separation for $^{214}$Bi $\beta$-$\alpha$ events in each AD.  The data of the far-hall ADs were combined in the bottom panel to increase statistics.  The differences are within $\pm$0.1\% at the criterion of 400~$\mu$s. }
\label{Fig:canv_ggh_Bi_diff}
\end{center}
\end{figure}

The AD-uncorrelated uncertainty of the 400~$\mu s$ criterion in the combined GdLS and LS volume was partly estimated using $\beta$-$\alpha$ events from the $^{214}$Bi-$^{214}$Po-$^{210}$Pb decay chain.  These events provided greater statistics than $n$H events and were used to determine the variation of the time measurements of the electronics.  
The lifetime of $^{214}$Po is 237~$\mu s$, which is close to the mean $n$H capture time in LS. 
The efficiency of the selection was determined relative to the number of double coincidences with a coincidence time window of [1, 1500]~$\mu$s.  
The relative differences of the efficiencies of all eight ADs are shown in Fig.~\ref{Fig:canv_ggh_Bi_diff}, and are within 0.1\% at the selection criterion of 400~$\mu s$.  

Similarly, the uncertainty associated with the 1~$\mu$s criterion was determined to be 0.1\% by comparing the relative number of events between 1 and 2~$\mu$s.  

Because the estimates of the uncertainties were performed using a source different from neutrons, additional uncertainties related to neutron-capture time were added.  The uncertainties considered were identified from an expression of the mean neutron-capture time: 
\begin{equation}
\label{eq:capTime}
\frac{1}{\tau} = \frac{v_n}{\lambda} = v_n \sum_i n_i \sigma_i(v_n), 
\end{equation}
where $v_n$ is the velocity of the neutron, $\lambda$ is the mean free-path of the neutron, $n_i$ is the number-density of nucleus $i$, and $\sigma_i$ is the neutron-nucleus cross section.  Isotopes other than Gd and H contributed less than 1\% of captures and were not considered.  
For the LS volume, the measured density differed by $<$~0.1\% among the ADs.  
In addition, the fluctuation in density caused by temperature changes uncorrelated among experimental halls was within 0.045\% during the data-recording period. These effects introduced a $<$~0.11\% uncertainty to the neutron-capture time $\tau$. This uncertainty was propagated through Eq.~(\ref{eq:EffTime_TimeFit}) to obtain an approximate 0.02\% AD-uncorrelated uncertainty.  

The uncertainties from the $^{214}$Bi $\beta$-$\alpha$ event comparisons and neutron-capture time-related quantities were combined to give a total AD-uncorrelated uncertainty of 0.14\% for the efficiency of the coincidence-time criterion.

\subsection{Delayed-Energy Selection}
\label{sec:Ecuts}

The efficiency of the delayed-energy selection was determined with simulation and defined as
\begin{equation}
\label{eq:delayedEff}
\varepsilon_{E_d} = \frac{N(E_d \pm 3\sigma; 1 < t_c < 400\ \mu s; E_p >\ 1.5\ \mathrm{MeV})}{N(1 < t_c < 400\ \mu s; E_p >\ 1.5\ \mathrm{MeV})}.  
\end{equation}
This definition does not preclude IBDs with the neutron captured by nuclei other than hydrogen; for example, $n$Gd IBDs comprised approximately 0.7\% of the IBDs after applying the delayed-energy criterion.  
For both simulation and data, the $\mu~\pm$~3$\sigma$ selection was applied using the mean $\mu$ and standard deviation $\sigma$ from a fit of the delayed-energy spectrum with the Crystal Ball function~\cite{CrystalBall}.  
The selection efficiency in the GdLS volume was about 15\% primarily because of neutron-capture by gadolinium.  
The efficiency in the LS volume was about 65\% primarily because of the outward escape of the $n$H $\gamma$'s.  

\begin{figure}[!t]
\includegraphics[width=\columnwidth]{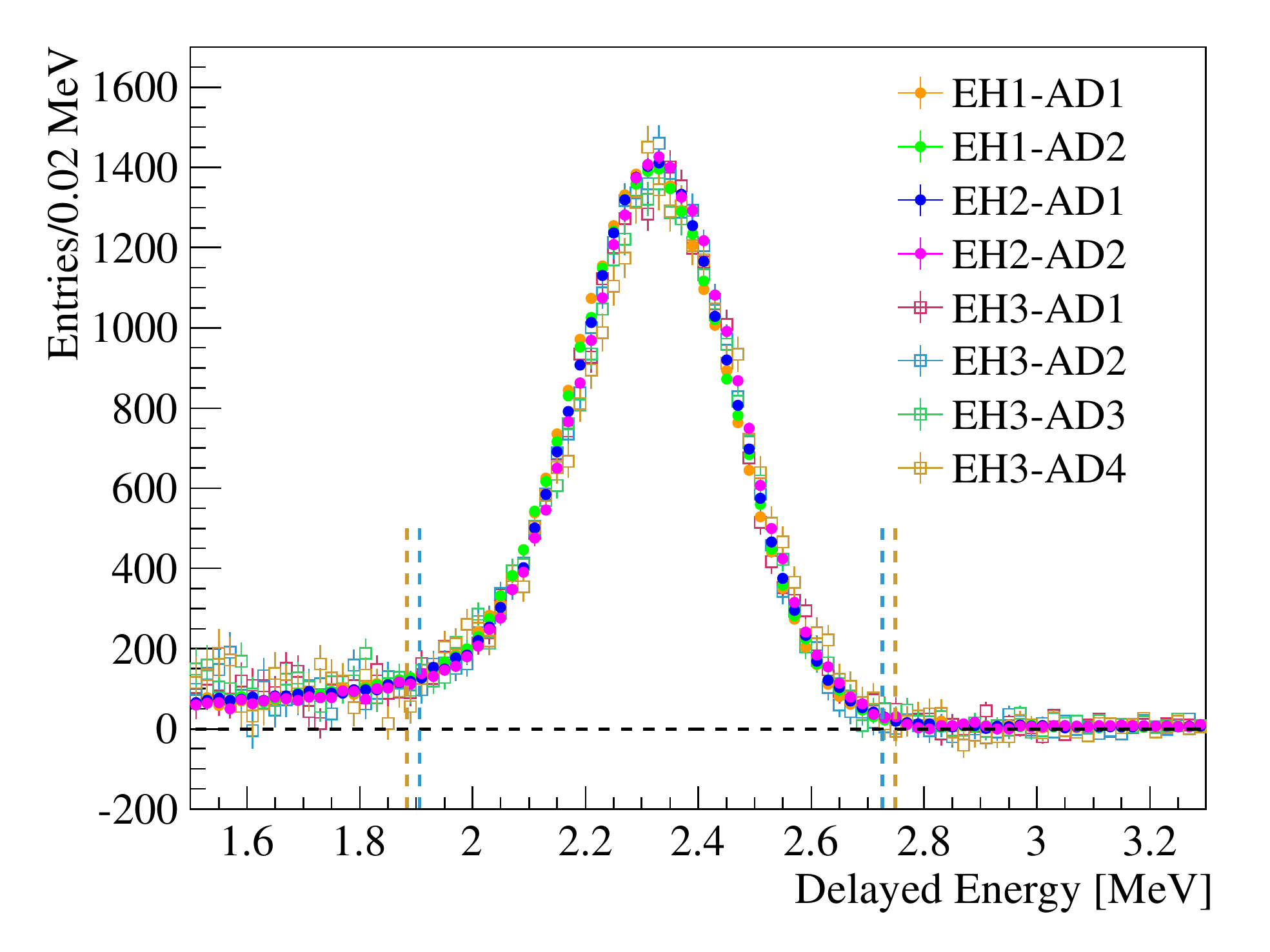}
\caption{Delayed energy spectra of $n$H-IBDs in all ADs.  The entries of each histogram  are normalized to the average number of IBDs in the far-hall ADs.  The fitted means are scaled to the average of the mean of the far-hall ADs.  The two pairs of vertical lines correspond to the largest and smallest 3$\sigma$ selections.}
\label{fig:nHspectra}
\end{figure}

Two methods were used to estimate the AD-uncorrelated uncertainty of the delayed-energy selection efficiency.  
One method is a relative comparison of the delayed-energy spectra of the ADs.  
The comparison was made after applying all the $n$H selection criteria and subtracting the accidental backgrounds (errors from accidental subtractions were included in the energy spectra).  
The method uses the number of events within two energy ranges: 
the first is the nominal selection of $\mu~\pm$~3$\sigma$, which is approximately [1.90, 2.74]~MeV, and the second is [1.50, 2.80]~MeV.  
These two ranges are visible for each AD in Fig.~\ref{fig:nHspectra}.  
The upper value of the latter range was chosen to include most of the $n$H IBDs with $E_d >$ 2.74~MeV (0.1\% of $n$H IBDs) while the lower value corresponds to the low-energy criterion (Section~\ref{sec:lowE}) and includes more of the tail of the spectrum (12\% more $n$H IBDs).  
The latter range contains both peak and tail portions of the spectrum and therefore is assumed to be sensitive to all factors that might influence the shape of the spectrum.  

For each AD $i$, the number of events in the nominal range $A$ ($N_{A,i}$) was plotted {\it vs.} the number of events in the extended range $B$ ($N_{B,i}$) and a linear relation was fit: 
\begin{equation}
\label{eq:lineFit}
\overline{N}_A(N_{B,i}) = a+bN_{B,i}. 
\end{equation}
This line represents the average behavior of all ADs, including differences in their spectral shape and backgrounds. Here, the efficiency of the delayed-energy selection $\varepsilon$ is defined as $N_A/N_\mathrm{Total}$, where $N_\mathrm{Total}$ is the number of events without the delayed-energy selection.  The fitted line was used to determine the relative variation of $\varepsilon$ for each AD: 
\begin{equation}
\label{eq:deltaEff}
\frac{\delta\varepsilon_i}{\varepsilon_i} = \frac{\delta N_{A,i}}{N_{A,i}} = \frac{N_{A,i}-\overline{N}_A}{N_{A,i}}=1-\frac{a+bN_{B,i}}{N_{A,i}}. 
\end{equation}
This determination assumes that there is no variation in $N_\mathrm{Total}$.  From studies with simulation, it was found that $N_A$ and $N_\mathrm{Total}$ are highly correlated under various scenarios that could modify the shape of the spectrum, including differences in OAV dimensions~\cite{DYB_Det} and the residual nonuniformity of $E_\mathrm{rec}$, making this assumption conservative.  This determination also assumes that variations in the spectrum outside range $B$ are not systematically different from those within.  Using simulation, differences in OAV dimensions or in the mean free path of the $\gamma$'s were found to have a greater impact on the shape of the spectrum at the low-energy end, but to contribute negligibly to $\delta\varepsilon_i/\varepsilon_i$.  In addition, comparing the high-statistics spectra of the near-hall ADs did not reveal any systematic trends in the differences among spectra above 1.5~MeV, suggesting that there may not be any such trends below 1.5~MeV.  
The statistical uncertainties of the data from the far-hall ADs were large, so they were excluded from the determination though they were conservatively used in the linear fit.  Comparing the four near-hall ADs, the half-range of the $\delta\varepsilon_i/\varepsilon_i$ was 0.33\%. 
This estimation directly includes AD-to-AD variations in the 3$\sigma$ selection, energy scale, and factors that may influence the shape of the spectrum; however, it does not include variations in the fraction of neutrons that capture on hydrogen (53\%) relative to other isotopes, such as Gd (46\%) and C (0.5\%), because such variations have an equivalent impact on $N_B$ and $N_A$.  

The fraction of neutrons that capture on isotope $x$ is expressed similarly as the mean capture time in Eq.~(\ref{eq:capTime}): 
\begin{equation}
\label{eq:capFrac}
f_x = \frac{n_x\sigma_x(v_n)}{\sum_i n_i \sigma_i(v_n)}. 
\end{equation}
Performing error propagation on Eq.~(\ref{eq:capTime}) and Eq.~(\ref{eq:capFrac}), and combining the results, the variation of $f_x$ among the ADs was expressed in terms of the variation of $\tau$ and one of the $n_i$.  In this way, the variation of the measured capture time in the GdLS was used to constrain the variation of $n_\mathrm{Gd}$.  
The variation of $n_\mathrm{H}$ was taken to be negligible 
because of the mixing of all production batches of scintillator~\cite{GdLS} and the filling procedures applied to the ADs~\cite{FillingSystem}.  
As a result, the AD-to-AD variation of the fraction of $n$H captures was estimated to be $<$ 0.01\% and 0.16\% in the LS and GdLS volumes, respectively.  
These two values correspond to approximately 0.03\% for the full volume.  

Combining the variations estimated from the spectral comparison and the $n$H capture-fraction calculation yields a total AD-uncorrelated uncertainty of 0.33\% for the delayed-energy selection efficiency.  

The second method used to evaluate the uncertainty of the delayed-energy selection efficiency is the ratio of the numbers of spallation $n$H to spallation $n$Gd ($N_{n\mathrm{H}}/N_{n\mathrm{Gd}}$), which utilizes the smaller variation of the $n$Gd delayed-energy selection efficiency and the larger sample of spallation neutrons.  
The energy spectrum of spallation-$n$H and -$n$Gd $\gamma$'s from each AD was obtained by subtracting a background spectrum recorded in a background time window from the spectrum recorded in a signal time window.  
Spallation neutrons generated by cosmogenic muons were identified as delayed-type events that followed WS- or AD-identified muons.  
These muons were identified with greater purity by augmenting the definitions of a $\mu_\mathrm{WS}$ and $\mu_\mathrm{AD}$: for both the IWS and OWS, $N_\mathrm{PMT}$ was required to be $>$ 20, and a $\mu_\mathrm{AD}$ was required to have $E_\mathrm{rec} >$ 50~MeV. A 20-$\mu s$ muon-event veto-time was applied to avoid the ``ringing'' of PMT signals that followed high-energy events~\cite{ringing}.  
The signal time window was between 20 and 700~$\mu s$ after a muon event.  The background time window was a similar length, however, given the different distributions of muon energy and trajectory among halls~\cite{muon} (which affected the characteristics of the spallation products), the background window was tuned to be slightly different in each hall.  By matching both the shape and population of the tail portions of the signal and background energy spectra, the background window was set to be between 700 and 1480, 1453, and 1384~$\mu s$, in EH1, EH2, and EH3, respectively.  
Both $n$Gd and $n$H delayed-energy criteria were nominal (see Table~\ref{tab:criteria}). 
 
\begin{figure}[!b]
\centering
\includegraphics[width=\columnwidth]{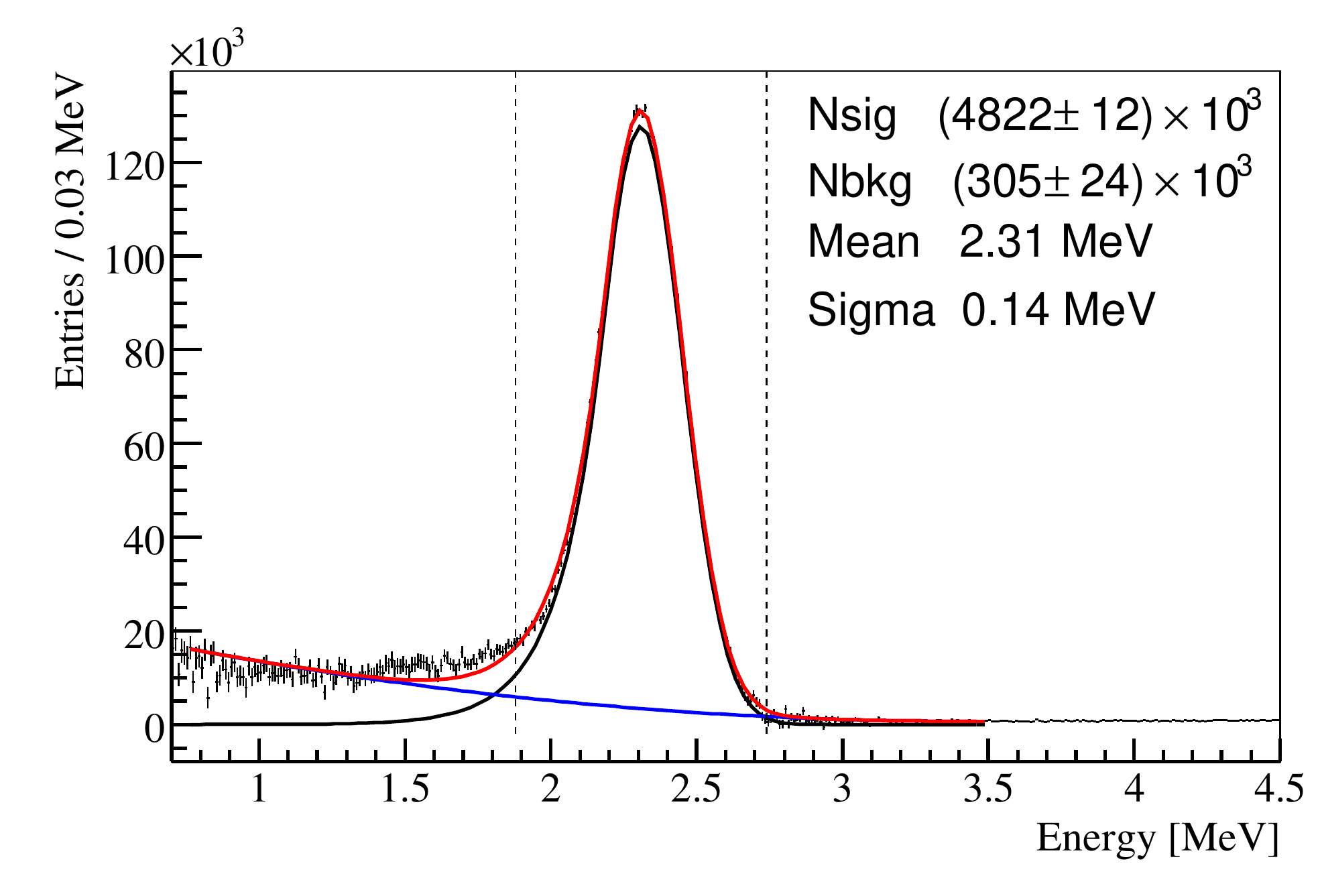}
\caption{Fit of the spallation-$n$H reconstructed energy spectrum (black points) with a Crystal Ball function (black line) and a second-order polynomial (blue line) in EH1-AD1.  The red line is the sum of the black and blue lines.  The vertical dashed lines represent the delayed-energy selection criteria (Mean $\pm$ 3Sigma) within which Nsig and Nbkg were counted.  }
\label{fig:nHpol2Fit}
\end{figure}

\begin{figure}[!b]
\centering
\includegraphics[width=\columnwidth]{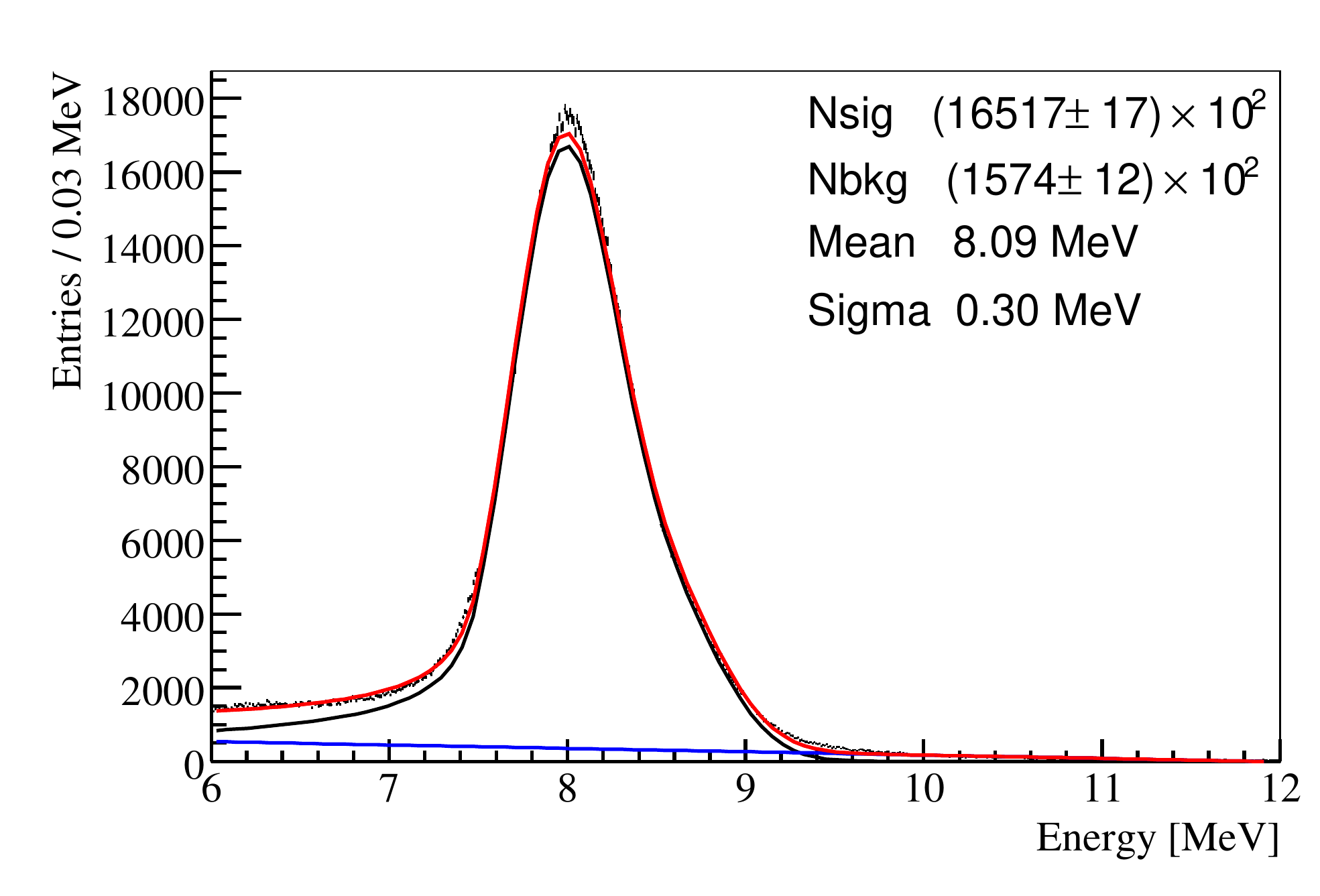}
\caption{Fit of the spallation-$n$Gd reconstructed energy spectrum (black points) with two Crystal Ball functions (black line) and a first-order polynomial (blue line) in EH1-AD1.  The red line is the sum of the black and blue lines.  }
\label{fig:nGdFit}
\end{figure}

The energy spectra of the spallation-neutron-capture $\gamma$'s were fitted with signal and background components.  
The background component accounted for residual spallation products that were not subtracted with the background window.  
For the $n$H spectrum, the signal component was a Crystal Ball function and the background function was a second-order polynomial. 
For $n$Gd, the signal component was two Crystal Ball functions as mentioned in Section~\ref{sec:Reconstruction}, and the background function was a first-order polynomial.  Fit results are shown in Figs.~\ref{fig:nHpol2Fit} and \ref{fig:nGdFit}, where the number of signal events defined as spallation neutrons are labeled as ``Nsig''.  

\begin{figure}[!hb]
\includegraphics[width=\columnwidth]{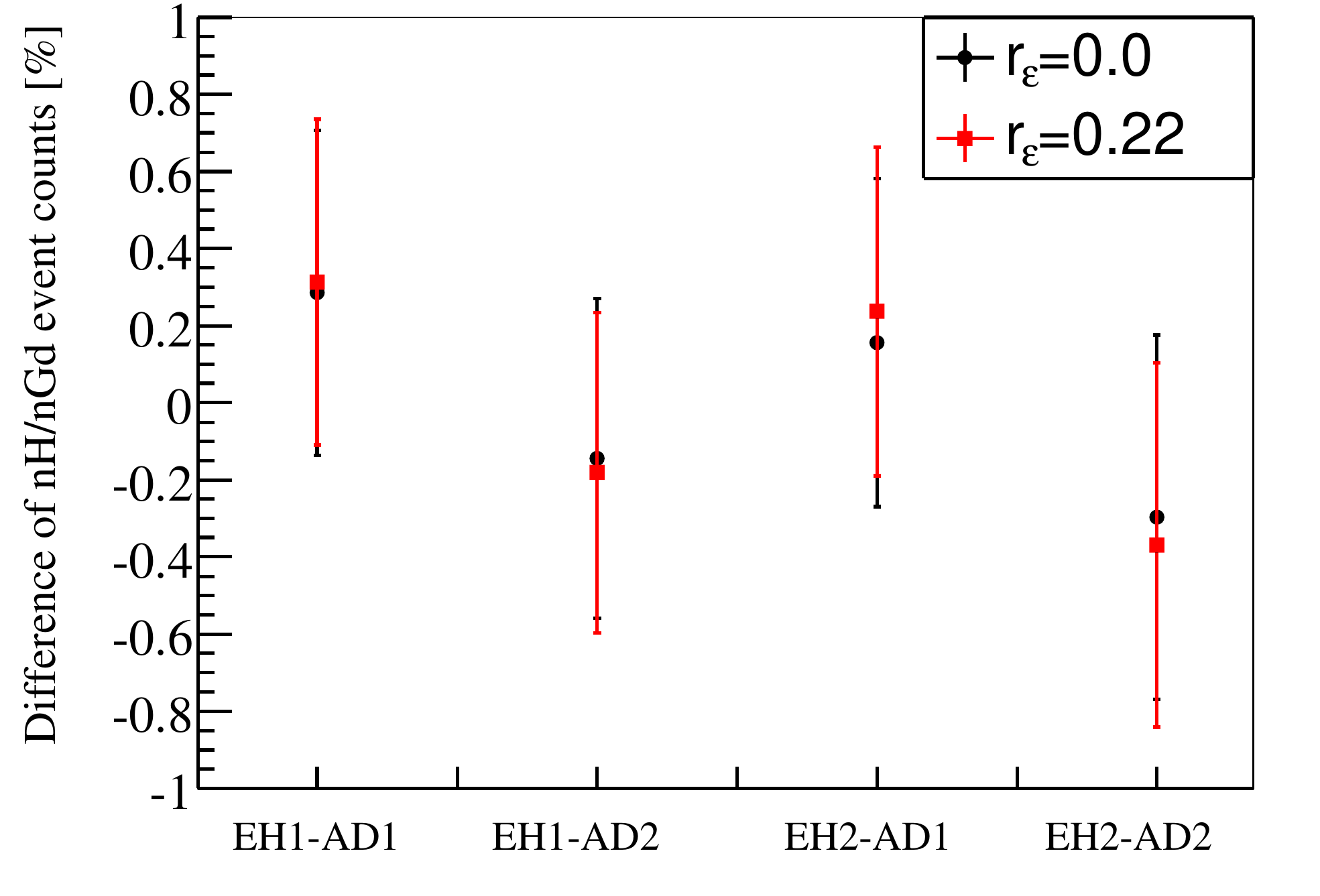}
\caption{Difference in the ratio of the number of spallation-$n$H/-$n$Gd events in the fitted energy peaks of each near-hall AD relative to the mean of all near-hall ADs.  See the text for details.}
\label{fig:nHnGdRD}
\end{figure}

Compared with the previous analysis~\cite{DYB_nH}, the spallation neutron ratio is updated in this article by normalizing the number of neutrons to the number of target protons $N_p$ (Section~\ref{sec:TargetProton}): 
\begin{equation}
\label{eq:spallationRatio}
\frac{N_{n\mathrm{H}}/(N_{p,\mathrm{LS}} + r_\varepsilon N_{p,\mathrm{GdLS}})}{N_{n\mathrm{Gd}}/N_{p,\mathrm{GdLS}}}, 
\end{equation}
where $r_\varepsilon$ is the ratio of efficiencies of selecting spallation $n$H in GdLS $vs.$ LS: 
$r_\varepsilon \equiv \varepsilon_\mathrm{GdLS}/\varepsilon_\mathrm{LS}$.  
Due to the non-uniform distribution of spallation neutrons, $r_\varepsilon$ is not precisely known; therefore, two extreme cases were considered: (a) the distribution is entirely within the LS ($r_\varepsilon$~=~0); (b) the distribution is uniform ($r_\varepsilon$~=~0.22 from simulated IBDs).  
Figure~\ref{fig:nHnGdRD} shows the difference in the ratio defined by Eq.~(\ref{eq:spallationRatio}) for each near-hall AD relative to the mean of the four ADs.  
The far-hall ADs were not used due to their lack of statistics.  
The choice of $r_\varepsilon$ is found to have little impact on the variation of the ratio.  
The half-range of the ratios of the near-hall ADs is approximately 0.35\%.  Due to the use of a ratio with $n$Gd events, this estimation inherently includes the variation of the $n$Gd delayed-energy criterion, which was estimated to be 0.12\% for IBDs~\cite{DYB3}.  This estimate also inherently includes the variation of the fraction of neutrons that capture on hydrogen.  

Given the 0.33\% and 0.35\% relative uncertainties from the two independent methods, 0.35\% was assigned for the total AD-uncorrelated uncertainty of the delayed-energy selection efficiency.  

To determine the correlation of the delayed-energy selection efficiency between the $n$H and $n$Gd analyses, the uncertainty was decomposed into three components: 3$\sigma$ variation, energy scale variation, and others.  The contributions of the first two components were estimated with simulation by applying the largest and smallest 3$\sigma$ ranges (see Fig.~\ref{fig:nHspectra}) and shifting the energy scale (see Section~\ref{sec:Eprompt}), respectively.  
The first component, which was dominant, does not exist for the $n$Gd-IBD analysis and thus, is uncorrelated.  The correlation of energy scale variations between the $n$H- and $n$Gd-IBD analyses was estimated to be 0.8 with a linear fit of the measured $n$H-IBD {\it vs.} $n$Gd-IBD delayed-energy peaks.  The latter component of ``others'' accounts for any contributions not directly evaluated, such as differences in OAV dimensions or the residual nonuniformity of $E_\mathrm{rec}$, and was assumed to be fully correlated.  The hydrogen capture fractions of the $n$H analysis were determined to be anticorrelated with the gadolinium capture fraction of the $n$Gd analysis: in the GdLS volume, if the fraction of captures on Gd increases, then naturally the fraction on H decreases.  
In the LS volume, the same anticorrelated relation exists via neutrons that are produced in GdLS or LS but capture in the other of the two volumes.  
Combining the correlation constants and corresponding component uncertainties from both the $n$H and $n$Gd analyses yields an overall correlation coefficient of 0.07 for the efficiency of the delayed-energy selection.

\subsection{Coincidence-Distance Selection}
\begin{figure}[!t]
\begin{center}
\includegraphics[angle=0,width=\columnwidth]{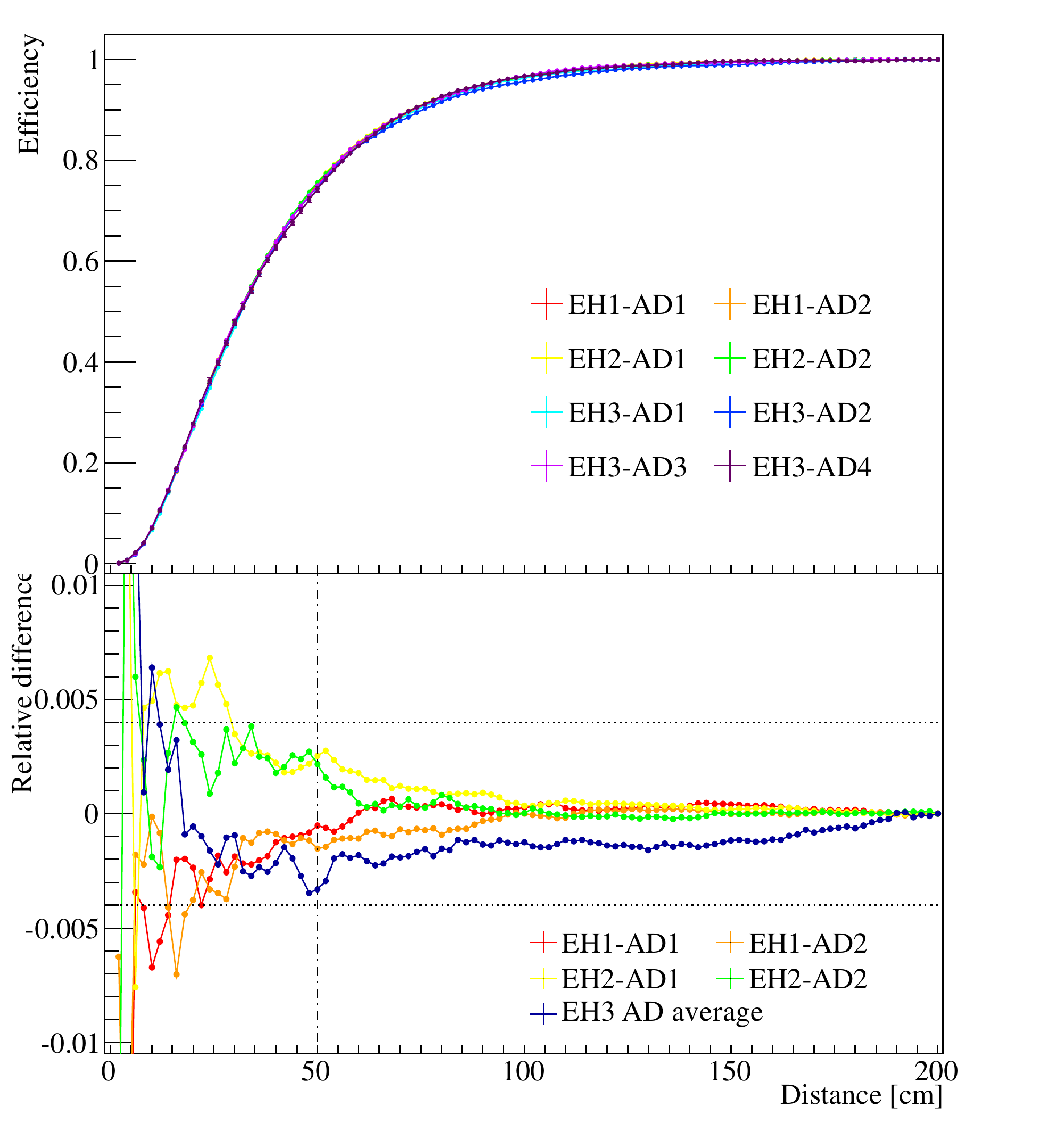}
\caption{Efficiency (top panel) and relative difference to the average (bottom panel) {\it vs.} coincidence-distance for correlated double coincidences ($N_\mathrm{Cor}$) in each AD.  The data of the far-hall ADs were combined in the bottom panel to increase statistics.  The differences are within $\pm$0.4\% at the criterion of 50~cm. }
\label{Fig:DistanceEffDiff}
\end{center}
\end{figure}

The efficiency of the coincidence-distance selection was measured with data and defined as 
\begin{equation}
\label{eq:EffDistance}
\varepsilon_{D} = \frac{N(d_c<50\ \mathrm{cm}; E_d \pm 3\sigma;~...~; E_p>1.5\ \mathrm{MeV})}{N(E_d \pm 3\sigma; 1<t_c<400\ \mu s; E_p>1.5\ \mathrm{MeV})}.
\end{equation}
The efficiency was determined relative to the number of DCs with $d_c <$ 200~cm using the data of all 8 ADs with accidental backgrounds subtracted as shown in Fig.~\ref{Fig:DistIBD}.    
The efficiency curves and relative differences with respect to the average are shown in Fig.~\ref{Fig:DistanceEffDiff}.  
The efficiency for $d_c <$ 50~cm was about 75\%.  
Because the total statistics of the far-hall ADs was only about half that of a single near-hall AD, the data of the four far-hall ADs were merged together when calculating the relative difference. All the differences were within $\pm$0.4\% at the 50-cm selection criterion.  
Therefore, the AD-uncorrelated uncertainty of the efficiency of the coincidence-distance criterion was assigned to be 0.4\%.

\subsection{IBDs in Acrylic and Mineral Oil}
\label{sec:spill}
The target materials were primarily liquid scintillator, however, the IAV, OAV, and acrylic-encased reflectors were in direct contact or close proximity with the scintillators such that an IBD positron originating in these elements could enter the scintillators and deposit sufficient energy to trigger an AD.  Such IBDs contributed an estimated 1.0\% of the $n$H-IBDs after selection.  

IBD positrons originating in the MO rarely reached the scintillator and generally produced an insufficient amount of light to trigger an AD.  
However, a few percent of the IBD positrons annihilated in-flight, producing a higher-energy $\gamma$ that was sometimes directed toward the scintillator with enough energy to pass the low-energy criterion.  Some fraction of the corresponding IBD neutrons propagated to the LS and captured on H.  
From simulation, it was estimated that approximately 0.06\% of the IBDs in the MO survived the selection criteria.  This ``spill-in'' effect from the MO was found to have a negligible impact on the determination of $\sin^{2}$2$\theta_{13}$ and was not included in this analysis.  

The impact of neutrons or $\gamma$'s (and their secondaries) that spill-out into the MO, or spill-in/out between the GdLS and LS, is naturally included in the prompt- and delayed-energy selection efficiencies and their uncertainties.

\subsection{Target Proton Number}
\label{sec:TargetProton}
The number of target protons $N_p$ was determined  for each AD from the measured target masses $M$ and hydrogen mass-fractions $w_\mathrm{H}$ of the GdLS, LS, and acrylic volumes $v$: 
\begin{equation}
\label{eq:TargetProton}
N_{p,v} = M_v\ w_{\mathrm{H},v}\ N_\mathrm{A}\ /\ m_\mathrm{H}\ ,
\end{equation}
where $N_\mathrm{A}$ is Avogadro's number and $m_\mathrm{H}$ is the molar mass of hydrogen.  

The mass-fractions of hydrogen were determined by combustion analysis to be about 12.0\% for both GdLS and LS (with uncertainties at the level of 0.1\%)~\cite{DYB_Det}. 
For acrylic (C$_5$H$_8$O$_2$), $w_\mathrm{H}$~=~8.05\%.  
The AD-uncorrelated uncertainties of these quantities were taken to be negligible as described for $n_\mathrm{H}$ in Section~\ref{sec:Ecuts}.  

The total masses of GdLS and LS were measured when filling each AD, using a load cell and Coriolis flow meter, respectively~\cite{FillingSystem}.  The masses of acrylic components were measured with an industrial scale before filling~\cite{AVs}.   The relevant masses are given for each AD~\cite{supp}.  
Only the uncertainties of target mass were propagated to the final uncertainty of target proton number.  

The average numbers of target protons in the GdLS, LS, and acrylic volumes are $1.43\times10^{30}$, $1.54\times10^{30}$, and $0.18\times10^{30}$, respectively.  Values for each AD are provided~\cite{supp}.  
The AD-uncorrelated uncertainties are listed in Table~\ref{tab:Eff}.

\subsection{Detector Leak}

Around the end of July, 2012, when data-recording was paused to install the final two ADs, a leak began between the LS and MO volumes of EH3-AD1.  The levels of GdLS and LS in the overflow tanks~\cite{TargetMass} (see Fig.~\ref{fig:AD}) of EH3-AD1 slowly decreased while the level of MO slowly increased, suggesting that the LS was leaking into the MO region. This hypothesis was supported by measurements using the MO clarity system~\cite{DYB_Det} which showed significant decreases in the transmission of shorter-wavelength light through the MO and an increase of MO light yield over time, consistent with a gradual addition of scintillator into the MO.  The hypothesis was further supported by the observation of an increased (decreased) rate of higher-energy (lower-energy) muons reconstructed in the MO volume.  
These observed trends stabilized after about two years with an estimated leakage of about 20~kg.  This loss of mass lowered the height of the LS level in the overflow tank and did not directly impact the number of target protons in the LS volume.  

No impact on the detector response  is expected in the LS volume due to the direction of the leak; however, in the MO volume, there is potential for an increase in trigger rate.  Given a 20-kg leakage into the 36-ton volume, and assuming the light yield of the LS is two orders of magnitude greater than that of the MO, one may naively estimate an average increase of the light yield in the MO volume on the order of 1\%.  In simulation, this increase was modeled as an increase in the energy scale, and was applied to prompt and delayed events of IBDs generated in the MO, resulting in a $O$(0.001)\% increase of the $n$H-IBD selection efficiency. 
Indeed, no impact of the leak to the $n$H-IBD analysis has been observed in comparisons of various quantities before and after the start of the leak.  These quantities included various event rates, neutron-capture energy peak and resolution, and IBD prompt and delayed event-position distributions.  
Given the observed stabilization of the leak, no impact is expected in the future.

\subsection{Summary}
\label{sec:EffSum}
The efficiencies of the PMT flash rejection, prompt- and delayed-energy selection, and coincidence-time selection criteria were determined with simulation, while the number of target protons, the muon-veto and multiplicity and coincidence-distance selection efficiencies were determined with data.  The AD-uncorrelated uncertainties of these quantities were determined by comparing data among the eight ADs.  

The efficiency of the PMT flash rejection criterion was $>$ 99.99\% (see Section~\ref{sec:PMTflash}) and had a negligible uncertainty.  Muon-veto and multiplicity selection efficiencies ($\varepsilon_{\mu}$ and $\varepsilon_m$) are listed in Table~\ref{tab:IBDsummary} and had negligible AD-uncorrelated uncertainties.  
The product of the efficiencies of the prompt- and delayed-energy, and time selection criteria were about 14\%, 50\%, and 5\% in the GdLS, LS, and acrylic volumes, respectively.  The efficiency of the coincidence-distance criterion was determined as an average for all volumes: 75\%.  The AD-uncorrelated uncertainties of these efficiencies are listed for each detector volume $v$ in Table~\ref{tab:Eff}.  The uncertainty of the delayed-energy selection efficiency reduced from 0.5\%~\cite{DYB_nH} to 0.35\% because of a new estimation and an update of the original estimation to scale the number of spallation neutrons with the number of target protons.  This reduced the uncertainty of the $n$H-IBD selection efficiency by 15\%.  
\begin{table}[!htb]
\begin{tabular}{ l c c }\hline\hline
  &  Uncertainty (\%)    &  Correlation \\ \hline
Target protons ($N_{p,\mathrm{GdLS}}$)     &  0.03  &  1  \\
Target protons ($N_{p,\mathrm{LS}}$)         &  0.13  &  0    \\
Target protons ($N_{p,\mathrm{acrylic}}$)  &  0.50  &  -    \\ 
Prompt energy ($\varepsilon_{E_p}$)           &  0.10  &  1    \\
Coincidence time ($\varepsilon_{T}$)               & 0.14  &  1    \\
Delayed energy ($\varepsilon_{E_d}$)           & 0.35  &  0.07 \\
Coincidence distance ($\varepsilon_{D}$)               &  0.40  &  0    \\ \hline
Combined ($N_{\varepsilon}$)                         & 0.57  & 0.07   \\ \hline \hline
\end{tabular}
\caption{The relative per-detector uncorrelated uncertainties for each detector-related quantity.  
The uncertainties of the $N_p$ are weighted when determining the combined uncertainty of $N_{\varepsilon}$ in the bottom row.  
The last column contains the estimated correlation coefficients between the $n$H- and $n$Gd-IBD analyses.}
\label{tab:Eff}
\end{table}

Table~\ref{tab:Eff} also gives the estimated correlation coefficients between the detector efficiencies of the $n$H- and $n$Gd-IBD analyses.   The number of target protons were fully correlated in the GdLS while uncorrelated in the LS due to their identical and independent methods of mass measurement, respectively.  
The efficiency of the prompt-energy criterion was correlated through a common dependence on energy scale, and was conservatively treated as fully correlated.  The coincidence-time criterion was also treated as fully correlated. 
The delayed-energy criterion was largely independent because the primary contribution to the uncertainty in the $n$H analysis was the variation of the 3$\sigma$ selection, which does not exist in the $n$Gd analysis.  
The coincidence-distance criterion was uncorrelated because there was no such selection in the $n$Gd-IBD analysis.  
The overall correlation between the IBD detection efficiencies of the $n$H- and $n$Gd-IBD analyses was about 0.07.  

The last row of Table~\ref{tab:IBDsummary} shows the ratio of the efficiency- and target proton-corrected rates of IBDs for the $n$H- and $n$Gd-IBD analyses, for each AD.  The errors are the statistical, background, and AD-uncorrelated systematic uncertainties of both analyses.  The consistency of the eight values with one another reflects the consistency of the selected number of IBDs, background estimates, and per-AD target proton and efficiency corrections, between the two analyses.  
 The consistency of the eight values with 1 reflects the accuracy of these values for both analyses.

\section{Results}
\label{sec:Results}

\begin{figure}[!b]
\begin{center}
\includegraphics[angle=0,width=\columnwidth]{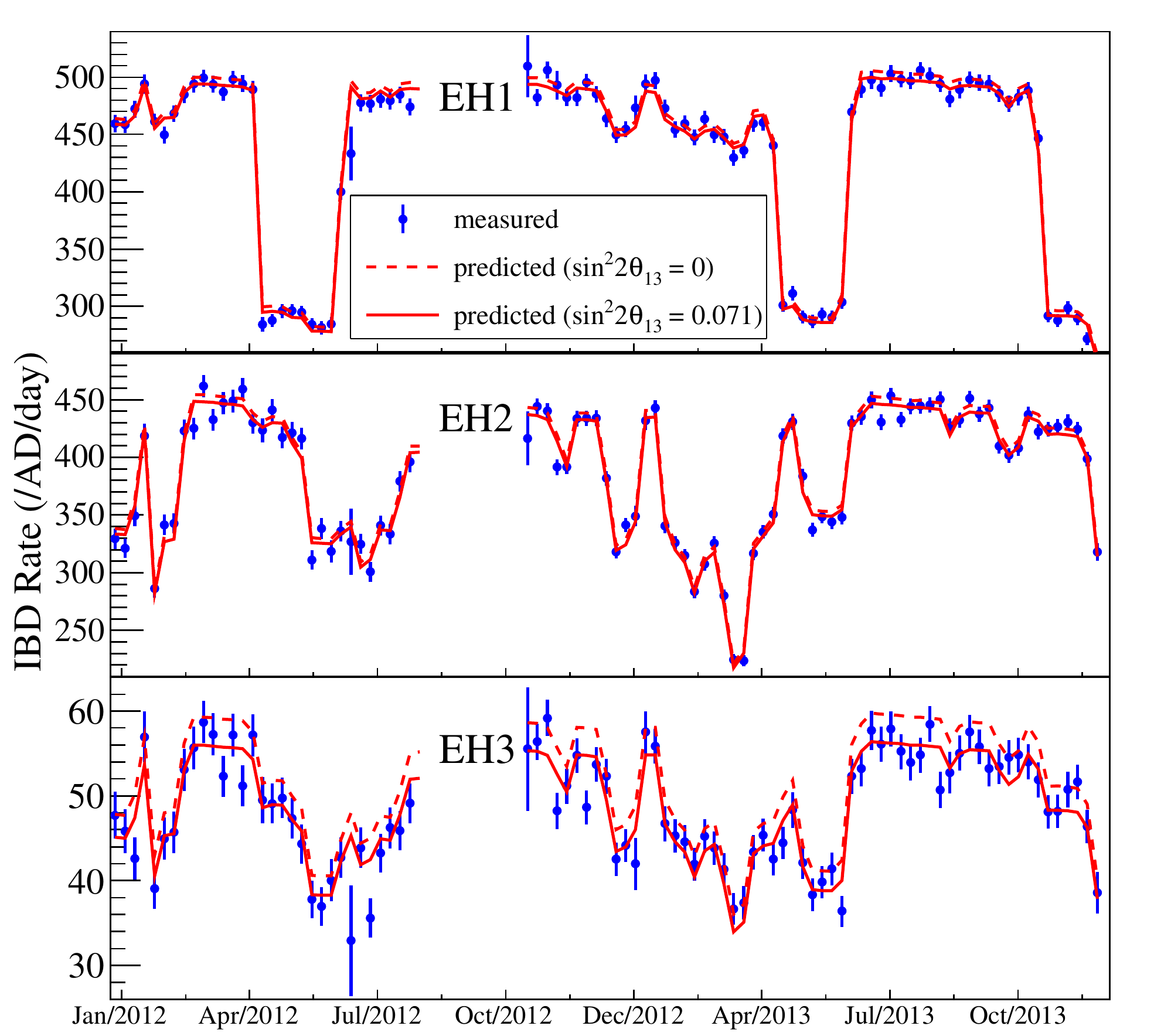}
\caption{Measured IBD rate {\it vs.} time for each experimental hall (blue points).  Each point spans one week and the error bars are purely statistical.  The dashed red lines are the expected IBD rates assuming no oscillation.  The solid red lines are the expected IBD rates with the best-fit value of $\sin^2$2$\theta_{13}$.  The final two of eight ADs were installed during the $\approx$12-week gap in all halls.}
\label{Fig:IBDRealTime}
\end{center}
\end{figure}

The measured and predicted IBD rates of each hall are shown over time in Fig.~\ref{Fig:IBDRealTime}.  The measured rates are background-subtracted and efficiency-corrected ($\varepsilon_{\mu}\varepsilon_m$).  The predictions are from Eq.~(\ref{eq:predIBD}) [$i.e.$, Eqs.~(\ref{eq:Phi}) and (\ref{eq:eff})], and are adjusted with the best-fit normalization factor $\epsilon$ from Eq.~(\ref{eq:Chi2Definition}).  
The six reactors are seen to have operated continually at their nominal power output.  The two reactors nearby EH1 were refueled every 16 months and the four reactors nearby EH2 were refueled every 8-12 months, each with 1-2 months downtime.  

\subsection{Antineutrino Disappearance}
\label{sec:RatioDeficit}
The disappearance of \nuebar~is quantified without invoking a model of neutrino oscillation and with minimal impact from models of reactor antineutrino spectra, by directly comparing the measured IBD rate at the far hall with the rate expected based on the measurements at the near halls. 
The expected number of IBDs in the far hall was expressed as a combination of the two near-hall measurements: 
\begin{equation}
\label{eq:FarNoOsc}
\overline{N}_\mathrm{EH3} \equiv \alpha N_\mathrm{EH1} + \beta N_\mathrm{EH2}, 
\end{equation}
where $N_\mathrm{EH1}$ and $N_\mathrm{EH2}$ are the measured numbers of IBDs after subtracting all the backgrounds and correcting for the muon-veto and multiplicity selection efficiencies ($\varepsilon_{\mu}$ and $\varepsilon_m$) in EH1 and EH2. 

Expressions for the weights $\alpha$ and $\beta$ were determined using Eq.~(\ref{eq:FarNoOsc}) with the number of measured IBDs replaced by the number of predicted IBDs assuming no oscillation.  
This number was calculated for experimental hall $i$ using Eq.~(\ref{eq:predIBD}) without oscillation: 
\begin{equation}
\label{eq:pred}
\overline{N}_i= 
\sum_{r=1}^6 \overline{N}_{ir} \equiv \sum_{r=1}^6 \sum_{d_i} \frac{N_{\varepsilon,d_i}}{4\pi L_{d_ir}^2} \iint_{\{t_{d_i}\}} \! \sigma_\nu \frac{d^{2}N_r}{dE dt} dE dt, 
\end{equation}
where $d_i$ denotes the $d$-th AD in experimental hall $i$ and the $N_{\varepsilon}$ do not include $\varepsilon_m$ and $\varepsilon_\mu$.  
The modified Eq.~(\ref{eq:FarNoOsc}) directly yields $\beta = (\overline{N}_\mathrm{3}-\alpha \overline{N}_\mathrm{1}) / \overline{N}_\mathrm{2}$.  
The weight $\alpha$ was obtained by operating on the difference between the two predictions for EH3: $\Delta\overline{N} = \overline{N}_\mathrm{3}-\alpha \overline{N}_\mathrm{1} - \beta \overline{N}_\mathrm{2}$.  The variance of $\Delta\overline{N}$ ($\sigma_\Delta^2$) was obtained via error propagation with respect to the reactor-uncorrelated relative uncertainty (which was taken to be identical for all reactors), and then its minimum was found with respect to $\alpha$, yielding  
\begin{equation}
\label{eq:alpha}
\begin{aligned}
\alpha = \frac{\sum_r{(\overline{N}_{3r}-\frac{\overline{N}_3}{\overline{N}_2}\overline{N}_{2r})(\overline{N}_{1r}-\frac{\overline{N}_1}{\overline{N}_2}\overline{N}_{2r})}}{\sum_r{(\overline{N}_{1r}-\frac{\overline{N}_1}{\overline{N}_2}\overline{N}_{2r}})^2}.  
\end{aligned}
\end{equation}
This expression minimizes the impact of the reactor-uncorrelated uncertainty.  

For the 621-day data set used in this analysis, $\alpha =$~0.054 and $\beta =$~0.216.  
These values are dominated by the baselines $L_{dr}$, and only slightly influenced by the integrated emission rates $d^{2}N_r(E, t)/dE dt$.  
Thus, $\beta$, which is associated with EH2, is four times larger than $\alpha$ primarily because of the shorter baselines between EH3 and the four reactors nearby EH2.  
The reactor-uncorrelated uncertainty is suppressed by a factor of about 20, which can be seen by evaluating the expression for $\sigma_\Delta^2$.  

Using Eq.~(\ref{eq:FarNoOsc}) and the values of $\alpha$ and $\beta$, the ratio of the observed to the expected number of IBDs at the far hall was
\begin{equation}
\label{eq:RatioDeficit}
\begin{aligned}
R \equiv \frac{N_{\mathrm{EH3}}}{\overline{N}_{\mathrm{EH3}}} = 0.950 \pm 0.005.
\end{aligned}
\end{equation}

\begin{figure}[ht]
\begin{center}
\includegraphics[angle=0,width=\columnwidth]{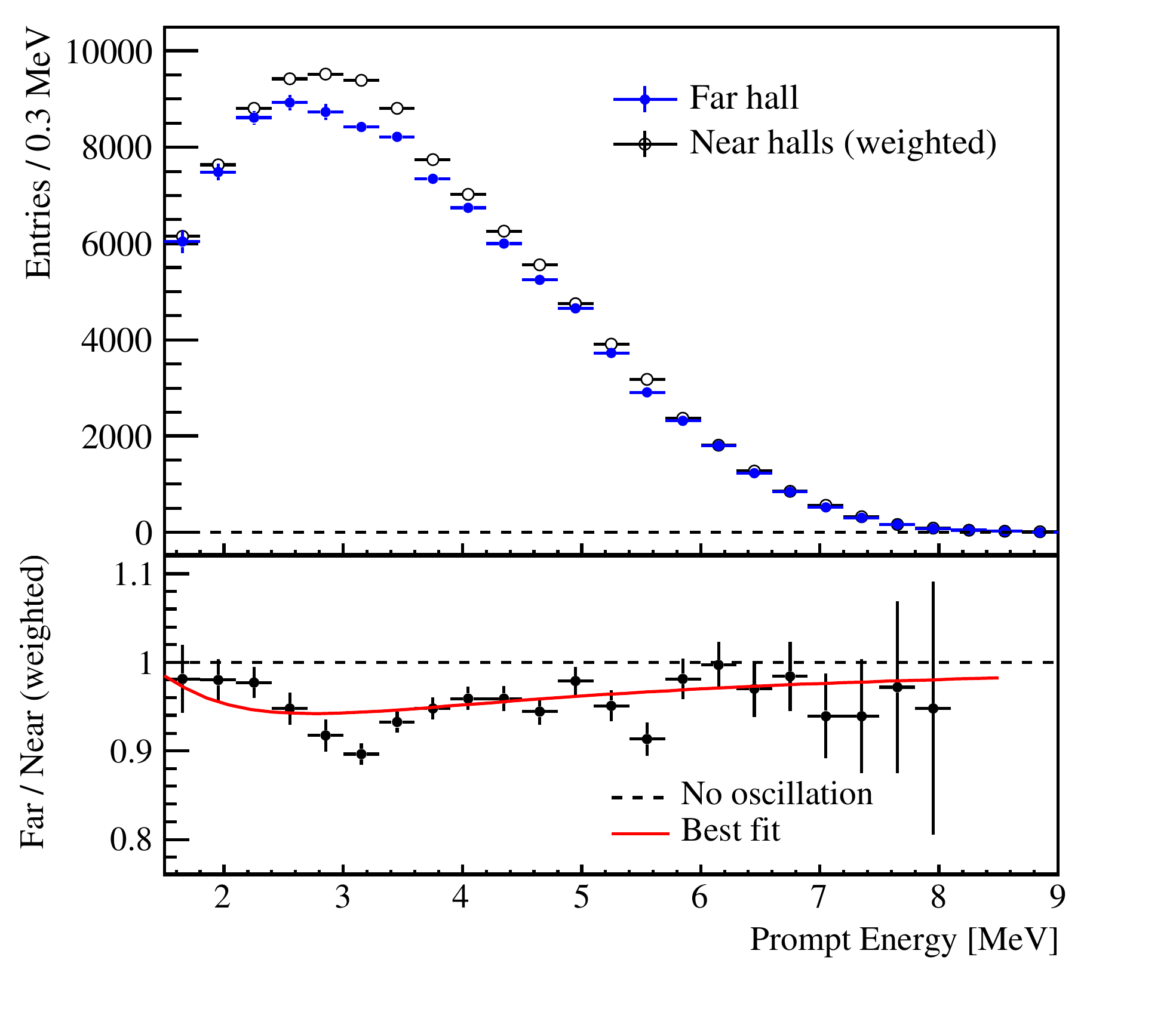}
\caption{Top: Reconstructed prompt-energy spectrum of the far hall (solid blue points) and the expectation based on the measurements of the two near halls (empty black points).
Spectra are background-subtracted.  Error bars are purely statistical.  
Bottom: Ratio of the Far/Near halls and the curve representing the best-fit value of $\sin^{2}$2$\theta_{13}=$~\nHresult.  }
\label{Fig:canv_sig_near}
\end{center}
\end{figure}

Figure~\ref{Fig:canv_sig_near} shows the measured prompt-energy spectrum at the far hall and that predicted with the near-hall measurements via Eq.~(\ref{eq:FarNoOsc}).  
The ratios $R$ of each energy bin are shown in the bottom panel and demonstrate the effect of \nuebar~disappearance as a function of energy.  The best-fit curve is the ratio of far-hall and normalized near-hall predictions using Eq.~(\ref{eq:predIBD}) and the result for $\sin^2$2$\theta_{13}$ presented in the next section.  

\subsection{Fit for \boldmath$\sin^{2}\!2\theta_{13}$\unboldmath}
\label{sec:Fit}
To determine $\sin^{2}$2$\theta_{13}$, a $\chi^2$ was constructed with pull terms for the background uncertainties and the AD- and reactor-uncorrelated uncertainties: 
\begin{equation}
\label{eq:Chi2Definition}
\begin{aligned}
\resizebox{\hsize}{!}{$\displaystyle\chi^2 = \sum_{d=1}^{8}\frac{[N_{\mathrm{DC},d}-\overline{N}_{\mathrm{IBD},d}(1 +\epsilon +\sum_{r=1}^{6} \omega^{d}_{r}\alpha_r +\epsilon_d)-(1+\eta_d)B_d]^2}{(\sigma_{\mathrm{DC},d})^2}$}\\
\resizebox{\hsize}{!}{$\displaystyle \hspace{0.9cm} +\sum_{r=1}^{6}\frac{\alpha_r^2}{\sigma_R^2} + \sum_{d=1}^{8}\left(\frac{\epsilon_d^2}{\sigma_D^2} +\frac{\eta_d^2}{(\sigma_{B,d})^2}\right), \hspace{4.2cm}$}
\end{aligned}
\end{equation}
where $N_{\mathrm{DC},d}$ is the number of measured double coincidences from the $d$-th AD given in Table~\ref{tab:IBDsummary},
$B_d$ is the sum of the accidental and correlated backgrounds derivable from Table~\ref{tab:IBDsummary},
$\sigma_{\mathrm{DC},d}$ is the statistical uncertainty of $N_\mathrm{DC}$,
and $\overline{N}_\mathrm{IBD}$ is the expected number of IBDs from Eq.~(\ref{eq:predIBD}), which contains the oscillation parameter $\sin^{2}$2$\theta_{13}$.  
The $\omega_r^d$~\cite{supp} are the fractions of IBDs in the $d$-th AD due to the $r$-th reactor, which were calculated using Eq.~(\ref{eq:predIBD}) without oscillation (including oscillation decreased the best-fit value of $\sin^{2}$2$\theta_{13}$ by less than 0.03\%).  
The reactor-uncorrelated uncertainty (0.9\%) is denoted as $\sigma_R$.
The parameter $\sigma_D$ is the AD-uncorrelated uncertainty of IBD detection efficiency from Table~\ref{tab:Eff}.
The parameter $\sigma_{B,d}$ is the combination of all background uncertainties, which are given in Table~\ref{tab:IBDsummary}.
There are twenty two corresponding pull parameters denoted as $\alpha_r$, $\epsilon_d$, and $\eta_d$.
The normalization factor $\epsilon$ was fit and accounted for any biases in 
the backgrounds $B_d$ that were common to all halls or detectors, 
and any biases in the predicted number of IBDs $\overline{N}_{\mathrm{IBD},d}$ that were common to all detectors; {\it i.e.}, in reactor-related models/quantities, the IBD cross section model, or IBD selection efficiencies.  

Iterating over $\sin^{2}$2$\theta_{13}$ with the efficiency correction factors as described in Section~\ref{sec:promptVar}, the best-fit value for both the normal and inverted neutrino-mass hierarchies was 
\begin{equation}
\begin{aligned}
\sin^{2}2\theta_{13}=0.071 \pm 0.011, 
\end{aligned}
\end{equation}
with a $\chi_\mathrm{min}^2$ per degree of freedom of 6.3/6.  

\begin{figure}[!b]
\begin{center}
\includegraphics[angle=0,width=\columnwidth]{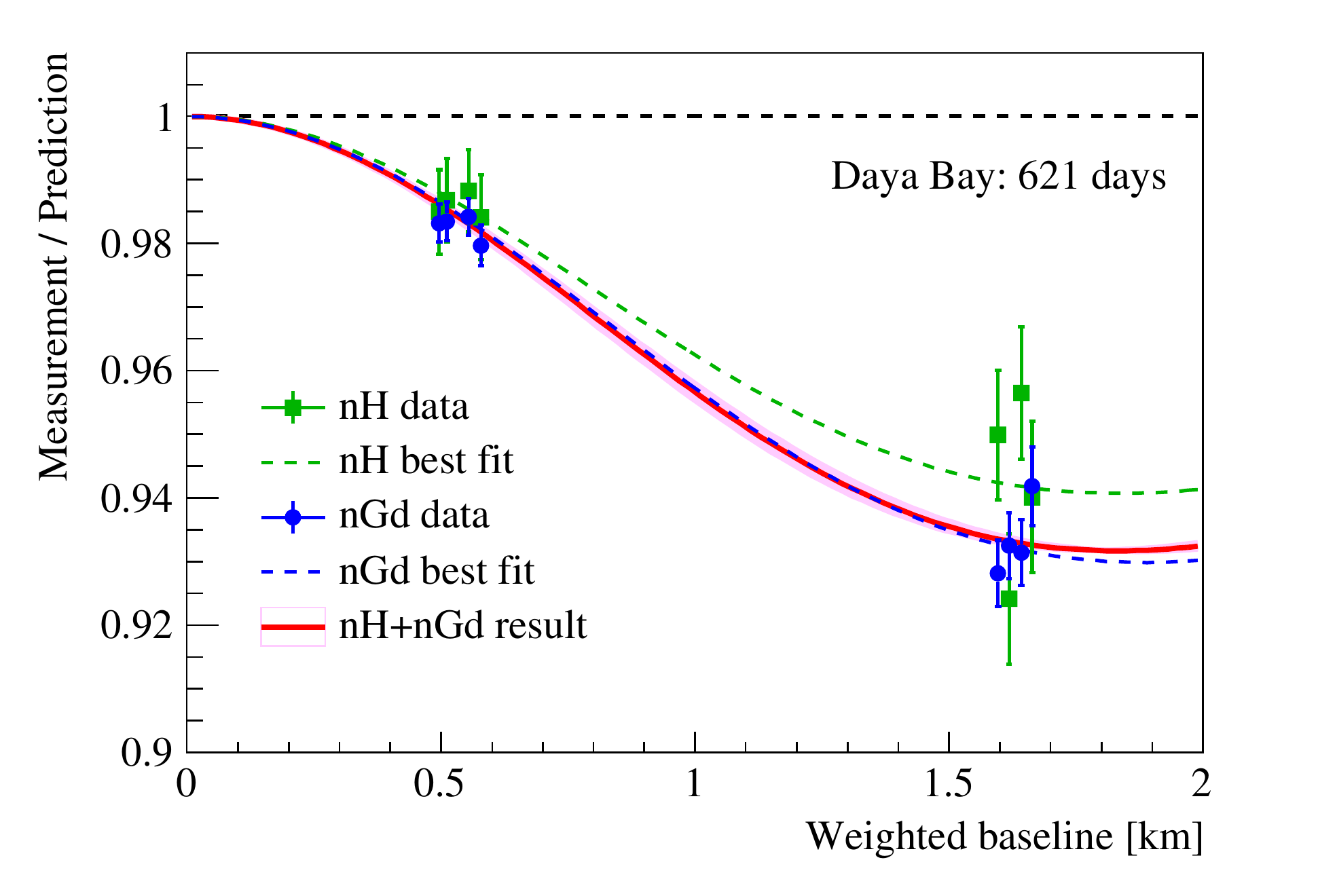}
\caption{Ratio of measured to predicted IBD rate in each detector assuming no oscillation {\it vs.} flux-weighted baseline.  Each detector is represented with a green square (blue circle) for the $n$H ($n$Gd) analysis.  Error bars include statistical, detector-related, and background uncertainties.  The dashed green (blue) curve represents the neutrino oscillation probability using the $n$H ($n$Gd) result for $\sin^{2}$2$\theta_{13}$ and the global fit value of $\Delta m_{32}^{2}$ (the $n$Gd result for $\Delta m^2_{ee}$~\cite{nGd8AD}).  The solid red curve represents the oscillation probability using the $n$H-$n$Gd combined result and $\Delta m_{32}^{2}$, and its magenta error band is from the uncertainty of $\Delta m_{32}^{2}$. 
The baselines of EH1-AD2 and EH2-AD2 are shifted by +20~m, and those of EH3-AD1, 2, 3, and 4 are shifted by -30, -10, +10, and +30~m, respectively, for visual clarity.}
\label{Fig:canv_th13_pred_measu_edit}
\end{center}
\end{figure}

Figure~\ref{Fig:canv_th13_pred_measu_edit} shows the ratio of the measured rate to the predicted rate assuming no oscillation, for each detector.  
The most recent $n$Gd result from Daya Bay~\cite{nGd8AD} is included for comparison.  
The 5.0\%-deficit of EH3 relative to the near halls given in Eq.~(\ref{eq:RatioDeficit}) is apparent.  For the $n$Gd-IBD analysis, this deficit was about 5.2\%, and the best-fit value was $\sin^2$2$\theta_{13}$ = 0.084.  
The red curve is the oscillation survival probability $P_\nu$ of Eq.~(\ref{eq:Psur}) with a value of $\sin^2$2$\theta_{13}$ = 0.082 from the combination of the $n$H- and $n$Gd-IBD analyses, which is described in the next section.  

The contributions of various quantities to the total uncertainty of $\sin^2$2$\theta_{13}$ ($\sigma_\mathrm{total}$) are listed in Table~\ref{tab:errorBudget}, where they are presented as fractions of $\sigma^2_\mathrm{total}$.  
The variance of a quantity was estimated as $\sigma^2_\mathrm{total}$ minus the square of the fit error when fixing the nuisance parameter of said quantity to its best-fit value.  
The sum of the fractions is not equal to 1 due to correlations.  
The statistical uncertainty is the largest individual component.  The second- and third-largest uncertainties are those of the coincidence-distance criterion and the delayed-energy criterion (see Table~\ref{tab:Eff} for the components of the detector contribution).  The reactor-uncorrelated uncertainty is reduced by a factor of 20, as in the relative expression of Eq.~(\ref{eq:RatioDeficit}).  
\begin{table}[!h]
\begin{tabular}{ l c c }\hline\hline
  &  Uncertainty Fraction (\%)    &  Correlation \\ \hline
Statistical  &  51.8  &  0  \\
Detector  &  39.2 &  0.07  \\
Reactor   &   4.2  &  1    \\
$^9$Li/$^8$He   &   4.4  &  0    \\
Accidental            &   0.4  &  0    \\
Fast neutron        &   0.3  &  0    \\
Am-C                    &   0.1  &  0.7    \\ \hline
Combined             &  100.4  &  0.02   \\ \hline \hline
\end{tabular}
\caption{Contributions of individual uncertainties to the total uncertainty of $\sin^2$2$\theta_{13}$.  See the text for details.  Detector uncertainties are characterized in Table~\ref{tab:Eff}.  The last column contains the estimated correlation coefficients between the $n$H- and $n$Gd-IBD analyses.}
\label{tab:errorBudget}
\end{table}

\subsection{\boldmath$n$\unboldmath H-\boldmath$n$\unboldmath Gd Combined Result}
\label{sec:nGdnHcombined}
The result for $\sin^{2}$2$\theta_{13}$ from the current analysis was combined with that from the most recent $n$Gd-IBD spectral analysis from Daya Bay~\cite{nGd8AD}.  The combination was performed both analytically and via a simultaneous fit of the $n$Gd-IBD and $n$H-IBD data sets.  Correlations between the two analyses were estimated for efficiencies, backgrounds, and reactor-related quantities.  

The correlation coefficients of the various uncertainty components are listed in Tables~\ref{tab:Eff} and \ref{tab:errorBudget}.  Reactor-related uncertainties are fully correlated and statistical uncertainties are uncorrelated.  The correlation of quantities with negligible uncertainty, such as DAQ time and muon-veto efficiency, had negligible impact. 
The correlation coefficients of the detector-related quantities are described in Section~\ref{sec:EffSum} and listed in Table~\ref{tab:Eff}. 
The accidental backgrounds were treated as uncorrelated because of the distinct methods and event samples used in the $n$H- and $n$Gd-IBD analyses.  
The Am-C background was estimated to have a correlation coefficient of 0.7, 
while the other backgrounds were uncorrelated (see Section~\ref{sec:CorrBkg}).  

The procedure to analytically combine the analyses is the same as that used for the previous combination~\cite{DYB_nH}.  Updated values for backgrounds, efficiencies, and the fraction of uncertainty due to statistics were taken from Ref.~\cite{nGd8AD}, for the $n$Gd-IBD analysis.  For the $n$H-IBD analysis, these values are listed in Tables~\ref{tab:IBDsummary}, \ref{tab:Eff}, and \ref{tab:errorBudget}, respectively.  

Using the correlation coefficients presented in this article, these values give an overall correlation coefficient of 0.02 between the two analyses, indicating essentially independent determinations of $\sin^{2}$2$\theta_{13}$.  Though the correlation will increase as the fraction of statistical uncertainty decreases, this value is smaller than the previous correlation coefficient of 0.05~\cite{DYB_nH} primarily because of the distinct estimation of the $n$H-$^{9}$Li background and the significant reductions in the systematic uncertainties of the Am-C backgrounds for both analyses.  

With the $n$Gd-IBD result of $\sin^{2}$2$\theta_{13}$ = \nGdResult~and the $n$H-IBD result of \nHresult, both the analytical calculation and simultaneous fit resulted in 
\begin{equation}
\begin{aligned}
\sin^{2}2\theta_{13}=\combin , 
\end{aligned}
\end{equation}
which is an 8\% improvement in precision.

\subsection{Independent Analysis}

The present $n$H-IBD analysis was cross-checked with an independent analysis based on a different analysis framework~\cite{LAF}.  
IBD candidates were independently selected using the same criteria (see Table~\ref{tab:criteria}) and the backgrounds and muon-veto efficiencies were independently evaluated.  
Using the $\chi^{2}$ in Eq.~(\ref{eq:Chi2Definition}), 
the best-fit value was $\sin^{2}$2$\theta_{13}=0.071 \pm 0.011$, 
with a $\chi_\mathrm{min}^{2}$ per degree of freedom of 6.4/6.

\section{Discussion}
\label{sec:Future}
The precision to which $\theta_{13}$ is determined is crucial to constraining the leptonic {\it CP} phase $\delta$~\cite{T2K_CPvsT13, NOvA, MINOS, LBNE}.  
The $n$H-IBD analysis in this article provides an independent determination of $\sin^2$2$\theta_{13}$ and improves the overall precision of $\theta_{13}$. 

Given that the uncertainty of the $n$H-IBD result is dominated by the systematic uncertainties of the delayed-energy and coincidence-distance criteria, improved precision is foreseen by reducing the uncertainties of the distance criterion with increased statistics, and the delayed-energy criterion with an optimization of the selection.  
In addition, improved precision will be achieved with a spectral analysis of the prompt-energy spectrum, which is underway.  This will also provide a new determination of the mass-squared difference $\Delta m^2_{32}$.  

The analysis of $n$H-IBDs has helped to maximize the fiducial volume of the ADs to supernova neutrinos~\cite{DYB_SN}.  
It should also provide an opportunity to reduce the dominant uncertainty of detection efficiency in the measurement of reactor antineutrino flux~\cite{DYB_reactor}, given the lesser sensitivity of the $n$H-IBD analysis to neutron spill-in/out effects.  
Furthermore, the data-driven techniques developed to study the accidental background and the IBD selection criteria may be useful for other experiments that use or plan to use $n$H-IBDs, such as JUNO~\cite{JUNO}, RENO-50~\cite{RENO50}, and LENA~\cite{LENA}.

\section{Conclusion}
A sample of about 780000 $n$H-IBDs was obtained with the 6-AD and full 8-AD configurations of the Daya Bay experiment and was used to compare the number of reactor antineutrinos at far and near halls, yielding a new independent determination of sin$^{2}2\theta_{13}$ = \nHresult.  The uncertainty is reduced by 40\% compared with the previous $n$H-IBD result primarily because of the factor of 3.6 increase in statistics, but also because of the 15\% and 30\% reductions in the uncertainties of the IBD selection efficiency and backgrounds, respectively.  
The new result is consistent with that from the $n$Gd-IBD analysis from Daya Bay, providing a valuable confirmation of the $n$Gd-IBD result.  Combining the $n$H- and $n$Gd-IBD results provides a new improved determination of 
sin$^{2}2\theta_{13}$~= $\combin$.

\section{Acknowledgments}

The Daya Bay Experiment is supported in part by 
the Ministry of Science and Technology of China,
the United States Department of Energy,
the Chinese Academy of Sciences,
the CAS Center for Excellence in Particle Physics,
the National Natural Science Foundation of China,
the Guangdong provincial government,
the Shenzhen municipal government,
the China General Nuclear Power Group,
the Key Laboratory of Particle \& Radiation Imaging (Tsinghua University), Ministry of Education, 
the Key Laboratory of Particle Physics and Particle Irradiation (Shandong University), Ministry of Education,
the Research Grants Council of the Hong Kong Special Administrative Region of China,
the MOST fund support from Taiwan,
the U.S. National Science Foundation,
the Ministry of Education, Youth and Sports of the Czech Republic,
the Joint Institute of Nuclear Research in Dubna, Russia,
the NSFC-RFBR joint research program,
the National Commission for Scientific and Technological Research of Chile, and 
the Tsinghua University Initiative Scientific Research Program.
We acknowledge Yellow River Engineering Consulting Co., Ltd.\ and China Railway 15th Bureau Group Co., Ltd.\ for building the underground laboratory.
We are grateful for the ongoing cooperation from the China Guangdong Nuclear Power Group and China Light~\&~Power Company.

\end{document}